\documentclass[aps,pre,twocolumn,a4paper,10pt,notitlepage,footinbib,superscriptaddress,longbibliography]{revtex4-1}
\usepackage[english]{babel}
\usepackage{lmodern}
\usepackage[latin9]{inputenc}
\usepackage{endnotes}
\usepackage{amssymb,amsmath,amsfonts}
\usepackage{textcomp}
\usepackage{braket}
\usepackage{graphicx}
\usepackage{dcolumn}
\usepackage{bm}
\usepackage{xcolor}
\usepackage{soul}
\usepackage{hyperref}
\usepackage{url}

\begin{document}

\title{Lattice Boltzmann simulations for soft flowing matter} 

\author{A. Tiribocchi}
\affiliation{Istituto per le Applicazioni del Calcolo, Consiglio Nazionale delle Ricerche, via dei Taurini 19, Roma, 00185, Italy}
\affiliation{INFN "Tor Vergata" Via della Ricerca Scientifica 1, 00133 Roma, Italy}
\author{M. Durve}
\affiliation{Center for Life Nano- \& Neuro-Science, Fondazione Istituto Italiano di Tecnologia, viale Regina Elena 295, 00161 Rome, Italy}
\author{M. Lauricella}
\affiliation{Istituto per le Applicazioni del Calcolo, Consiglio Nazionale delle Ricerche, via dei Taurini 19, Roma, 00185, Italy}
\author{A. Montessori}
\affiliation{Department of Civil, Computer Science and Aeronautical Technologies Engineering, Roma Tre University, via Vito
Volterra 62, Rome, 00146, Italy}
\author{J.-M. Tucny}
\affiliation{Center for Life Nano- \& Neuro-Science, Fondazione Istituto Italiano di Tecnologia, viale Regina Elena 295, 00161 Rome, Italy}
\affiliation{Department of Civil, Computer Science and Aeronautical Technologies Engineering, Roma Tre University, via Vito
Volterra 62, Rome, 00146, Italy}
\author{S. Succi}
\affiliation{Istituto per le Applicazioni del Calcolo, Consiglio Nazionale delle Ricerche, via dei Taurini 19, Roma, 00185, Italy}
\affiliation{Center for Life Nano- \& Neuro-Science, Fondazione Istituto Italiano di Tecnologia, viale Regina Elena 295, 00161 Rome, Italy}
\affiliation{Department of Physics, Harvard University, 17 Oxford St, Cambridge, MA 02138, United States}

\begin{abstract}
Over the last decade, the Lattice Boltzmann method has found major scope for the simulation of a large spectrum of problems in soft matter, from multiphase and multi-component microfluidic flows, to foams, emulsions, colloidal flows, to name but a few.
Crucial to many such applications is the role of supramolecular interactions which occur whenever mesoscale structures, such as bubbles or droplets, come in close contact, say of the order of tens of nanometers. Regardless of their specific physico-chemical origin, such near-contact interactions are vital to preserve the coherence of the mesoscale structures against coalescence phenomena promoted by capillarity and surface tension, hence the need of including them in Lattice Boltzmann schemes.
Strictly speaking, this entails a complex multiscale problem, covering about six spatial
decades, from centimeters down to tens of nanometers, and almost twice as many in time.
Such a multiscale problem can hardly be taken by a single computational method, hence the
need for coarse-grained models for the near-contact interactions.
In this review, we shall discuss such coarse-grained models and illustrate their application
to a variety of soft flowing matter problems, such as soft flowing crystals, strongly
confined dense emulsions, flowing hierarchical emulsions, soft granular flows, as 
well as the transmigration of active droplets across constrictions.
Finally, we conclude with a few considerations on future developments in the direction of 
quantum-nanofluidics, machine learning, and quantum computing for soft flows applications.
\end{abstract}

\maketitle


\tableofcontents

\section{Introduction}

\begin{figure*}[htbp]
\includegraphics[width=1.0\textwidth]{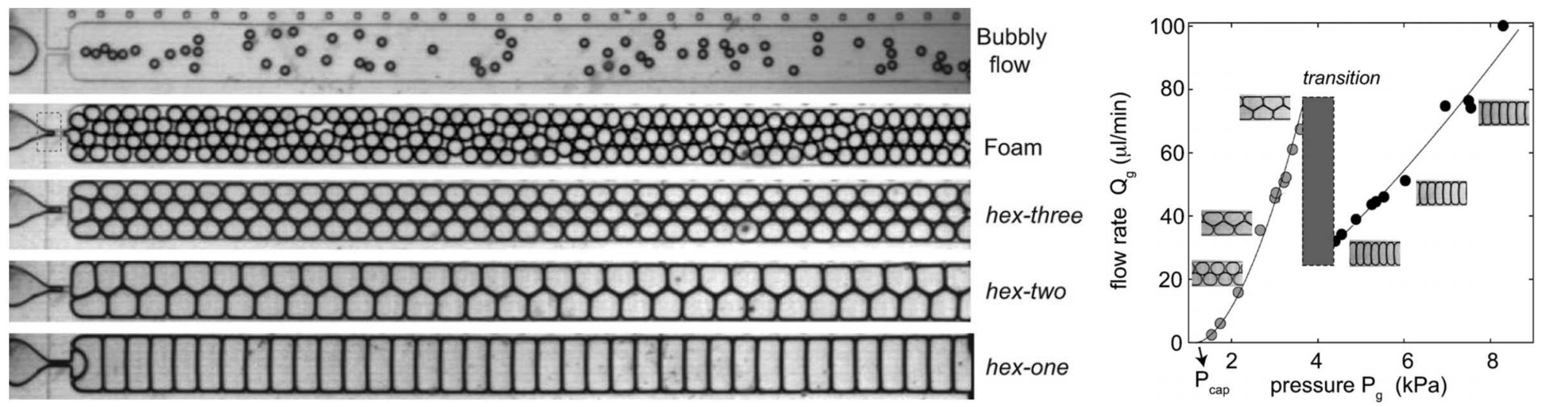}
\caption{Droplet-based states of matter.  Top left panel: a dilute "gas" of droplets at low oil/water ratios.
Second from top left: a "liquid-like" collection of disordered 
droplets in near-contact. Third from top left: a "solid-like" configuration of droplets 
with honeycomb topology. Fourth from top left: another "solid-like" configuration of droplets 
with hexagonal symmetry (flowing crystal).
Bottom: a one-dimensional flowing crystal.
Right panel: Flow curve of a soft flowing foam. The figures are reproduced with permission from Ref.\cite{marmottant_soft}.}
\label{marmottant_fig}
\end{figure*}

Soft matter lives at the intersection between the three fundamental states 
of matter: gas, liquid, and solid. Much of its fascination draws from the  fact 
that such intersection is all but a linear superposition of the three
fundamental states, and supports instead genuinely new mechanical and rheological behavior, with important consequences for many applications in 
science and engineering 
\cite{piazza,masao_doi,fern-nieves,lavrentovich}. 
The demise of linear superposition can be traced to the nonlinear interactions associated
with the configurational degrees of freedom: for instance, foams are typically disordered assemblies of a gas (vapor) phase into a liquid water matrix and 
even though both phases are Newtonian, i.e. they respond to external
load in direct proportion to the load intensity, once combined together, they
develop a nonlinear and often non-local response.
Given its paramount role in modern science and engineering, the quantitative study of 
soft matter has witnessed a burgeoning growth in the last decades \cite{nagel,yodh,hamley}.
This interest is further accrued by considering situations in which soft materials 
flow through confined geometries, since this gives rise to entirely new regimes 
and driven non-equilibrium steady states. 
In this review, we shall be concerned mostly with soft flowing systems 
characterized by a dispersed phase (droplets or bubbles) into a 
continuum matrix, say oil droplets in a water continuum, under strong
geometrical confinement.
Such specific systems have witnessed major development in 
the last decades mostly on account of
progresses in experimental microfluidics,  whereby one can 
control the spatial configurations of the
droplets by simple experimental handles,  typically 
the ratio of oil to water mass flow and the 
geometrical setup of the microfluidic channel.
By properly tuning such parameters,  one can seamlessly 
move from dilute systems (droplet gas) to denser 
disordered states (droplet liquids) and finally even denser ordered
ones, i.e.  droplet solids with various topologies 
\cite{marmottant_prl,marmottant_soft} (see Fig.\ref{marmottant_fig}).
Even though one can pass from one phase to another by simply 
increasing the oil/water mass influx, the corresponding rheology does 
not respond linearly and sometimes not even smoothly.
For instance, instabilities may arise whereby an increase in pressure 
no longer results in a corresponding increase in mass
flow, but triggers instead a collapse of the mass flow, due to the inability of the droplets to undergo additional deformations 
to absorb the effects of an increasing pressure gradient. 
A new mechanism is needed to adjust the increasing pressure drive, a
mechanism which materializes in the form of topological 
rearrangements of the droplet configurations such as to match interfacial
dissipation at both external and internal boundaries with the pressure load.

\begin{figure*}[htbp]
\centering
\includegraphics[width=1.0\textwidth]{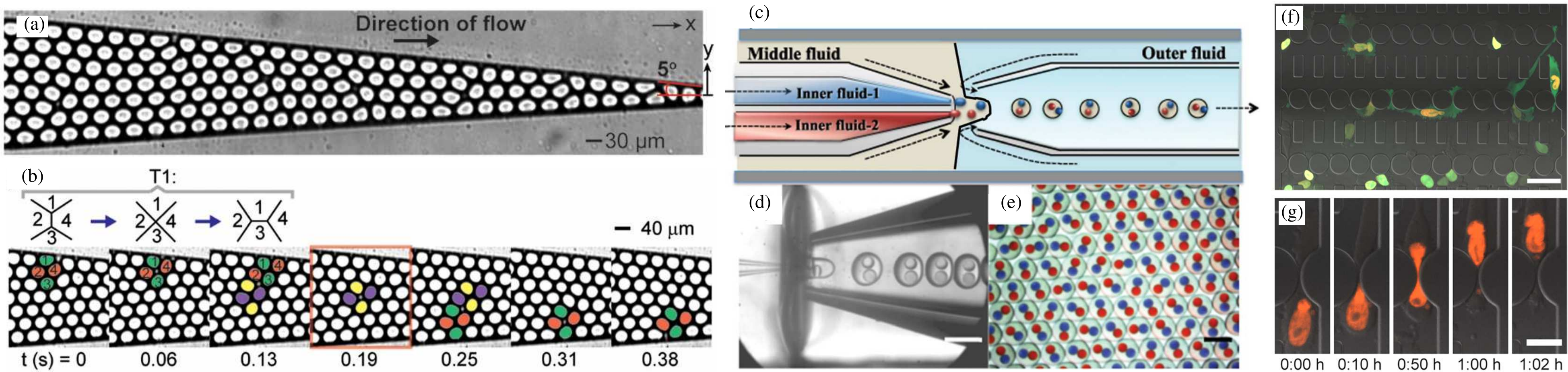}
\caption{Left panel (a): A concentrated emulsion flowing in a tapered channel. (b): A cascade of T1 events (equivalent to a dislocation) involving
converging and diverging droplets. 
From $t=0$ to $t=0.13$ red droplets diverge and green ones converge; from $t=0.13$ to $t=0.25$ purple droplets diverge and yellow ones converge; from $t=0.25$ to $t=0.38$ red droplets diverge and green one converge. The figures are reproduced with permission from Ref.\cite{tang}. (c) Sketch of a microfluidic device used for the production of multiple emulsions. (d)-(e) Experimental images of monodisperse double emulsions with two different inner drops. The figures are reproduced with permission from Ref.\cite{adams}. (f) Microfluidic chip modeling a sequence of constrictions typical of a porous tissue. (g) Close-up view of a cell nucleus crossing a narrow constriction. The figures are reproduced with permission from Ref.\cite{elaqua_plosone}.}
\label{weitz_fig}
\end{figure*}

Many different experiments can be performed with these droplet-based
states of matter, by changing the geometrical and physical setup.
For instance, Tang and collaborators studied the intriguing 
non-equilibrium pattern formation phenomena that occur in dense 
emulsions flowing in tapered channels \cite{tang} (see Fig.\ref{weitz_fig}a,b).
These authors report an unexpected order in the flow of a 
concentrated emulsion in a tapered microfluidic channel, in the form of orchestrated sequences of dislocation nucleation and migration events giving rise to a highly ordered deformation mode. 
This suggests that nanocrystals could be made to deform 
in a more controlled manner than previously expected, and also hints
at possible novel flow control and mixing strategies in droplet microfluidics.
On a similar although distinct vein, using a 
microfluidic velocimetry technique, Goyon et al. \cite{goyon}
characterized the flow of thin layers of concentrated emulsions confined in gaps 
of different thicknesses by surfaces of different roughness. 
These experiments show clear evidence of finite size effects in the flow behavior, 
which can be interpreted in terms of a non-local viscosity 
of practical relevance for applications involving thin layers, e.g. chemical
or geometrical coatings. 
Yet another important area of experimental microfluidics concerns
the realization of hierarchical emulsions (droplets within droplets).
For instance, double emulsions are highly structured 
fluids consisting of emulsion drops that contain smaller droplets inside. 
Although double emulsions are potentially of 
commercial value, traditional fabrication by means 
of two emulsification steps leads to very ill-controlled structuring. 
Using a microcapillary device, experimentalists managed to fabricate
double emulsions that contained multiple internal droplets in a core-shell geometry.  
By manipulating the properties of the fluid that makes 
up the shell, it is possible to manufacture encapsulation structures 
with a high degree of control and reproducibility
\cite{utada,datta} (see Fig.\ref{weitz_fig}c,d,e). 
These microfluidic technologies have also opened intriguing avenues in microbiology,  
for the detection of pathogens, antibiotic testing, cell migration, and motility \cite{garstecki_rev,bray_book}. 

The last two, in particular, are crucial to a number of physiological and pathological processes, such as wound healing, embryonic development, and cancer metastasis. 
In many of such instances, cells are found to migrate through 
highly confined environments, such as dense tissues and interstices, dramatically 
impacting their morphology and mechanics. 
For instance, in Refs.\cite{elaqua_plosone}, the authors studied the dynamics 
of a cancer cell crossing a narrow constriction (whose design is inspired 
by physiologically-relevant conditions) 
and quantified shape deformations as the cell migrates through tiny pores 
(see Fig.\ref{weitz_fig}f,g). Similar dynamic behaviors have been observed in Refs.\cite{davidson1,davidson2} using fibroblast cells moving in confined environments.  
In these respects, double emulsions may serve as a simplified model to study 
biological cells, where the innermost droplet would be a representation of a nucleus 
and the layer would mimic the cell cortex containing  motor proteins, 
such as the actomyosin complex \cite{weitz_cell1,weitz_cell2}.

The overall rheology of these complex states of soft flowing matter emerges 
from the competition/cooperation of multiple concurrent mechanisms 
acting across a broad spectrum of scales in space and time.
Typically these encompass supramolecular interactions between near-contact 
surfaces at a scale of a few nanometers, all the way up to 
the overall size of the device, of the order of centimeters, 
spanning six decades in space and nearly twice as many in time in the process.
Owing to the highly complex and nonlinear coupling between such mechanisms, analytical 
methods can only supply qualitative information, hence 
falling short of providing the degree of accuracy required
by engineering design. 
At the same time, experimental methods have limited 
accessibility to the spectrum of the 
scales in action, such as the finest details of the flow 
configuration in the interstitial regions between interfaces or inside 
the droplets and bubbles.     

Under such conditions, computer simulations emerge as an invaluable tool for gathering information that is otherwise inaccessible through theoretical approaches or experimental methods.
Microfluidic flows are typically simulated using the principles of continuum fluid dynamics, as it is established that within the bulk flow, distant from sources of heterogeneities, the continuum assumptions remain valid down to nanometric scales \cite{bocquet1}.
In the vicinity of external surfaces, such as solid boundaries, or in the interstitial region
between approaching interfaces, the continuum assumption may either go 
under question on sheer physical grounds, or simply become computationally unwieldy
on account of the high surface/volume ratios characterizing microfluidic multiphase or multicomponent
flows, such as the ones discussed in this review. 

Until some two decades ago, the standard and only option out of the 
continuum was the resort to atomistic simulation, e.g. Molecular Dynamics (MD). 
Unfortunately, notwithstanding major sustained progress in the field, MD still falls 
short of reaching the scales of interest for microfluidics, both in space
and more so in time. As a figure of reference, even a multi-billion MD simulations can
only cover three spatial decades, say from 1 nanometer to 1 micrometer, which is clearly far below the size of
most experimental devices. Multiscale MD-Fluid procedures have been developed in the last decades but they are still laborious in day-to-day operations.
In the last three decades a third, alternative avenue, based on the intermediate level described
by kinetic theory, has generated a number of very appealing and 
useful {\it mesoscale} computational methods, either based on suitably 
discretized lattice versions of 
Boltzmann kinetic theory \cite{succi2018lattice} or coarse-grained versions of 
stochastic particle dynamics, such as Dissipative Particle Dynamics \cite{Warren} and 
Stochastic Rotation methods \cite{gompper}. 
Even though each of these methods comes with its 
strengths and weaknesses, we believe it
is fair to state that the Lattice Boltzmann (LB) stands out as 
the one featuring an especially high degree
of physical flexibility and computational efficiency 
across the full spectrum of scales of motion. 
This is why this review is focused on recent 
developments of the LB method
specifically aimed at capturing the complexity of soft flowing 
matter states, such as the ones described above. 

The review is organized as follows.
In Section II we provide a discussion of the basic physical scenario addressed here, namely 
soft flowing matter in microfluidic devices.
In Section III we review the basic ideas behind the LB method for ideal and non-ideal
fluids, including recent works accounting for near-contact interactions at nanometric scales, which are
key for the description of dense systems under strong geometric confinement.
In Section IV we provide an account of recent implementations of the above 
schemes on massively parallel computers, mostly aimed at minimizing the 
costs of memory access, an increasingly pressing topic for large-scale simulations.
In Section V we discuss a list of selected applications, namely soft flowing crystals,
soft granular media, dense emulsions with heat exchange, hierarchical emulsions under confinement,
migration of active droplets through geometrical constrictions and flow through deep-sea
sponges. The choice of such applications, which is inevitably partial and subjective, simply
serves the purpose of conveying an idea of the broad spectrum of applications that can be
handled by comparatively minor variants of the basic LB scheme.
Finally, in Section VI we conclude with an outlook of potential future directions, namely the extension of current LB methodology to nanofluids with
quantum interfacial effects, the use of machine learning to 
enhance LB simulations, and finally a few considerations on the prospects of
quantum computing for soft flowing matter.

\section{Basic physics of microfluidic soft flowing matter}\label{basic_physics}

\begin{figure}[htbp]
\centering
\includegraphics[width=0.9\linewidth]{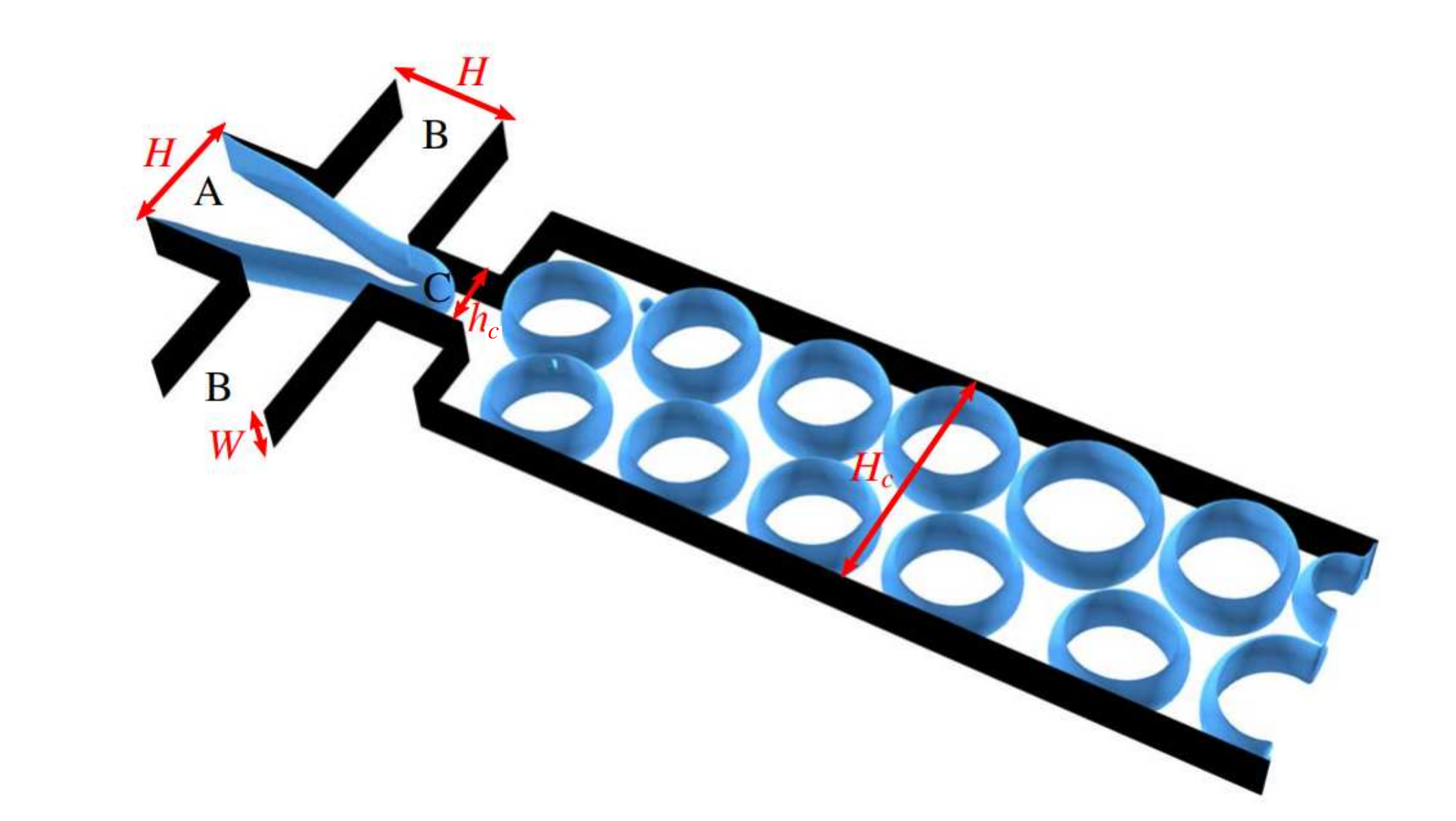}
\caption{Typical structure of flow focuser and emulsion confined within. The device consists of three inlet channels (of height $H\simeq 200\mu$m and width $W\simeq 100\mu$m) supplying dispersed (A) and continuous phases (B), a narrower orifice (C) (of height $h_c\simeq 100\mu$m) and large outlet channel ($H_c\simeq 500\mu$m). The droplets emerging from the outlet of the microfluidic channel 
are evaporated away while the continuum phase is solidified by 
shining UV light into the fluid, so that one is left with a solid substrate with
a regular sequence of holes (scaffold).
The challenge is to preserve the stability of the soft flowing crystal 
at increasing speed, so as to maximize the production of scaffolding material. 
The figure is reproduced with permission from Ref.\cite{montessori_jfm}.}
\label{fig_flow}
\end{figure}

We shall refer to a confined microfluidic flow composed of two immiscible fluids A and B, say water and oil for the sake of concreteness. 
An example of the system under study is shown in Fig.\ref{fig_flow}, where a collection 
of monodisperse droplets (generated by the breaking of the jet of the 
dispersed phase (A) by the flow of the continuous one (B) in the orifice) 
flows in the exit channel of a flow focuser. 

The physics of this system is usually described by the standard equations of 
continuum mechanics, i.e. the Navier-Stokes equation
\begin{equation}
\label{NSE}
\rho D_t {\bf u} \equiv \rho\left(\partial_t {\bf u} + {\bf u} \nabla {\bf u}\right) = 
{\bf F}_{pre} + {\bf F}_{dis} + {\bf F}_{cap} + {\bf F}_{nc} 
\end{equation}
supplemented with the incompressibility constraint
\begin{equation}
\label{INC}
\nabla \cdot {\bf u} = 0. 
\end{equation}
In the above ${\bf u}$ is the barycentric flow speed, $p$ the pressure and $\nu$ the kinematic viscosity. The total density $\rho$ is set to one for convenience, i.e.
$\rho=\rho_A+\rho_B=1$.

The momentum equation contains four types of forces per unit volume. 
The first three, which represent the standard treatment, are 
the pressure gradient
\begin{equation*}
{\bf F}_{pre}  = -\nabla p, 
\end{equation*}
the dissipative force 
\begin{equation*}
{\bf F}_{dis} = \rho\nu \nabla {\bf u}, 
\end{equation*}
and the capillary force 
\begin{equation*}
{\bf F}_{cap} = \sigma \nabla 
\cdot\left[ \bigg(\nabla^2 \phi - (\nabla \phi)^2\bigg) {\bf I} -\frac{1}{2} \nabla \phi \nabla \phi\right],
\end{equation*}
where ${\bf I}$ is the unit matrix, $\sigma$ is the surface tension and 
\begin{equation}\label{order_param}
\phi = \frac{\rho_A - \rho_B}{\rho_A + \rho_B}
\end{equation}
is a phase field (the order parameter) ranging between $-1$ and $+1$ telling the two fluids apart: namely $\phi=1$ in the A phase, $\phi=-1$ in the B phase,  $\phi=0$ at
the A/B interface.  

The fourth term calls for some specific comments.
It is intended to model "near-contact" interactions, which arise whenever two interfaces 
come to a distance within the range of supramolecular forces, say around 10 nanometers.
The physical origin of such near-contact forces may range from steric and depletion interactions to even Casimir-like interactions \cite{derjaguin,verwey,lamoreaux}. However, in our case, we shall be agnostic regarding their physical basis and focus instead on the task of efficiently incorporating them into the LB solvers for soft flowing matter.
Prior to turning to the numerics, a brief survey of basic physics is in order. 

It is often customary in microfluidics to define a reduced set of dimensionless groups 
accounting for the mesoscale physics of the system under scrutiny. 
The first one is the Reynolds number which measures the ratio between inertial and viscous forces and is defined as
\begin{equation}
Re=uH_c/\nu.
\end{equation}
Since we are dealing with slow flows whose speeds are of the order of $u=1$ mm/s,  inertial effects are usually small under strongly confined geometries, say 1 mm in the crossflow direction $H_c$, thus the  
Reynolds number is of the order of $Re \sim 1$. 
The next,  possibly most important dimensionless group, is the capillary number
\begin{equation}
Ca = F_{vis}/F_{cap} = u \mu/\sigma,
\end{equation}
where $\mu=\rho \nu$ is the dynamic viscosity.
Typical values in microfluidic experiments are $Ca \sim 10^{-4}$, indicating that 
capillary forces are three-four orders above the dissipative ones, which are in 
turn comparable with inertial forces.
Another way of rephrasing this is to say that the capillary speed $u_{cap} = \sigma/\mu$ is 
three-four orders of magnitude larger than the flow speed. 
Capillarity promotes shape changes towards sphericity and ultimately coalescence, which is definitely an unwanted effect for the design of droplet-based materials and applications in general. 
To this purpose, microfluidic experiments generally cater for a third 
dissolved species, namely surfactants, which prevent nearby interfaces from merging.
The same role is played by repulsive near-contact interactions.

\begin{figure}[h]
\centering
\includegraphics[width=1.0\linewidth]{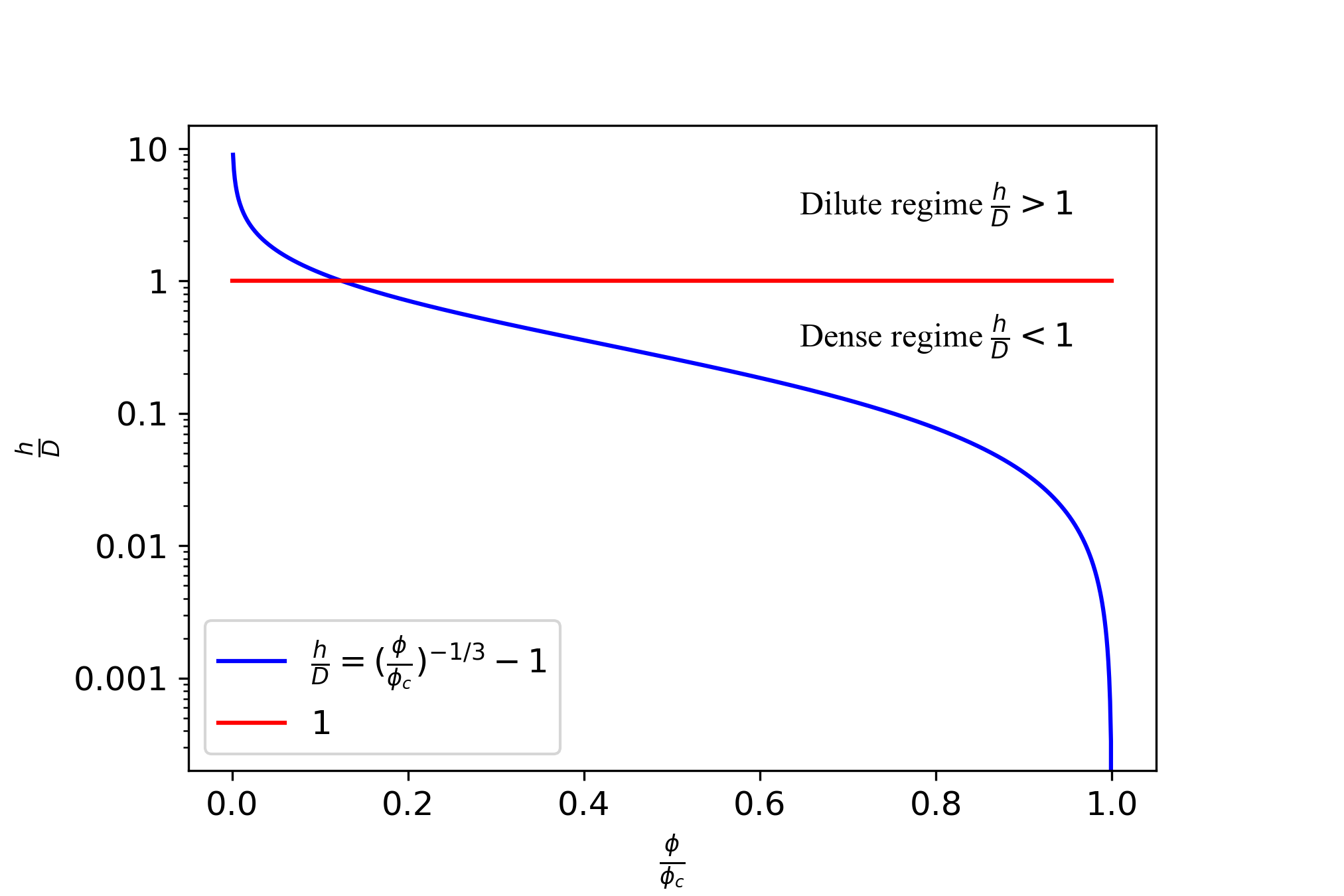}
\caption{
The $h/D$ parameter as a function of the relative volume fraction $\phi/\phi_c$. 
If $\phi/\phi_c\lesssim 0.2$, the mixture is in a dilute regime, thus near-contact 
forces can be neglected, whereas if $\phi/\phi_c> 0.2$, these forces 
significantly affect the mechanical properties of the system.}\label{nc_fig}
\end{figure}

The competition between coalescence-promoting interactions (capillarity) and coalescence-frustrating ones (near-contact repulsion) lies at the heart of the complexity of these states of confined soft flowing matter.
Such competition is measured by a nameless number that we dub 
$\mathcal{N}$ accordingly,  defined as the ratio between near-contact and capillary forces.
On dimensional grounds, we write the corresponding forces per unit volume
as $$F_{cap} \sim \sigma/D^2$$ and $$F_{nc} = E_{nc}/h^4,$$ where $D$ is the droplet diameter, $E_{nc}$ is a typical energy scale of near-contact interactions and $h$ is the film thickness separating the two interfaces.  
As a result, we obtain
\begin{equation}
\mathcal{N} = \frac{E_{nc}}{\sigma D^2} \frac{D^4}{h^4}.
\end{equation}
By taking $D \sim 100$ $\mu$m, $h \sim 10$ nanometers and $E_{nc} \sim kT$, we obtain
${\cal N} \sim  \frac{0.25 \cdot 10^{-20}} {70 \cdot 10^{-3} \cdot 10^{-16}}\times 10^{8}  \sim 10^3$,
indicating strong dominance of near-contact forces.
However,  a more realistic estimate is $E_{nc}/kT \sim 10^{ -3}$,  leading to  ${\cal N} \sim O(1)$ at $h = 10$ nanometers. 
The above relation can also be written as
\begin{equation}
\mathcal{N} = \alpha \frac{k_BT}{\sigma D^2} \left(\frac{h}{D}\right)^{-4},   
\end{equation}
where we have defined $\alpha = E_{nc}/k_BT$.
By choosing a reference value of $\sigma= 0.1$ N/m and $D= 100$ $\mu$m, we obtain
\begin{equation}
\mathcal{N} = 4 \times 10^{-12}\alpha \left(\frac{h}{D}\right)^{-4}.    
\end{equation}
Hence, the condition for the prevalence of near-contact interactions over capillary ones reads as
\begin{equation}
\frac{h}{D} < (4\alpha)^{1/4}10^{-3}. 
\end{equation}
Due to the weak dependence on $\alpha \simeq 10^{-3}$ \cite{karniadakis2006,wenbing}, the above formula shows that 
near-contact forces prevail below $h/D\sim 10^{-4}$. 
For the case in point, this occurs for $h$ below 10 nanometers (since 
$D\sim 100\mu m$).
It is also of interest to observe that such near-contact interactions require dense
regimes pretty close to the maximum packing fraction $\phi_c$ (see Fig.\ref{nc_fig}).
Based on the expression
\begin{equation}
\frac{h}{D} = \left(\frac{\phi_c}{\phi}\right)^{1/3} -1,  
\end{equation}
one concludes that $h/D \sim 10^{-4}$ implies $\phi/\phi_c \sim 1 - 10^{-4}/3$ 
which is indeed extremely close to the maximum packing fraction, a regime in which
the material is basically a soft flowing solid.
It should be observed that while such values might be much smaller than the average, they
can nonetheless occur as a result of strong dynamic fluctuations of the flowing 
multi-droplet configuration.

\subsection{The multiscale scenario and Extended Universality}

Since near-contact forces prevail at scales roughly below ten nanometers,  predicting the
rheological behavior at scales of experimental interest, say one centimeter,  presents 
a complex nonlinear and nonlocal multiscale problem, spanning six decades in space
and nearly twice as many in time.
Whence the paramount role of advanced computational methods.
Many methods are available in principle, but all of them
face severe difficulties in handling the simulation of 
such a wide range of scales.
LB methods are no exception in this respect. 
However, as anticipated earlier on, in this review we shall focus 
on LB mostly on account of its flexibility to incorporate non-equilibrium physics 
beyond hydrodynamics in realistically complex geometries, including 
coarse-grained interactions representing the effects of the 
unresolved scales on the resolved ones.

\begin{figure}
\centering
\includegraphics[width=1.0\linewidth]{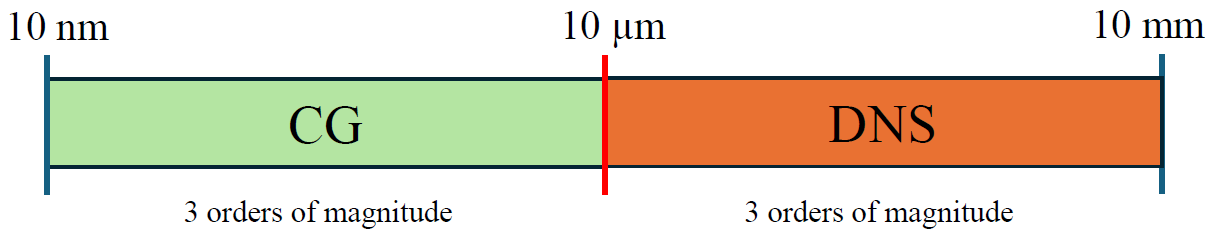}
\caption{A schematic view of the typical scales accessible to 
coarse-grained (CG) and direct numerical simulations (DNS). 
The figure portrays a 3:3 balance between DNS and CG, which 
is the one adopted in most simulations discussed in this review.
As computer power goes up, one can progressively extend the
DNS bandwidth and shrink the CG one accordingly.
}\label{fig5}
\end{figure}

The usual computational strategy consists of splitting the six
decades into a range of scales that are suitable to direct numerical 
simulation and a range of subgrid scales that need to be modeled.
Current LB simulations can easily handle one billion lattice
sites, thus covering three spatial decades, say from centimeters to tens of 
microns. Assuming molecular specificity to set in
at about ten nanometer scale (which we dub {\it supra-molecular}
scale), the remaining three decades can be covered either by grid-refinement 
or handled by coarse-grained models (see the sketchy Fig.\ref{fig5}).
The former strategy requires about ten levels of grid 
refinement ($2^{10} = 1024$) which involves a very substantial
implementation effort, especially to secure a sustainable
parallel efficiency. The latter is far simpler from the 
computational standpoint, but it is subject to issues of physical
inaccuracy. Fortunately, at least for the (broad) spectrum of
applications discussed in this work, it was found that 
the latter route provides a satisfactory description
of the large-scale rheological behavior.

In hindsight, this is attributed to a sort of {\it Extended
Universality} (EU) of the physics in point, meaning by
this that the rheology of the soft matter systems 
under investigation is to a large extent insensitive to
the details of the supra-molecular scales.
This can be formally evaluated by looking, for example, at the Cahn number $Cn=w/D$ (where $w$ is the width of droplet interface) which controls the interface dynamics.   
For a droplet with diameter $D \simeq 100$ $\mu$m and a typical width $w \simeq 1$ nm, the physical Cahn number is $Cn \simeq 10^{-5}$. Since in  LB simulations the interface
is of the order of a few lattice spacings, hence a few microns, the LB Cahn number is around $Cn_{LB}\sim 10^{-2}$, much larger than its physical counterpart.
However, evidence in the literature (see Ref.\cite{malgaletti}) shows that the interface dynamics is largely insensitive to the specific value of the Cahn number, as long as this number is well below $1$. In other words, the Cahn number is a "sloppy" parameter. Hence, even though the prime scope of the LB simulations is the global rheology, we expect that they can provide a reasonable account also of the small-scale interface dynamics.

A related question regards the validity of our NCI treatment in the near-interface region
where $w/h \sim 1$. Here we observe that this ratio takes comparable values both in actual
reality and in the LB simulations, since both $w$ and $h$ are upscaled by three orders of
magnitude. This is where our coarse-graining procedure makes a strong call to EU, namely that by imposing
values of $w/h$ and ${\cal N}$ similar to the experimental ones, we expect the effects of the NCI physics on the rheological behavior to be
captured to a satisfactory degree of accuracy.

More precisely, the essential effects of the near-contact interactions on large-scale rheology are largely dictated by
their intensity relative to capillary forces, namely
the dimensionless near-contact number $\mathcal{N}$ defined in the previous section.
Values of $\mathcal{N} \sim 0.1$ are sufficient to prevent coalescence and secure
a regular "crystal-like" macroscale structure \cite{montessori_jfm}. 
Upon increasing $\mathcal{N}$, disorder sets in and the ordered 
crystal tends to evolve into a disordered emulsion.
Of course, one cannot expect EU to hold as a general 
rule, but for all the applications discussed in this
work, it was found to provide a pretty satisfactory
description of the complex rheology of the systems under
investigation. In a broad sense, one could state that the LB
simulations help unveiling the EU properties of soft 
flowing matter where they are. 

\section{The Lattice Boltzmann methods}

In this section, we provide a short survey of the three major families of Lattice Boltzmann 
methods, namely color-gradient (also termed chromodynamic), free energy, and pseudo-potential models. 
For an extensive discussion, we refer to specific textbooks, such as Refs. \cite{succi2018lattice,halim,sukop}. 

We begin by discussing some fundamental aspects of the theoretical background of non-ideal fluids, which are grounded in density functional theory (DFT). Subsequently, we will present the aforementioned numerical methods.

\subsection{Density functional and lattice kinetic theory}

The kinetic theory of gases, as devised by Ludwig Boltzmann,
was restricted to binary collisions of point-like particles hence formally limiting its application to dilute gases in the high Knudsen regimes.
\cite{huang,pathria}.  Over the years, many efforts have been made to broaden the range of applicability of the kinetic theory to the characterization of dense gases and liquids, most of them facing significant challenges mainly connected with the emergence of infinities while handling higher-order, many body collisions.
Although several strategies have been deployed to cope with such problems, the kinetic theory of dense, heterogeneous fluids remains a difficult subject to this day. A similar situation holds in the field of complex flows with interacting interfaces, often encountered in engineering, soft matter, and biology. In this context, a particularly interesting framework to deal with such complex flows is provided by the DFT \cite{KOHN,KOHN2,KOHN3}.

The idea behind DFT is that much of the complexity connected with the physics of interacting, many-body fluid systems can be elucidated by tracking the dynamics of the fluid density, namely a single one-body scalar field. Of course, such a dynamics is subject to self-consistent closures, stemming from physically-compliant guesses of
the generating functional from which the effective one-body equation for the density can be derived via minimization of the free-energy functional.
DFT has been successfully applied to the description of quantum many-body systems, resulting in the formulation of fundamental theorems and associated computational techniques that form the basis of modern computational quantum chemistry
\cite{mourik}. The same framework can be safely applied to classical systems encompassing interacting fluids with interfaces, and in the following, we proceed to illustrate such a picture in some more details.

The starting point of DFT is the definition of the free-energy functional
\begin{equation}
\label{FEF}
\mathcal{F}[\rho] = \int_V [f_b(\rho) + \frac{k}{2} (\nabla \rho)^2 ] dV,
\end{equation}
where $f_b(\rho)$ is the bulk free energy depending on the local fluid density $\rho({\bf x})$, while the second term, multiplied by the constant $k$, represents the cost associated with the build 
up of interfaces within the fluid.  
The bulk component of the free energy determines the non-ideal equation of state via the Legendre 
transform $p=\rho df_b/d\rho-f_b$, while the interface term fixes the surface tension.

In the case of a binary mixture (two components A and B), it proves expedient to define an order parameter as in Eq.(\ref{order_param}),
which varies between $1$ in phase A and $-1$ in phase $B$ and takes value zero at the interface.
Since the order parameter is conserved,  the associated continuity  equation reads
\begin{equation}
\label{CON}
\partial_t \phi = - \nabla\cdot \left( M \nabla \frac{ \delta \mathcal{F}[\phi]}{\delta \phi}\right),
\end{equation}
where $M$ is the mobility, $\delta$ stands for the functional derivative and ${\cal F}[\phi]$ is a suitable free energy often expressed as 
a quartic polynomial in $\phi$ with square-gradient terms.

The order parameter is convected by the barycentric velocity of the two species, ${\bf u}({\bf x},t)$,
which obeys the standard Navier-Stokes equations of fluids, augmented with an extra  non-ideal pressure tensor, known as Korteweg tensor \cite{landau}, formally given by  
$K_{\alpha\beta}(\rho) = \frac{ \delta^2 \mathcal{F}}{\delta g_{\alpha} \delta g_{\beta}}$,  where $g_{\alpha} \equiv  \partial_{\alpha} \rho$ (Greek subscripts denote Cartesian components). The explicit form of the Korteweg tensor is as follows
\begin{equation}
K_{\alpha\beta} =  \left(p +  \frac{k}{2}(\partial_{\gamma} \phi)^2 - k\phi\partial_{\gamma}\partial_{\gamma}\phi\right) \delta_{\alpha\beta} - k \partial_{\alpha} \phi \partial_{\beta} \phi,
\end{equation}
where $p=p(\rho)$ is the ideal pressure.
Finally, the divergence of the full pressure $\Pi_{\alpha\beta}$ tensor at the interface determines the mechanical force acting upon the fluid interfaces,
 $F_{\alpha}[\rho] = - \partial_\beta \Pi_{\alpha\beta}$ and the
condition $F_{\alpha}=0$ selects the density profile realizing the mechanical equilibrium of the interface.

The coupling between the continuity and the Navier-Stokes equations with non-ideal pressure tensor
provides a self-consistent mathematical framework describing the dynamics of the binary mixture.
In such a framework, the lattice kinetic theory serves as a natural bridge between the microscopic physics and the macroscopic hydrodynamic interactions.
Indeed, in the kinetic theory the main ingredient is represented by the non-ideal force term 
$S=F[\rho]\partial_u f$ (we omit subscripts and vector notation for simplicity), where $f$ is 
the probability density function.
Such a forced-streaming term can be brought to the right-hand side of the kinetic 
equation and treated as a soft-collision term. The possibility to handle the partial derivative in velocity space by integrating it by parts permits to move the distribution function along characteristics $\Delta {\bf x}_u={\bf u}\Delta t$ and include the effect of soft forces as a perturbation to the free-streaming motion of $f$.

The above statement can be formalized as
\begin{equation}
    f({\bf x}+{\bf u}\Delta t,{\bf u},t+\Delta t)-f({\bf x},{\bf u},t)= (C-S)dt,
\end{equation}
where $C$ is the short-range collisional term which can be approximated by the Bhatnagar-Gross-Krook operator \cite{bhatnagar1954model}
\begin{equation}
    C=(f^{eq}-f)/\tau.
\end{equation}
Here,  $f^{eq}$ is the local equilibrium distribution function and $\tau$ is a relaxation time.

The source collision term can also be turned into the following algebraic form
\begin{equation}
    S=F[\rho]\sum_l s_{l}H_l(v),
\end{equation}
where $s_{l}$ is the $l^{th}$ Hermite coefficient and $H_l$ the $l^{th}$ tensor Hermite basis.
To note, the above relation can be obtained by performing an integration by parts in the velocity space of the continuous source term and by exploiting the recurrence relations of Hermite polynomials.

The advantage of this approach is that the streaming step (left-hand side of the equation) is exact (zero round-off error), being the distribution convected along linear characteristics, in stark contrast with the hydrodynamic formulation in which information moves along space-time dependent
material lines defined by the 
fluid velocity. 
Thus, the above mathematical procedure allows the condensation of all the complex physics of moving interfaces into a purely local source term $S$. It is worth noting that, this "perturbative" treatment works numerically as long as the magnitude of the source term is small enough to guarantee that the local Froude number is sufficiently small, i.e. $Fr=|\frac{a \Delta t}{u}|<<10^{-3}$ where $a=F/m$, a condition which may be broken in the presence of large density gradients. 

Lattice density functional kinetic theories, as discussed above, are currently being used over an amazingly broad spectrum of soft-fluid problems, well beyond the original 
realm of rarefied gas dynamics. 
Indeed, in the last decades, kinetic theory has developed into a very elegant and effective
framework to handle a broad spectrum of problems involving complex states of flowing matter \cite{chapman1990mathematical,lifshitz_book,riboff_book}. 
In particular, the LB method has emerged as an alternative and computationally efficient way to capture the physics of fluid dynamic phenomena, even in the presence of complex geometries and interacting fluid interfaces \cite{succi2018lattice,halim}. The next paragraph is dedicated to presenting the working principles of this method.

\subsection{Basic ideas of the Lattice Boltzmann method}\label{basic_LB}

\begin{figure}[htbp]
\centering
\includegraphics[width=1.0\linewidth]{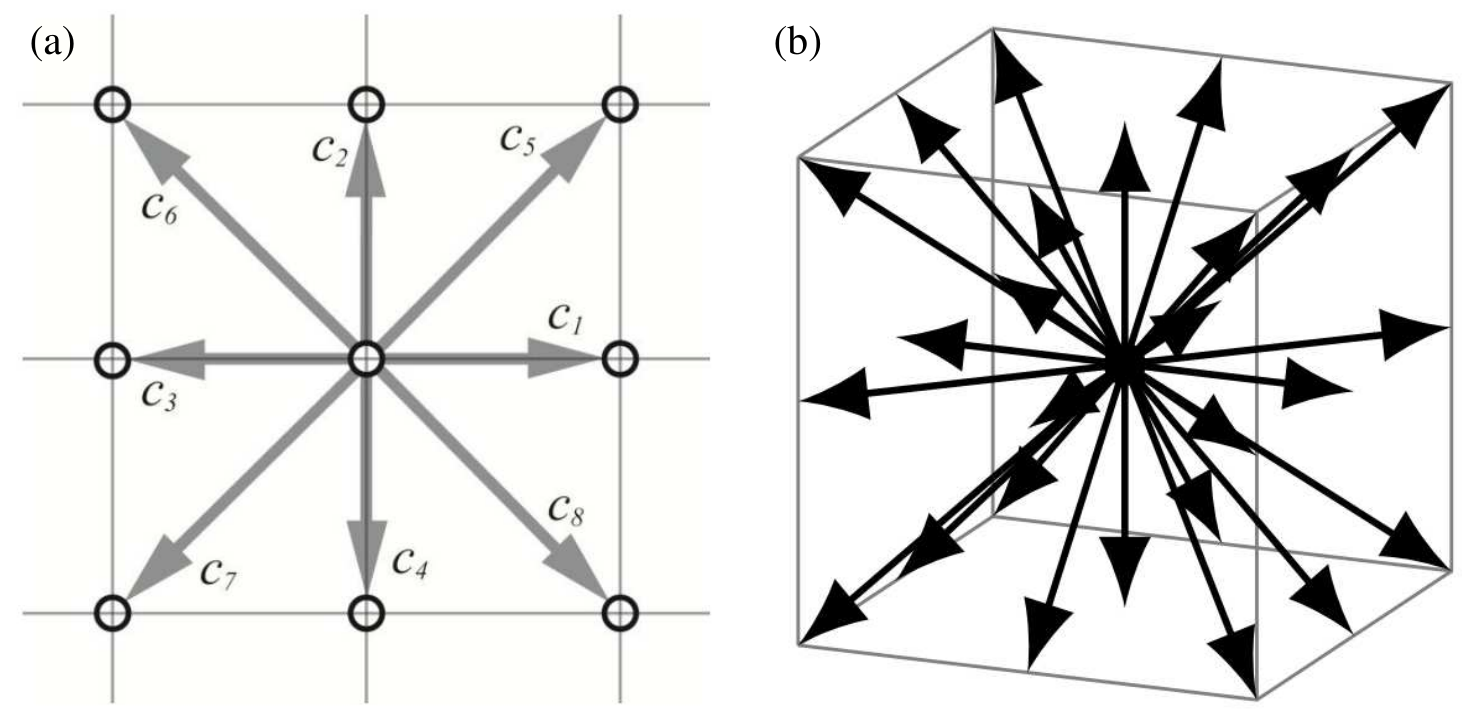}
\caption{Two typical lattice meshes that are often used in LB simulations. (a) A two-dimensional lattice with nine discrete velocities ($i=0,...,8$). (b) A three-dimensional lattice with twenty-seven discreet velocities ($i=0,...,26$).}
\label{lattice_vel}
\end{figure}

Let us now introduce the fundamental components that enable the functionality of the LB method. The LB equation states as follows
\begin{equation}
    f_i({\bf x}+{\bf c}_i \Delta t,t+\Delta t)-f_i({\bf x},t)=-\Omega_{ij}(f_j({\bf x},t)-f^{eq}_j), 
\label{eq:lbm}
\end{equation}
where  $f_i({\bf x},t)$ stands for the probability of finding a particle at lattice site ${\bf x}$ and time $t$ with a molecular velocity ${\bf u} = {\bf c}_i$ along the {\it i}-th direction of the lattice support in use. In particular, the set of discrete velocities ${\bf c}_i$ with $i \in [0,b]$ is chosen to guarantee sufficient symmetry and to comply with the principles of conservation of mass-momentum. Typical examples of LB lattices are shown in Fig.\ref{lattice_vel}, where Fig.\ref{lattice_vel}a represents a two-dimensional mesh with nine discrete velocities (also termed as D2Q9 model) and Fig.\ref{lattice_vel}b represents a three-dimensional one with twenty-seven discrete velocities (D3Q27 model).
The term $\Omega_{ij}$ denotes a collision operator that can take the form of a single relaxation term $\Omega=1/\tau$ ($\tau$ being a relaxation time) \cite{bhatnagar1954model} or more complex structures, such as multiple-relaxation time operators \cite{lallemand2000theory,d2002multiple} and central-moment-based ones \cite{fei2018modeling,fei2018three}, to name notable examples.

The LB equation reported in Eq.(\ref{eq:lbm}) can be split into two parts: the left-hand side represents the free-flight of the distributions along lattice characteristics, also called the streaming step, while the right-hand side stands for a collisional relaxation step of the set of probability distribution functions towards a local thermodynamic, Maxwell-Boltzmann equilibrium $f^{eq}_i$.  
In particular, the local equilibrium is obtained by performing a Taylor expansion to the second order of the Mach number $Ma = u/c_s$ 
\begin{equation}
f^{eq}_i( {\bf x},{\bf u})=w_i\rho\Big[ 1+ \frac{ {\bf c}_i\cdot{\bf u}}{c_s^2}+\frac{({\bf c}_i {\bf c}_i- c_s^2 \mathbf{I}):{\bf u} {\bf u} }{2 c_s^4} \Big],
\label{disteq}
\end{equation}
where $c_s$ is the lattice sound speed, ``:'' denotes a tensor contraction between two second-order tensors, and $w_i$ represents a set of normalized weights.
It is worth noting that the truncation is an outcome of lattice discreteness, allowing the recovery of Galilean invariance only up to a limited expansion in the Mach number.

The relevant hydrodynamic quantities can be obtained by computing the statistical moments of the distribution functions $f_i$ up to an order compliant with the moment isotropy of the lattice. At the Navier-Stokes level, the moment isotropy of the lattice must guarantee the retrieval of at least the first three moments of the distributions, 
namely density $\rho$, linear momentum $\rho{\bf u}$, and momentum flux tensor ${\bf \Pi}$
\begin{equation}
\rho= \sum_i f_{i}\left({\bf x},\,t\right),
\label{eq:rho}
\end{equation}
\begin{equation}
\rho {\bf u} = \sum_i   f_{i}\left({\bf x},\,t\right) {\bf c}_{i},
\label{eq:u}
\end{equation}
\begin{equation}
{\bf \Pi} = \sum_i   f_{i}\left({\bf x},\,t\right) {\bf c}_{i} {\bf c}_{i}.
\label{eq:Pi}
\end{equation}

The link between LB equation and Navier-Stokes equation can be found through a Chapman-Enskog analysis \cite{chapman1990mathematical}, which essentially consists of a multiscale expansion of the distribution function about equilibrium (and of spatial and temporal derivatives) in the Knudsen number $\epsilon=\lambda/L$, where $\lambda$ is the molecular mean-free path and $L$ is a characteristic macroscopic length of the system.  For small values of $\epsilon$ ($\epsilon \ll 1$),
$\lambda$ is much smaller than $L$, hence a continuum theory would provide a reliable description of the fluid dynamics. It can be shown that this technique allows to recover both continuity and Navier-Stokes equation \cite{halim,livio_epje}, in which the kinematic viscosity is given by
\begin{equation}
 \nu=c^2_s\left(\tau-\frac{\Delta t}{2}\right).   
\end{equation}

The theory described so far is generally common to different LB methods simulating multiphase and multicomponent fluids at the microscale. However, the details of their implementation, as well as the way in which the physics of the fluid interface (i.e. pressure and surface tension) is computationally modeled, depends significantly on the type of LB adopted. In the next paragraph, we describe three widely used LB approaches to simulate soft-flowing systems, namely color-gradient, free energy, and pseudopotential models.

\subsection{Modern Lattice Boltzmann methods for soft matter}\label{modernLB}

\subsubsection{Color-gradient Lattice Boltzmann}\label{color-gradient}

In the color gradient LB for multicomponent flows, two sets of distribution functions are employed to track the evolution of two interacting fluid components. Following Eq.(\ref{eq:lbm}), 
the streaming-collision algorithm becomes
\begin{equation} \label{CGLBE}
f_{i}^{k} \left({\bf x}+{\bf c}_{i}\Delta t,\,t+\Delta t\right) =f_{i}^{k}\left({\bf x},\,t\right)+\Omega_{i}^{k}( f_{i}^{k}\left({\bf x},\,t\right)),
\end{equation}
where $k$ runs over the fluid components. 
The density $\rho_k$ of the $k^{th}$ component is given  by the zeroth moment of the distribution functions
\begin{equation}
\rho_k\left({\bf x},\,t\right) = \sum_i f_{i}^{k}\left({\bf x},\,t\right),
\end{equation}
while the total fluid density is 
\begin{equation}
\rho=\sum_k \rho_k.    
\end{equation}
Also, the total momentum of the mixture is
defined as the sum of the linear momentum of the two components
\begin{equation}
\rho {\bf u} = \sum_k  \sum_i f_{i}^{k}\left({\bf x},\,t\right) {\bf c}_{i}.
\end{equation}

The collision operator $\Omega_i^k$ on the right hand side of Eq.(\ref{CGLBE}) can be split into three components \cite{gunstensen,leclaire2012,leclaire2017}
\begin{equation}
\Omega_{i}^{k} = \left(\Omega_{i}^{k}\right)^{(3)}\left[\left(\Omega_{i}^{k}\right)^{(1)}+\left(\Omega_{i}^{k}\right)^{(2)}\right],
\end{equation}
where $\left(\Omega_{i}^{k}\right)^{(1)}$ is the standard BGK operator $\frac{1}{\tau}(f_i^{k,eq}-f_i^k)$ \cite{succi2018lattice}, $\left(\Omega_{i}^{k}\right)^{(2)}$ performs the perturbation step \cite{gunstensen}, which builds up the surface tension and $\left(\Omega_{i}^{k}\right)^{(3)}$ is the recoloring step \cite{gunstensen,latva2005}, which promotes the separation between components, in order to minimize the mutual diffusion.

The correct form of the stress tensor can be obtained by imposing the constraints
\begin{eqnarray} \label{consconstr}
\sum_i \left(\Omega_{i}^{k}\right)^{(2)}&=&0 \\\label{consconstr2}
\sum_k \sum_i \left(\Omega_{i}^{k}\right)^{(2)} {\bf c}_i&=&0,
\end{eqnarray}
which stems from the idea of continuum surface force \cite{brackbill1992}, where the surface tension is interpreted as a continuous transport coefficient across an interface, rather than as a boundary value condition on the interface.

By using a Chapman-Enskog expansion of the distribution functions, it can be shown that the hydrodynamic limit of Eq.(\ref{CGLBE}) leads to the continuity and Navier-Stokes equations
\begin{eqnarray} 
\partial_t \rho+ \nabla \cdot (\rho {\bf u})&=& 0 \label{cont_eq} \\
\partial_t (\rho {\bf u}) + \nabla \cdot (\rho {\bf u}{\bf u})&=&-\nabla p + \nabla \cdot [\rho \nu (\nabla {\bf u} + \nabla {\bf u}^T)] + \nabla \cdot \bm{\Sigma},\nonumber\\\label{Nav_Stok_eq}
\end{eqnarray}
where $p=\sum_k p_k$ is the pressure and ${\bf \Sigma}$ is the non-ideal stress tensor. The latter 
can be written in terms of the perturbation operator and is given by
\begin{equation}
\bm{\Sigma}=-\tau\sum_i \sum_k\left(\Omega_{i}^{k}\right)^{(2)} {\bf c}_i {\bf c}_i.
\end{equation}

An explicit expression of the surface tension $\sigma$, which is generated by $(\Omega_i^k)^{(2)}$, can be obtained through the relation
\begin{equation} \label{SeqF}
\nabla \cdot \bm{\Sigma}= {\bf F},
\end{equation}
where ${\bf F}$ is a force inducing a stress jump across the interface. Following Refs.\cite{brackbill1992,liu2012}, a general expression for the force is given by
\begin{equation}\label{force}
{\bf F}({\bf x},t)= 
\nabla\cdot[\sigma (\mathbf{I} - {\bf n}\otimes {\bf n})\delta_I],
\end{equation}
where $\delta_{I}=\frac{1}{2} | \nabla \phi |$ is an index function which
localizes the force on the interface, $\phi=\frac{\rho_A - \rho_B}{\rho_A + \rho_B}$ is 
the phase field associated to the components $A$ and $B$, and
the normal ${\bf n}$ at the interface can be approximated by the gradient of the phase field, i.e. $\mathbf{n}= \nabla \phi/|\nabla \phi|$ \cite{liu2012}. Note that Eq.(\ref{force}) is nothing but 
the stress jump across an interface given by 
${\bf T}_A\cdot{\bf n}- {\bf T}_B\cdot{\bf n}=-\nabla\cdot[\sigma({\bf I}-{\bf n}\otimes{\bf n})]$, where ${\bf T}=-p{\bf I}+\rho\nu(\nabla{\bf u}+\nabla{\bf u}^T)$ is the stress tensor.

By choosing  
\begin{equation}
\left(\Omega_{i}^{k}\right)^{(2)}= \frac{A_k}{2} |\nabla \phi|\left[w_i \frac{(\vec{c}_i \cdot \nabla \phi)^2}{|\nabla \phi|^2} -B_i \right]
\end{equation}
and substituting it into Eq.(\ref{consconstr}), Eq.(\ref{consconstr2}) and Eq.(\ref{SeqF}), one gets the surface tension \cite{leclaire2012} 
\begin{equation}\label{sigmaA}
\sigma=\frac{2}{9}(A_A+A_B)\frac{1}{\bar{\omega}},
\end{equation}
where $\bar{\omega}=2/(6 \overline{\nu} -1)$ and $\overline{\nu}^{-1}=\frac{\rho_A}{\rho_A+\rho_B}\nu_1^{-1}+\frac{\rho_B}{\rho_A+\rho_B}\nu_2^{-1}$.
In this model it is customary to assume $A_A=A_B$, thus $\sigma=\frac{4}{9}A\frac{1}{\bar{\omega}}$.
In practical terms, the coefficients $A_A$ and $A_B$ are computed once the values of viscosity and surface tension are set. Finally, the coefficients $B_i$ in $(\Omega_i^k)^2$ can be obtained by imposing  the following isotropy constraints
\begin{eqnarray}
\sum_i B_i= \frac{1}{3}, \hspace{0.5cm} \sum_i B_i {\bf c}_i=0, \hspace{0.5cm} \sum_i B_i {\bf c}_i {\bf c}_i= \frac{1}{3} \mathbf{I}.
\end{eqnarray} 

As previously mentioned, the perturbation operator $(\Omega_i^k)^2$ allows for the building of the surface tension but does not guarantee the immiscibility of the fluid components. The latter condition is achieved by means of the recoloring step, which enables the interface to remain sharp and prevents the two fluids
from mixing. Following Ref.\cite{latva2005}, the recoloring operators are defined as follows
\begin{eqnarray}
(\Omega_i^1)^{(3)}(f_i^A)&=&\frac{\rho_A}{\rho} f_i^* + \beta \frac{\rho_A\rho_B}{\rho^2} \cos{\theta_i} f_i^{eq,0}\label{ome3_1}, \\
(\Omega_i^2)^{(3)}(f_i^B)&=&\frac{\rho_B}{\rho} f_i^* - \beta \frac{\rho_A\rho_B}{\rho^2} \cos{\theta_i} f_i^{eq,0}\label{ome3_2},
\end{eqnarray}
where $f_i^*=\sum_k f_i^{k,*}$ denotes the set of post-perturbation distributions, $\rho=\rho_A + \rho_B$,  $\theta_i$ is the angle between the phase field gradient and the $i^{th}$ lattice vector, $f_i^{eq,0}=\sum_k f_i^k(\rho,{\bf u}=0)^{eq}$ is the total zero-velocity equilibrium distribution function \cite{leclaire2012}
and $\beta$ is a free parameter tuning the interface width, thus playing the role of an inverse diffusion length scale \cite{latva2005}.
In \cite{montessori2018regular}, it has been also shown that the  color gradient LB scheme can be further stabilized by filtering out the high-order non-hydrodynamic (ghost) modes emerging in the under-relaxed regime (i.e. $\tau \geq \Delta t$)
\cite{montessori2015lattice,zhang2006efficient,latt2006,coreixas2017recursive,mattila2017high,hegele2018high,montessori2016effects}. This improvement essentially allows the model to operate at higher values of droplet speed minimizing the effects due to spurious currents and ghost modes \cite{benzi1995,benzi1992,montessori2018regular,montessori2018elucidating}. 

\begin{figure}[h]
\centering
\includegraphics[width=1.0\linewidth]{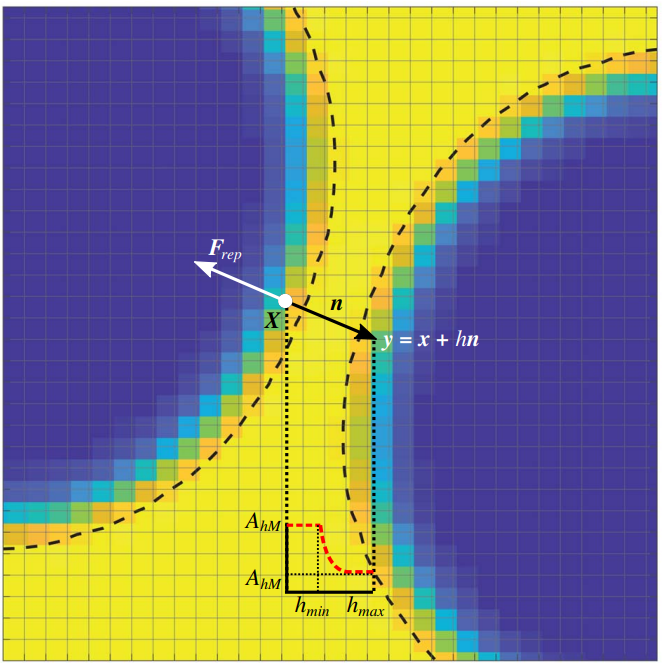}
\caption{The figure shows a representation of the near-contact forces between two fluid interfaces (highlighted by a dotted curve). Once two droplets are in close contact, a repulsive force ${\bf F}_{rep}$ is switched on 
to prevent droplet merging.
The blue region is the dispersed phase (droplets) while the yellow one is the continuous phase. 
The figure is reproduced with permission from Ref.\cite{montessori_jfm}.}
\label{sketchrep}
\end{figure}

\paragraph{Near-contact interactions}
In the color-gradient scheme, the effect of repulsive near-contact forces can be modeled, at the mesoscale, by including an additional term in the stress-jump condition \cite{montessori_jfm}, which reads as follows:
\begin{equation}\label{stress_jump}
\mathbf{T}_A\cdot {\bf n} - \mathbf{T}_B \cdot {\bf n}=-\nabla\cdot[\sigma(\mathbf{I} - {\bf n}\otimes {\bf n})] - \pi {\bf n}.
\end{equation}
The last term $\pi[h({\bf x})]$ is responsible for the repulsion between neighboring fluid interfaces and 
$h({\bf x})$ is the distance, along the normal ${\bf n}$, between positions
${\bf x}$ and ${\bf y}={\bf x}+ h {\bf n}$ of the two interfaces (see Fig.\ref{sketchrep}).

By neglecting surface tension variations along the interface (thus the stress tensor ${\bf T}$ can be approximated as $\mathbf{T}=-p\mathbf{I}$ \cite{brackbill1992} and 
by projecting Eq.(\ref{stress_jump}) along the normal at the interface, an  extended Young-Laplace equation can be obtained \cite{chan2011, williams1982}
\begin{equation}
(p_B - p_A)=\sigma (\nabla \cdot \vec{n}) - \pi.
\end{equation}

The additional repulsive term can be readily included within the LB framework through a 
forcing contribution acting only between fluid interfaces in near-contact
\begin{equation}\label{near_cont_force}
{\bf F}_{rep}= \nabla \pi := - A_{h}[h({\bf x})]{\bf n} \delta_I,
\end{equation}
where $A_h[ h({\bf x})]$ is  a parameter tuning the strength
of the near-contact interactions  (see Fig.\ref{sketchrep}).
A reasonable choice for $A_h$
suggests taking it constant for $h\leq h_{min}$ (where $h_{min}$ could correspond to two/three lattice sites separating opposite interfaces) and decreasing as $h^{-3}$ for $h>h_{min}$.
Also note that, although the force in Eq.(\ref{near_cont_force})  depends solely on the distance between two fluid interfaces, its expression can be extended to account for local variations of distance caused by the spontaneous migration of the surfactant along the fluid interface \cite{gupta2017}.

We finally mention that, over the years, other methods have been proposed 
to model the physics of near-contact forces in these systems
\cite{nekovee2000,love2001,chen2000}. However, unlike the approach discussed here where their effect is taken into account by introducing a repulsive force (as a mesoscale representation of supramolecular interactions), in these alternative methods 
two additional BGK-like equations are used, one for the distribution functions of a surfactant (which essentially prevents droplet coalescence)  
and another one for the relaxation of
the average dipole vector. The latter is because the interaction among surfactant molecules depends on their relative distance and on their dipolar orientation. An important difference between these two approaches is that the color-gradient method augmented with near-contact forces only requires two sets of distribution functions regardless of the number of droplets, thus allowing for the simulations of  mixtures with high droplet volume fractions (see Section \ref{Applications}) at a dramatically reduced computational cost.

\subsubsection{Free-energy model}\label{free_energy}

The free-energy approach is based on the idea that macroscopic physical quantities (such as chemical potential and pressure tensor) are derived from a free energy functional capturing the equilibrium properties of the system under study \cite{yeomans_pre,yeomans_prl,halim,succi2018lattice}. Unlike other models (such as the Shan-Chen approach \cite{shan_chen,shan_doolen}, see next paragraph), this prescription allows to preserve the thermodynamic consistency of the model, since the fluid attains its thermodynamic equilibrium under the guidance of the free energy.
Here we begin with a short description of the thermodynamics of the free energy LB of a multicomponent system, such as a water-oil mixture \cite{yeomans_pre,yeomans_prl}, and then we discuss a few details about its numerical implementation. 

In this model, a free energy functional can be generally written as an expansion of suitable order parameters, such as fluid density or relative concentrations describing the bulk properties, plus gradient terms capturing the physics of the interfaces.
Hence, on a general basis, a free energy can be defined as follows
\begin{equation} \label{Psi}
    {\cal F}=\int_V(f_b+f_g)dV,
\end{equation}
where $f_b$ is the bulk free energy leading to an equation of state in which different phases coexist, and $f_g$  is the interfacial term accounting for the energetic cost due to variations of order parameters.

The Landau theory allows us to derive these contributions analytically, namely  
\begin{equation}
f_b=c_s^2ln\rho + \frac{a}{4}(\phi^2-1)^2
\end{equation}
and 
\begin{equation}
f_g=\frac{k}{2}(\nabla\phi)^2,
\end{equation} 
where $\phi$ is a scalar phase field which distinguishes the bulk of the two interacting fluids. In $f_b$, the first term is the ideal gas free energy while the second one ensures the existence of two coexisting minima at $\phi=1$ and $\phi=-1$ for $a>0$, where the latter condition guarantees the immiscibility of the two fluid components.  The interface free energy $f_g$ determines
the surface tension, which can be calculated by integrating
the free energy density across an interface. This leads to
\begin{equation}\label{surf_tens_free}
\sigma=\int_{-\infty}^\infty \left[\frac{a}{4}(\phi^2-1)^2 + \frac{k}{2}(\nabla\phi)^2\right]dx=\sqrt{\frac{8ak}{9}}.
\end{equation}

A further thermodynamic quantity controlling the dynamic of the system is the chemical potential, which is defined as the functional derivative of the free energy with respect to the phase field. It is given by 
\begin{equation}
\mu_{\phi}\equiv\frac{\delta {\cal F}}{\delta \phi}=-a\phi+a\phi^3-k\nabla^2\phi,
\end{equation}
which is constant at equilibrium. In this condition, the equilibrium profile between the two coexisting bulk components
is given by
\begin{equation}
\phi_{eq}=\mathrm{tanh}\left(\frac{2x}{\xi}\right), 
\end{equation}
where 
\begin{equation}\label{int_w_free}
\xi=\sqrt{\frac{2k}{a}}
\end{equation}
is the interface width. 

This physics is linked to the Navier-Stokes equations through a pressure tensor (i.e. the Korteweg tensor) $\Pi_{\alpha\beta}$, which can be derived upon imposing the condition $\partial_{\beta}\Pi_{\alpha\beta}=\phi\partial_{\alpha}\mu_{\phi}$. The latter leads to the following expression
\begin{equation}\label{pressure_bf}
\Pi_{\alpha\beta}=\left(p_b - \frac{k}{2}(\partial_\gamma\phi)^2-k\phi\partial_\gamma\partial_\gamma\phi\right)\delta_{\alpha\beta} +k(\partial_\alpha\phi)(\partial_\beta\phi),
\end{equation}
where $p_b=c^2_s\rho+a\left(-\frac{1}{2}\phi^2+\frac{3}{4}\phi^4\right)$ is the equation of state. Once the thermodynamics is defined,  
the continuity and Navier-Stokes equations for a multicomponent fluid are equivalent to Eq.(\ref{cont_eq})-(\ref{Nav_Stok_eq}) with $ \Pi_{\alpha\beta}=p\delta_{\alpha\beta}+\Sigma_{\alpha\beta}$.

Alongside the continuum equations, in the multicomponent free energy model one also needs an evolution equation for the phase field $\phi$. This one is governed by the Cahn-Hilliard equation \cite{cahn}
\begin{equation}\label{ch_eq}
\partial_t\phi+\nabla\cdot(\phi {\bf u})=\nabla\cdot(M\nabla\mu_{\phi}),
\end{equation}
where the second term on the left-hand side represents an advection contribution due to the fluid velocity and the diffusive-like term (where $M$ is the mobility) on the right-hand side is the current driving the system towards equilibrium.

\paragraph{Multiphase field model}
In the same spirit of the color-gradient LB approach, the free-energy model can also be extended to simulate systems, such as dense emulsions \cite{foglino1}, with more than two components and where coalescence is inhibited.  An emulsion of $N$ immiscible droplets, for example, can be described as a set of scalar fields $\phi_n$  ($n=1,....,N$) capturing the density of each droplet, whose dynamics obeys a corresponding set of advection-diffusion equation
\begin{equation}\label{multi_adv}
\partial_t\phi_n+\nabla(\phi_n{\bf u})=\nabla(M\nabla\mu_{\phi_n}).
\end{equation}
Once again, the chemical potential of each droplet is defined as $\mu_{\phi_n}=\delta{\cal F}/\delta\phi_n$, where ${\cal F}=\int_V fdV$. In this system, a suitable form of the free energy density $f$ is
\begin{equation}\label{free_mp}
f= \sum_{n=1}^N\left[\frac{a}{4}\phi_n^2(\phi_n-\phi_0)^2+\frac{k}{2}(\nabla\phi_n)^2\right]+\sum_{n,m,n<m}\epsilon\phi_n\phi_m,
\end{equation}
where the first term guarantees the existence of the minima $\phi_n=\phi_0$ (for example inside the $n$-th droplet) and $\phi_n=0$ (outside), and the second one accounts for the fluid interface.
The positive constants
$a$ and $k$ control surface tension and interface thickness, given by Eq.(\ref{surf_tens_free}) and Eq.(\ref{int_w_free}),
respectively. 
Finally, the last term models a repulsive effect hindering droplet merging.  This contribution slightly modifies the stress tensor of Eq.\ref{Nav_Stok_eq},  which includes a term proportional to $-\sum_n\phi_n\nabla\mu_n$.

\paragraph{Extension to active fluids} 
In recent years, research on active matter in general and active fluids in particular has known a burgeoning growth \cite{marchetti}.
Active matter presents a few distinctive features that set it apart from passive one, such as lack of reciprocity of pair interactions and the resulting breaking of time-reversal symmetry which
triggers spontaneous motion.
More often than not, however, the actual physical mechanisms responsible for the above features 
can be described in terms of additional internal degrees of freedom, coupled to the external ones
(position and momentum) describing passive matter.
As a result, from an operational standpoint, active systems can be regarded as passive
systems with internal degrees of freedom coupled to the external ones in such a way as to
sustain feedback loops with the environment. 
Therefore, the LB method can be readily extended to handle active 
flowing matter as well.
Indeed, over the last two decades, the free-energy LB method has been 
successfully adapted to the simulation the hydrodynamics of active fluids and self-propelled 
droplets, in which the active fluid is segregated \cite{livio_epje,cates_softmatter}. 
As compared to an isotropic passive fluid, active fluids are made of individually oriented 
units supporting spontaneous symmetry breaking of rotational order, i.e. capable of developing 
orientationally ordered structures on macroscopic scales. 
This feature is generally captured by introducing an order parameter, such as the polar vector ${\bf p}({\bf x},t)$ or 
the tensor $ {\bf Q}({\bf x},t)$, describing the ordering 
properties of the symmetry-broken phase, such as liquid crystals \cite{degennes}.

The thermodynamics of these systems  is governed by a quantity akin to the free energy of Eq.(\ref{free_mp}) with further 
contributions depending on the additional fields, ${\bf p}({\bf x},t)$ or ${\bf Q}({\bf x},t)$. 
For instance, in the case of a polar active fluid droplet, 
the additional terms take the following form
\begin{equation}
f_{LC}(\phi, {\bf p})=-\frac{\alpha}{2}\frac{\phi-\phi_{cr}}{\phi_{cr}}|{\bf p}|^2
+\frac{\alpha}{4}|{\bf p}|^4+\frac{\kappa}{2}(\nabla{\bf p})^2,
\end{equation}
where $\alpha$ and $\kappa$ are positive constants and $\phi_{cr}$ is the critical 
concentration for the transition from the unbroken isotropic phase 
$(|{\bf p}|=0)$ to the broken polar one $(|{\bf  p}|>0)$. 
In the above, the first two terms represent the bulk free energy of the polar phase, while the last 
one describes the elastic penalty of the liquid crystal distortions in the single elastic constant approximation \cite{degennes}. 

The polar field evolves according to an advection-relaxation equation of the form 
\begin{equation}
D_t{\bf p}=-\frac{1}{\Gamma}\frac{\delta{\cal F}}{\delta {\bf p}}, 
\end{equation}
where, besides the usual Lagrangian derivative, $D_t$ includes additional terms accounting for 
rigid rotations and deformations of the fluid element \cite{livio_epje} and $\Gamma$ is the rotational
inertia of the liquid crystal. 

Finally, the fluid velocity obeys the Navier-Stokes equations, augmented with 
an active stress tensor of the form
\begin{equation}
\Pi_{\alpha\beta}^{act}\propto -\zeta\phi p_{\alpha}p_{\beta}, 
\end{equation}
where $\zeta$ gauges the active strength, negative for contractile particles 
and positive for extensile ones. These two classes distinguish particles where the fluid is pulled towards the center of mass (i.e. contractile) from the ones where the fluid is pushed away from it (i.e. extensile). The active stress tensor is responsible for the coupling between the polar droplet and the 
surrounding fluid, leading to flow patterns that sustain the active motion of the droplet
\cite{ramaswamy_prl,pedley_arfm}.

\paragraph{Basic implementation}
Historically, two main free-energy LB  schemes 
have been used to simulate binary fluids.
In the first one, the Cahn-Hilliard equation is solved by introducing a set of distribution functions $g_i({\bf x},t)$ connected to the phase field and the Navier-Stokes equation via the usual distribution functions $f_i({\bf x},t)$ defining zeroth, first and second order momenta. In the second one the approach is hybrid, meaning that the Cahn-Hilliard equation is integrated via finite difference algorithms, while the Navier-Stokes one through a standard LB method. 
Both approaches have been widely used to simulate a variety of soft matter systems under confinement, such as liquid-vapor mixtures \cite{sofonea}, binary and ternary fluids \cite{lamura,xu_pre,xu_pre2,gonnella_pre,tiribocchi_pattern,tiribocchi_pre2,tiribocchi_pof1,tiribocchi_pof2,tiribocchi_nat}, liquid crystals \cite{cates_softmatter,stratford,henrich2,marenduzzo_lc,sulaiman,perm_yeomans,mare_jnfm,tirib_bistable,tirib_nemat,tirib_soft2,tirib_soft3} and active matter \cite{yeomans_act1,fielding,tyler1,yeomans_act2,dema_softmatter,negro_soft,bonelli,bonelli2,tiribocchi_shape,tirib_spont}. 
Since some applications discussed in Section \ref{Applications} are simulated using the second algorithm, we shortly recap its main features, while a detailed description of the first one can be found, for example, in Refs.\cite{halim,livio_epje}. 

In the hybrid approach, the phase field $\phi$ 
obeys Eq.(\ref{ch_eq}) and it is numerically updated in the following two steps. 

\begin{enumerate}
\item An explicit Euler algorithm is used to integrate the convective term
\begin{equation}
\phi^*({\bf x})=\phi-\Delta t_{FD}(\phi\partial_{\alpha}u_{\alpha}+u_{\alpha}\partial_{\alpha}\phi),
\end{equation}
where $\Delta t_{FD}$ is the time step of the finite difference scheme and the velocity ${\bf u}$ stems from the LB equation. Note that the terms on the right-hand side are computed at time $t$ and lattice position ${\bf x}$.
\item The diffusive part is then updated
\begin{equation}
\phi({\bf x},t+\Delta t_{FD})=\phi^*+\Delta t_{FD}\left(\nabla^2 M\frac{\delta {\cal F}}{\delta \phi}\right)_{\phi=\phi^*},
\end{equation}
\end{enumerate}
where finite difference operators are calculated using a stencil representation to ensure isotropy of the lattice \cite{tiribocchi_pre,thampi}. 
Although more complex schemes have been used (such as predictor-corrector ones \cite{marenduzzo_lc,henrich}), this one combines satisfactory numerical stability and easy computational implementation.

The resolution of continuity and Navier-Stokes equations follows the prescriptions outlined in subsection \ref{basic_LB} properly extended to a binary fluid,  
basically meaning that the pressure tensor of the form of Eq.(\ref{pressure_bf}) must be implemented in the model. This can be done, for example, by
modifying the constraints on the second moment of the distribution functions according to the following relation
\begin{equation}
\sum_if_i^{eq}({\bf x},t)c_{i\alpha}c_{i\beta}=-\Pi_{\alpha\beta}+\rho u_{\alpha}u_{\beta}.
\end{equation}

Note that, due to the symmetry of the left-hand side, this approach can be applied to systems in which the stress tensor is symmetric, such as a binary fluid. In the presence of anti-symmetric contributions (like in liquid crystals), the algorithm can be modified by introducing, on the right-hand side of Eq.(\ref{eq:lbm}), a suitable forcing term $p_i$ fulfilling the following relations
\begin{equation}
\sum_ip_i=0, \hspace{0.5cm} \sum_ip_i c_{i\alpha}=\partial_{\beta}\Pi_{\alpha\beta}^{anti}, \hspace{0.5cm}\sum_ip_i c_{i\alpha} c_{i\beta}=0,
\end{equation}
where $\Pi_{\alpha\beta}^{anti}$ represents the anti-symmetric component of the stress tensor. Further details on this model can be found in Refs.\cite{halim,livio_epje}.

\subsubsection{Lattice pseudo-potential approach}\label{lattice_pp}

An alternative route to simulate the hydrodynamics of multi-component mixtures is represented by the lattice pseudo-potential model, initially proposed in Refs. \cite{shan_chen,shan_doolen}.
Unlike the  free-energy method, this one follows a "bottom-up" approach by postulating a microscopic interaction between fluid elements from which a non-ideal equation of state 
as well as macroscopic observables, such as the surface tension and disjoining pressure, emerge (or can be eventually incorporated). 

In this model, the crucial difference with the free-energy LB is that the pressure tensor is built from a force 
describing precisely such microscopic interactions. 
Assuming that the force between pairs of molecules is additive, one can expect that the interaction between fluid elements at positions ${\bf x}$
and ${\bf y}$ depends on the product $\rho({\bf x})\rho({\bf y})$. A suitable choice is \cite{shan_chen} 
\begin{equation}\label{force_sc}
{\bf F}({\bf x})=-\int ({\bf y}-{\bf x})G({\bf x},{\bf y})\psi({\bf x})\psi({\bf y})d^3{\bf y},
\end{equation}
where $\psi$ is a density functional, called pseudo-potential, and $G$ is a kernel function accounting for the spatial dependence of the force.

In their work \cite{shan_chen}, Shan and Chen introduced the pseudo-potential as follows
\begin{equation}
    \psi(\rho)=\rho_0[1-e^{-\rho/\rho_0}],
\end{equation}
where $\rho_0$ is a reference density usually set to unity. The above term is bounded between $0$ and $\rho_0$ for any value of $\rho$, allowing the magnitude of the interaction force to be finite, even in the presence of large densities. This secures the stability of the model by switching the interaction off progressively as density increases, thereby smoothing the build-up of density divergences. 

The implementation on the lattice follows a procedure partially akin to both color-gradient and free-energy LB, where different population sets represent a fluid component which evolves via a separate LB equation.
The integral of Eq.(\ref{force_sc}) 
can be discretized by considering the interaction force to be short-ranged, so that the fluid element at the lattice site ${\bf x}$ can interact only with other neighboring elements at ${\bf y}={\bf x}+{\bf c}_i\Delta t$. Moreover, $G({\bf x},{\bf y})$ should be isotropic, thus depending on $|{\bf x}-{\bf y}|$ only. A common choice is $G({\bf x},{\bf y})=w_iG$ for ${\bf y}={\bf x}+{\bf c}_i\Delta t$ and zero otherwise.

The simplest form of the discretized Shan-Chen force for a multicomponent fluid is represented by a sum of pseudopotential interactions with nearest lattice neighbours
\begin{equation}
    {\bf F}_{k}({\bf x})= -\psi_{k}({\bf x}) \sum _{\Tilde{k}} G_{k \Tilde{k}} \sum_ i w_i\psi_{\Tilde{k}}({\bf x}+{\bf c}_{i}\Delta t) {\bf c}_{i}\Delta t,
\end{equation}
where the sum runs over the lattice links, $k$ and $\Tilde{k}$ are the two interacting components and $G_{k\Tilde{k}}$
is a parameter setting the strength of their interaction and controlling the surface tension in the model. Unlike the free energy LB, here the surface tension is an emergent effect and can be explicitly calculated using the Young-Laplace test \cite{halim}.

\paragraph{Thermodynamic consistency} One may finally wonder to what extent the Shan-Chen model is supported by a consistent thermodynamic description. In the following, we show that
an expression of the pressure tensor akin to that of the free-energy model can be obtained by properly expanding the pseudo-potential. 
 
Following the derivation of Ref.\cite{halim} we consider, for simplicity, a multiphase mixture, where the forcing term is given by
\begin{equation}\label{force_single}
    {\bf F}({\bf x})=-\psi({\bf x}) G \sum_i w_i\psi({\bf x}+{\bf c}_{i}\Delta t) {\bf c}_{i}\Delta t.
\end{equation}
A Taylor expansion of  $\psi({\bf x}+{\bf c}_{i}\Delta t)$  gives
\begin{eqnarray}\label{exp_pseudo}
 \psi({\bf x}+{\bf c}_{i}\Delta t) = \psi({\bf x}) &+& c_{i\alpha}\Delta t \partial_{\alpha}\psi({\bf x})\nonumber\\ &+& \frac{1}{2} c_{i\alpha}c_{i\beta}\Delta t^2\partial_{\alpha}\partial_{\beta}\psi({\bf x})\nonumber\\ &+& \frac{1}{6} c_{i\alpha}c_{i\beta}c_{i\gamma}\Delta t^3\partial_{\alpha}\partial_{\beta}\partial_{\gamma}\psi({\bf x}) + ....\nonumber\\
\end{eqnarray}
Substituting  Eq.(\ref{exp_pseudo})  into Eq.(\ref{force_single}) leads to
\begin{eqnarray}
{\bf F}({\bf x})=&-&G\psi({\bf x})\sum_i w_i {\bf c}_{i} \Delta t \biggr(\psi({\bf x}) + c_{i\alpha}\Delta t \partial_{\alpha} \psi({\bf x})\nonumber\\&+&\frac{1}{2} c_{i\alpha}c_{i\beta}\Delta t^2\partial_{\alpha}\partial_{\beta}\psi({\bf x}) + .... \biggl).
\end{eqnarray}
Due to the symmetry of the lattice set, odd terms (i.e. $\sum_i w_i c_{i\alpha}$ and $\sum_i w_i c_{i\alpha}c_{i\beta}c_{i\gamma}$) vanish, thus leading to
\begin{equation}
{\bf F}({\bf x})= - G\psi({\bf x})\left(c_s^2\Delta t^2 \nabla \psi({\bf x}) + \frac{c_s^4 \Delta t^4}{2}\nabla \nabla^2 \psi({\bf x})\right).
\end{equation}
The first term on the right-hand side is a gradient (of the form $-\frac{Gc_s^2\Delta t^2}{2}\nabla\psi^2({\bf x})$) that can be readily included in the equation of state  
of a multiphase fluid
\begin{equation}\label{pseudopoteos}
p(\rho)=\rho c_s^2 +\frac{c_s^2\Delta t^2 G}{2}\psi^2(\rho),
\end{equation}
while the second one resembles the surface tension contribution  $k \rho \nabla \nabla^2\rho$ usually employed in diffuse interface models.
Since, by definition,  $$\partial_{\beta}P_{\alpha\beta}=\partial_{\beta}(c^2_s\rho)-F_{\alpha},$$
it is finally possible to show that the Shan-Chen pressure tensor takes the following form
\begin{eqnarray}\label{sc_pres}
P_{\alpha\beta}&=&\left(\rho c_s^2 + \frac{c_s^2G}{2}\psi^2 + \frac{c_s^4G}{4}(\nabla\psi)^2 + \frac{c_s^4G}{4}\psi\nabla^2\psi\right)\delta_{\alpha\beta}\nonumber\\&-& \frac{c_s^4G}{2}(\partial_{\alpha}\psi)(\partial_{\beta}\psi),
\end{eqnarray}
which differs from the thermodynamically consistent expression of the pressure tensor.  However, in many practical situations where the density ratio between phases remains small, the surface tension and the density profiles obtained from Eq.(\ref{sc_pres}) are acceptable and numerical stability is preserved.
On the contrary, if the density ratio increases (typically higher than $10$), the stability can be improved by considering a Van der Waals-like equation of state (EOS), which dictates the pseudopotential according to
\begin{equation}
\psi(\rho)=\sqrt{\frac{2}{c_s^2\Delta t^2 |G|}(p_{EOS}(\rho)-c_s^2\rho)},
\end{equation}
where $p_{EOS}(\rho)$ is the desired equation of state.
Such an approach, combined with a recent version of the entropic LB \cite{entr_review} and a higher-order isotropic discretization of the gradient of the pseudo-potential, allows for the simulation of multiphase flows at Weber and Reynolds numbers above $10^2$ and $10^3$ respectively \cite{monte_entropic}, 
unattainable by standard pseudo-potential methods.

\subsection{Boundary conditions}

We conclude this section with a short discussion on the boundary conditions used in the applications presented in Section \ref{Applications}. 
Since the systems under study are either unbounded or confined within a microchannel, we focus on periodic boundaries and on solid ones modeling parallel flat walls, while we refer to specific textbooks \cite{succi2018lattice,halim} for a general overview. It is worth noting that, despite the complexity of the systems under study often involving confined deformable interfaces in close contact, the implementation of the boundary conditions (at least for the two aforementioned cases) follows well-known standard procedures. This is actually a key strength of the method, which proves capable of capturing the mesoscale physics of different examples of confined soft matter
with minimal modifications of existing algorithms. 

\begin{figure}[htbp]
\centering
\includegraphics[width=1.0\linewidth]{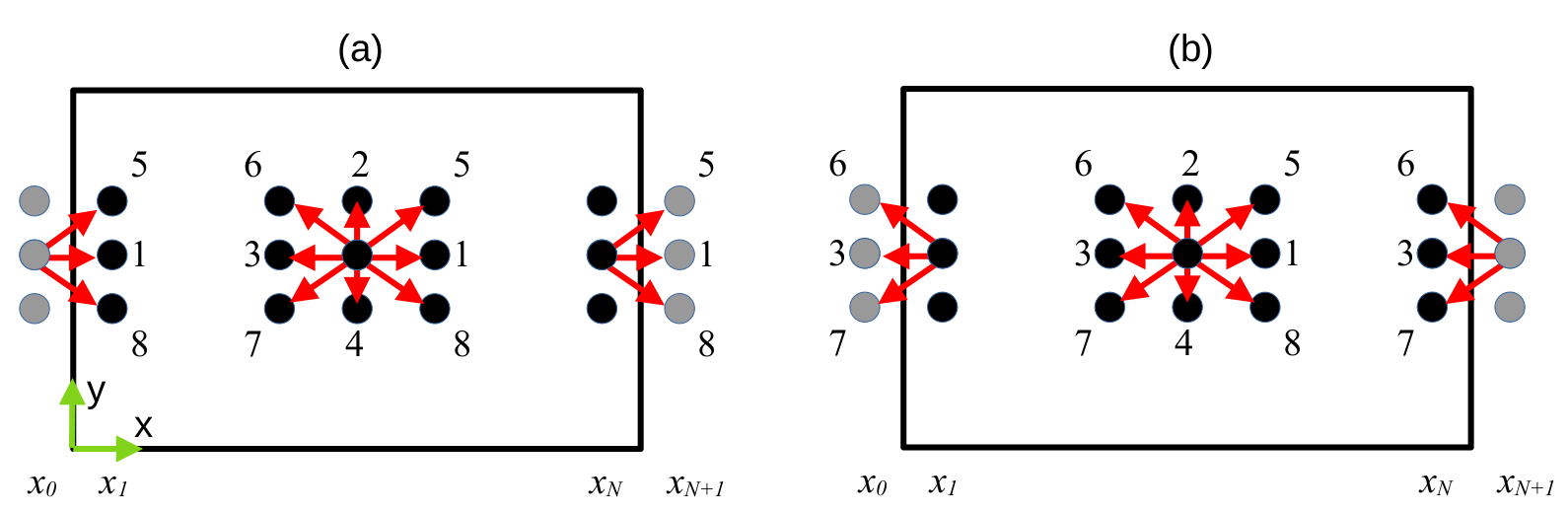}
\caption{Periodicity along the $x$ direction. 
Black spots denote real nodes while grey ones indicate the buffer. Red arrows represent lattice velocity vectors, numbered according to Fig.\ref{lattice_vel}.
(a) The distribution functions leaving from the leftmost column at $x_N$ fill in the virtual nodes on their rights, giving the populations on the opposite side at $x_1$. (b) A similar prescription holds for distribution functions leaving $x_1$ and giving the populations at $x_N$. The figure is inspired by Refs.\cite{succi2018lattice,halim}.}
\label{peridic_b}
\end{figure}

\paragraph{Periodic boundary conditions.}
We consider a two-dimensional one-component fluid (hence a single set of distribution functions), where periodic boundary conditions apply along the $x$ direction (see Fig.\ref{peridic_b}) of a rectangular simulation box of size $L_x\times L_y$. We refer to the $D2Q9$ mesh numbered from $0$ to $8$, as shown in Fig.\ref{lattice_vel}.
These conditions are typically used to isolate the bulk physics from actual boundaries (such as a wall) and essentially mean that the fluid leaving the lattice nodes located at $x=x_N$ re-enters at the ones located at $x=x_1$ and vice versa.

In terms of distribution functions, this prescription  reads
\begin{eqnarray}
f_5(x_0,t)=f_5(x_N,t)\hspace{1cm} f_6(x_{N+1},t)=f_6(x_1,t)\nonumber\\
f_1(x_0,t)=f_1(x_N,t)\hspace{1cm}
f_3(x_{N+1},t)=f_3(x_1,t)\nonumber\\
f_8(x_0,t)=f_8(x_N,t)\hspace{1cm}
f_7(x_{N+1},t)=f_7(x_1,t)\nonumber
\end{eqnarray}
for the inwards populations (left equations and Fig.\ref{peridic_b}a) and for the
outwards ones (right equations and Fig.\ref{peridic_b}b).
Note that, alongside the lattice nodes (black dots) of the  simulation box, an additional layer of virtual nodes (grey dots) is included. During the boundary procedure, the distribution functions are copied in this buffer following the above equations and then they are updated for the streaming step.

\paragraph{No-slip boundary conditions.}
A further typical boundary condition, often realized in a microchannel, is the one in which the fluid velocity at a solid surface is zero, thus there is no net motion of the fluid relative to the wall.  This no-slip effect can be modeled by imposing that the distribution functions hitting a boundary node just reverse to where they start. The complete reflection ensures that normal and tangential components of the velocity at the wall vanish, thus the wall is actually impenetrable and the fluid does not slip on it. 
In terms of numerical implementations, 
no-slip conditions can be generally realized assuming that either the physical boundary lies exactly on a grid line or the boundary lies in between two grid lines. For illustrative purposes, here we shortly describe the first approach following Refs.\cite{zou1997}.

\begin{figure}[htbp]
\centering
\includegraphics[width=0.8\linewidth]{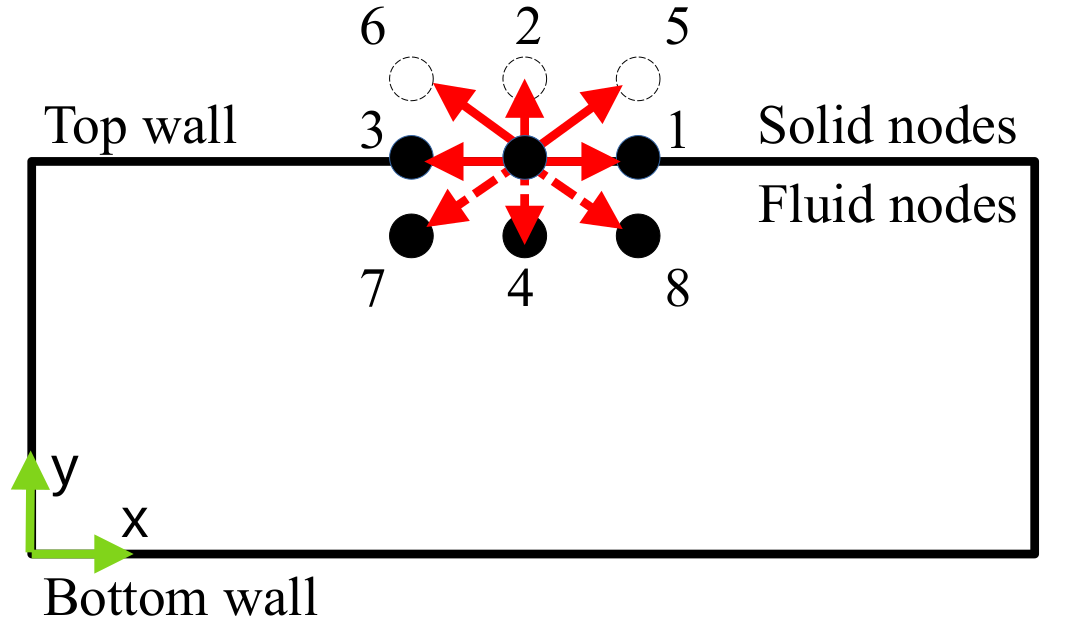}
\caption{In this figure, dotted spots represent solid nodes located above the top wall. Red dotted arrows indicate unknown distribution functions. The figure is inspired by Ref.\cite{succi2018lattice,halim}.}
\label{no_slip_w}
\end{figure}

We consider a one-component fluid confined within a microchannel consisting of two flat parallel walls and focus, for example, on the top one (see Fig.\ref{no_slip_w}). Similar considerations hold for the opposite wall. 

Once the propagation step is completed, distribution functions $f_0$, $f_1$, $f_5$, $f_2$, $f_6$ and $f_3$ (continuous red arrows) are known 
while $f_7$, $f_4$ and $f_8$ (dotted red arrows) are unknown. These ones can be determined by using Equations (\ref{eq:rho}) and (\ref{eq:u}). 
Requiring that, at the wall, the fluid velocity is zero, the following equations hold
\begin{eqnarray}
 \rho&=&\sum_if_i\rightarrow 
 f_7(t)+f_4(t)+f_8(t)=\nonumber\\&&\rho-[f_0(t)+f_1(t)+f_5(t)+f_2(t)+f_6(t)+f_3(t)],\nonumber\\
 \rho u_{x,w}&=&\sum_i f_ic_{ix}=0\rightarrow\nonumber\\&& f_8(t)-f_7(t)=f_6(t)-f_5(t)+f_3(t)-f_1(t),\nonumber\\
 \rho u_{y,w}&=&\sum_i f_ic_{iy}=0\rightarrow\nonumber\\&& f_7(t)+f_4(t)+f_8(t)=f_5(t)+f_2(t)+f_6(t).\nonumber
\end{eqnarray}
One can thus find the density in terms of known populations as
\begin{equation*}
\rho=f_0(t)+f_1(t)+f_3(t)+2[f_5(t)+f_2(t)+f_6(t)]
\end{equation*}
and the unknown terms using the bounce-back rule for the distribution functions normal to the wall. The final set of equations reads
\begin{eqnarray}
f_4(t)&=&f_2(t)\nonumber\\
f_7(t)&=&f_5(t)+\frac{1}{2}[f_1(t)-f_3(t)]\nonumber\\
f_8(t)&=&f_6(t)-\frac{1}{2}[f_1(t)-f_3(t)]\nonumber.
\end{eqnarray}
Further extensions of this approach aimed at guaranteeing improved numerical stability have been also proposed 
\cite{lamura2001,tiribocchi_pattern}.

\paragraph{Boundary conditions of the phase field.} In confined systems studied by means of the hybrid free energy LB,  it is also necessary to impose boundary conditions for the scalar field $\phi$.
It is often customary to adopt either neutral wetting or no-wetting conditions, where the former are achieved by setting 
\begin{equation}
 \frac{\partial\mu}{\partial z}=0, \hspace{0.5cm} \frac{\partial\nabla^2\phi}{\partial z}=0   
\end{equation}
at the walls ($z$ being the vertical direction). 
The first one ensures density conservation while
the second one guarantees the wetting to be neutral. These conditions are implemented by imposing that $\mu_{|z=\{0,L_z\}}=\mu_{|z=\{1,L_{z-1}\}}$ and $\nabla^2\phi_{|z=\{0,L_z\}}=\nabla^2\phi_{|z=\{1,L_{z-1}\}}$, where a stencil representation of finite difference operators is generally used.
No wetting boundaries are enforced by substituting the second equation with the condition $\phi = \phi_0$ at the first and second lattice nodes along the horizontal direction, where $\phi_0$ is the value of one of the coexisting densities near the walls. In an emulsion, for example, $\phi_0$ would be the density of the dispersed phase. 

\section{Recent Lattice Boltzmann HPC implementations}

In recent years, much work has been directed to developing advanced computational tools for the simulations of complex fluids on high performance computing (HPC) architectures. 
Indeed, the investigation of the flowing properties of confined soft materials often poses major multiscale challenges, due to the need to describe multi-component systems in centimeter-sized devices while retaining the physics of near-contact interactions.
In this respect, standard LB approaches suffer a series of issues, the first being the relatively low arithmetic intensity, which varies between 1 and 5 Flops/Byte ratios depending on the specific implementation. Consequently, the LB method is often recognized as a memory-bound problem, as its efficiency is constrained by the memory access time on computing systems \cite{succi2019towards}. This problem is due to
the necessity to directly store probabilistic distributions involving a set of approximately $b \sim 20-30$ discrete populations $f_i$ (where $i=0,...,b$ for each lattice point), clearly outnumbering the number of related hydrodynamic fields, which are a single scalar density $\rho$, the flow velocity $u_{\alpha}$ ($\alpha=1,3$), and momentum-flux tensor $\Pi_{\alpha\beta}$ (a $3\times 3$ symmetric tensor for an isotropic binary fluid), amounting to 10 independent fields in total. Therefore, the LB method demands roughly double the memory compared to a corresponding computational fluid dynamics method based on Navier-Stokes equations.  
This redundancy buys several major computational advantages, primarily the fact that streaming is exact and diffusion is emergent (no need for second-order spatial derivatives to describe dissipative effects), which proved invaluable assets in achieving outstanding parallel efficiency across virtually any HPC platforms, also in the
presence of real-world complex geometries \cite{succi2018lattice}.
These extra-memory requirements put a premium on strategies
aimed at minimizing the cost of data access in massively parallel LB codes.

Considering the extensive application of the LB method across various scales and regimes, it is crucial to devise strategies that mitigate the effect of data access on upcoming (exascale) LB simulations. Numerous such strategies
have been previously established, which rely on hierarchical memory access \cite{succi2019towards}, data arrangement in the form of an array of structures and structure of arrays enhancing the data-contiguity \cite{ansumali_intjc}, and crystallographic lattice schemes that double the resolution for a specific level of memory usage \cite{namburi2016crystallographic}.

A number of other approaches 
focus on reducing memory occupancy without significantly compromising the simulation quality. 
They can be categorized into three groups: the first one centers on the algorithmic implementation of the streaming step, the second one leverages diverse numerical representations to rounding in floating-point arithmetic, and the third one avoids the direct storage of the probabilistic information by exploiting a reconstruction procedure of distributions from the hydrodynamics fields via Hermite projection. 
Below we describe some features of each of these methods and then we compare the computational performance of a recent version of the LB pertaining to the third group with Navier-Stokes solvers.

\begin{figure*}[htbp]
\centering
\includegraphics[width=1.0\linewidth]{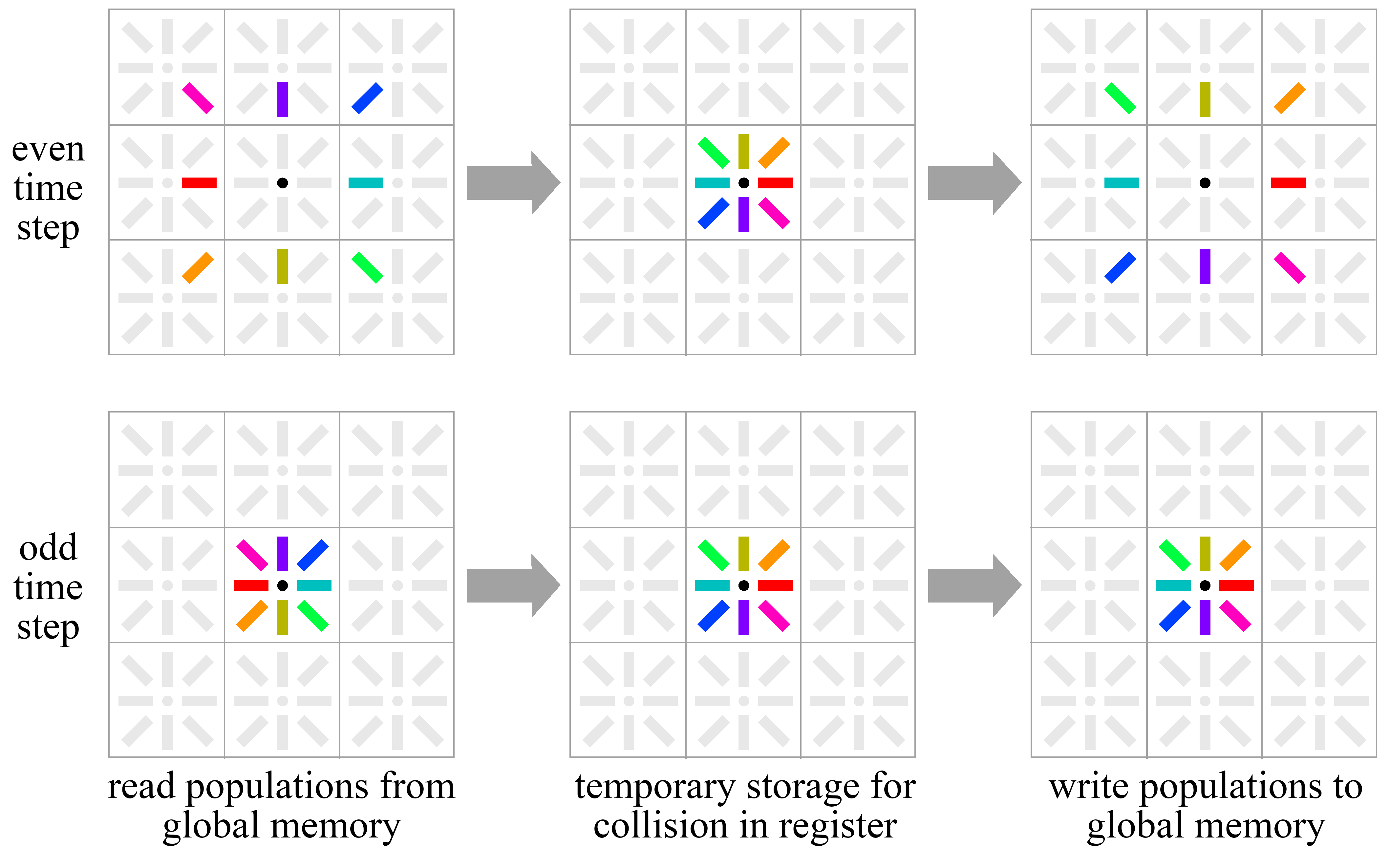}
\caption{A sketch showing the AA-Pattern in-place streaming scheme by Bailey et al. \cite{bailey2009accelerating}. Note that the populations are always stored in the same memory location from where they were fetched. Hence, only a single copy of the population needs to be stored. The figure is reproduced with permission from Ref. \cite{lehmann2022esoteric}.}
\label{aa-patter}
\end{figure*}

\subsection{First group} Notable instances of the first strategy are the so-called ``in-place'' streaming, such as AA-Pattern and esoteric twist \cite{lehmann2022esoteric,geier2017esoteric}. These approaches avoid using a double memory space to store the probabilistic populations; rather, they are aimed to read, compute, and write the results of the LB algorithm on the same memory space, thus halving the memory requirement. However, a race condition problem emerges when adjacent lattice points are handled simultaneously on parallel HPC clusters based on GPUs because the exact order of the floating-point operation execution is random \cite{lehmann2022esoteric}. Consequently, more threads could read a value from a memory address already updated by another concurrent thread.
To circumvent this issue,
Bailey et al. \cite{bailey2009accelerating} proposed an implementation strategy in which a streaming step, followed by a collision and another streaming step, are performed
at even time steps while the collision step and a direction shift are executed at odd time steps only (see Fig.\ref{aa-patter}). 
Thus, the algorithm, unlike the classical AB-pattern, reads the populations from copy A and writes back to the same copy A in place, whence the name the AA-pattern. 
Similarly, the esoteric twist in-place streaming scheme \cite{geier2017esoteric} and its variants (such as esoteric pull and esoteric push \cite{lehmann2022esoteric}), split the time integration into even and odd steps, performing the streaming of  half of the set of populations (e.g. populations with negative directions only) in the even time step, while the set of the remaining directions is streamed in the odd time step. Hence, a precise set of shifting rules is applied at the end of the even and odd time steps to resemble the correct pattern every two time steps \cite{lehmann2022esoteric}. However, the implementation of esoteric schemes is generally less straightforward 
since the populations are not always stored in the right positions, a drawback that further complicates a correct design of the boundary conditions.
Nonetheless, both approaches, AA-Pattern and esoteric twist, halve the memory requirements alongside the number of data accesses to the global memory.

\subsection{Second group} The second strategy employs 
a single precision (instead of a double one) or a mixed single precision/half-precision representation
which reduces the total memory usage by a factor of two without compromising the simulation quality \cite{lehmann2022accuracy,gray2016enhancing}. This approach was initially introduced by  
Skordos \cite{skordos1993initial}, 
who showed that if the distribution functions are adjusted by negating the equilibrium zero-velocity distribution function, the probabilistic populations have the potential to maintain more significant bits during the execution of floating-point operations, thus increasing the computational precision. This idea was further developed in Ref.\cite{gray2016enhancing}, where the dependence on the velocity of the accuracy of LB calculations, affecting the previous approach, was removed. 
More specifically, this method requires a modification of distribution functions and moments as follows.
Using  the symbols (float32) and (float64) to indicate the cast operators forcing a data type to be converted into single (32 bits/4 bytes) and double (64 bits/8 bytes) floating-point precision, and writing
\begin{equation}
f_i({\bf x},t)=f_i^{eq}(\rho_0,{\bf u}_0)+\delta f_i({\bf  x},t),
\label{eq:shift_f}
\end{equation}
where $\rho_0$ and ${\bf u}_0$ are reference values (taken for simplicity equal to one and zero, respectively),
the populations can be stored on the global memory saving only the extra term $[\delta f_{i}({\bf x},t)]_{\text{32bit}}= (\text{float32}) (f_i({\bf x},t)-f_i^{eq}(\rho_0,{\bf u}_0))$. Here the subscript  $[\cdot]_{\text{32bit}}$ remarks that the values were saved in single floating-point precision  (FP-32), while the precision in the floating-point operations is retained in double precision (FP-64). Thus, in a mixed-precision paradigm (FP-64/FP-32), the LB Eq.(\ref{eq:lbm}) with the BGK collision operator can be rewritten as
\begin{eqnarray}
&&[\delta f_i({\bf x}+ {\bf c}_i \Delta t,t+\Delta t)]_{\text{32bit}}= (\text{float32}) \biggl\{\omega \; \delta f_i^{eq}\nonumber\\&& + (1-\omega) (\text{float64}) [\delta f_i( {\bf x},t)]_{\text{32bit}}\biggr\},
\label{eq:shift-lbm}
\end{eqnarray}
where $[\delta f_i({\bf  x},t)]_{\text{32bit}}$ is first converted to double precision by the cast operator $(\text{float64})$. Hence, all the floating point operations were carried out in double precision with  $\delta f_i^{eq}$ defined as
\begin{eqnarray}
&&\delta f_i^{eq} =f_i^{eq}(\rho,{\bf u})-f_i^{eq}(\rho_0,{\bf u}_0) =\nonumber\\&& w_i \Big[ \frac{{\bf c}_i\cdot{\bf u}}{c_s^2}+\frac{({\bf c}_i 
{\bf c}_i- c_s^2 \mathbf{I}):{\bf u} {\bf u} }{2 c_s^4} \Big] + w_i (\rho-\rho_0 ),
\label{eq:shift-eq}
\end{eqnarray}
with ${\bf u}_0=0$. Lastly, the result of Eq.(\ref{eq:shift-lbm}) is again converted in single precision by applying the cast operator $(\text{float32})$ and stored in the global memory. The macroscopic hydrodynamics fields can be computed as
\begin{equation}
\rho = \sum_i  (\text{float64}) [\delta f_i( {\bf x},t)]_{\text{32bit}} + \rho_0,
\label{eq:rho-shift}
\end{equation}
\begin{equation}
\rho \vec{u} = \sum_i   (\text{float64}) [\delta f_i({\bf x},t)]_{\text{32bit}} {\bf c}_{i}.
\label{eq:u-shift}
\end{equation}
Note that 
the distributions $\delta f_i(\vec x,t)$ are centered around zero and shifted by $f_i^{eq}(\rho_0,{\bf u}_0)$, which means that the accuracy of the summation of small differences is not limited by the order of magnitude of the density $\rho_0$. Further, the summation can be performed with double precision accuracy. 

\begin{figure*}[htbp]
\centering
\includegraphics[width=1.0\linewidth]{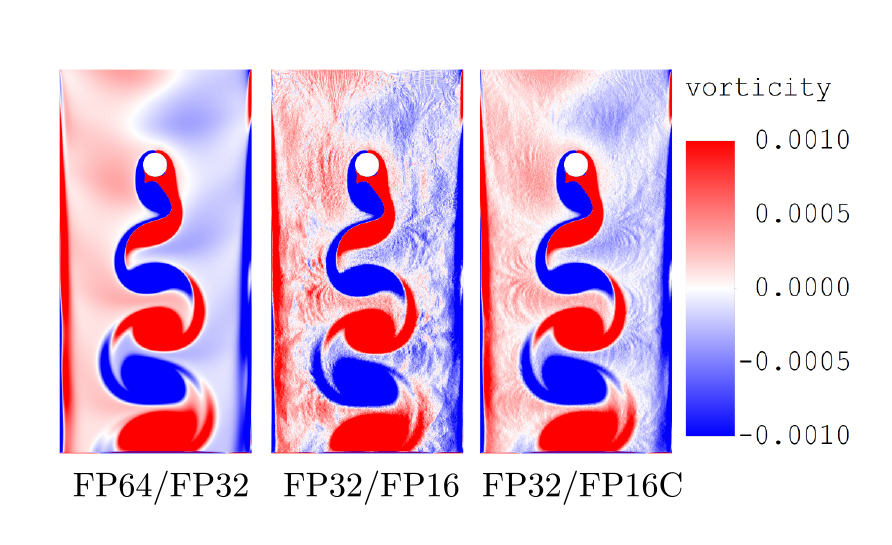}
\caption{LB simulations of the Karman vortex street in two dimensions using the population shifting approach on various mixed floating-point precision platforms.  The vorticity lies in the zoomed range of $\pm 0.001$, after $10^5$ LB time steps.
For the 16-bit formats, noise is present in low vorticity regions. The figure is reproduced
with permission from Ref. \cite{lehmann2022accuracy}.}
\label{mixed-precision}
\end{figure*}

Recently, Lehmann et al. \cite{lehmann2022accuracy} implemented the same strategy exploiting a mixed precision approach (FP-32/FP-16), where the floating-point operations are carried out in single precision, while the extra population terms are saved in half-precision (16 bits/2 bytes) $[\delta f_{i}(\vec x,t)]_{\text{16bit}}$. In particular, they introduced a customized half-precision number format (FP16C) that halved truncation error compared to the standard IEEE-754 half-precision floating-point format FP16 in LBM applications \cite{lehmann2022accuracy}. However, benchmarks on the Karman vortex street in two dimensions have shown the presence of numerical noise in the third decimal digits, where the vorticity is compared between the mixed precision platforms FP64/32 and FP32/16C (see Fig.\ref{mixed-precision}). 

In concrete numbers, it is possible to quantify the LB performance in terms of billion lattice sites per second (GLUPS), namely the number of updated lattice nodes per second reported in billions. For instance, Lehmann et al. \cite{lehmann2022accuracy} measured a performance equal to 8.5 GLUPS on a single GPU NVIDIA A100 PCIe with 40GB RAM implementing the AA-Pattern in-place streaming scheme for a single component BGK LB model in a single precision floating-point, 
and a performance peak of 16 GLUPS by adopting the mixed precision approach (FP-32/FP-16).
A list of benchmarks carried out using their code (named FluidX3d, freely available on GitHub \cite{FluidX3D})
on several GPU and CPU-based devices are reported in Table \ref{benchmark-lehmann}. 
They were performed using a standard BGK operator with a D3Q19 scheme on a three-dimensional cubic box of 256 nodes per side, for different mixed floating-point representations. 

\begin{table}[h]
\small
\resizebox{\columnwidth}{!}{
\begin{tabular}{llll}
\hline
CPU hardware & FP32/FP32  &  FP32/FP16 & FP32/FP16C \\ \hline
 2x Intel Xeon CPU Max 9480 & 2.037 & 1.520  & 1.464  \\
 2x Intel Xeon Platinum 8380 & 1.410 & 1.159 & 1.298  \\
 Intel Core i7-13700K & 0.504 & 0.398 & 0.424  \\
 Intel Core i7-9700 & 0.103 & 0.062 & 0.095 \\ \hline 
GPU hardware & FP32/FP32  &  FP32/FP16 & FP32/FP16C \\ \hline
 Nvidia H100 NVL & 20.018 & 32.613 & 17.605 \\ 
 Nvidia A100 PCIe 40GB & 8.526 & 16.035 & 11.088 \\
 Nvidia V100 PCIe 16GB & 5.128 & 10.325 & 7.683 \\  
 Nvidia GeForce RTX 4090 & 5.624 & 11.091 & 11.496 \\ \hline
\end{tabular}
}
\caption{Benchmarks performed by a standard BGK operator with the D3Q19 scheme on a three-dimensional cubic box of 256 nodes per side for different mixed floating point-representations. The results are reported in GLUPS  for several Intel CPU and Nvidia GPU architectures. The table is reproduced with permission from Ref. \cite{FluidX3D}.}
\label{benchmark-lehmann}
\end{table}

\subsection{Third group} The third strategy for managing memory usage is grounded on the ideas of Ladd and Verberg \cite{ladd2001lattice}, in which a memory reduction could be achieved by solely storing hydrodynamic fields and their gradients. This concept hinges on the fact that the probabilistic data needed to execute any LB method can be dynamically reconstructed using the available hydrodynamic quantities, thus eliminating the need for population storage. This is especially beneficial for multi-component and multi-species applications transported by a common flow field in which 
a single hydrodynamic field (i.e., the density) is needed rather than a full kinetic representation with $O(30)$ populations.
A similar logic applies to flows that are far from equilibrium or involve relativistic hydrodynamics, which require higher-order lattices, sometimes involving hundreds of discrete populations per species. All methods based on this third approach typically rely on a moment-based representation of the LB, bypassing the direct storage of the probabilistic populations along each direction of the lattice. 
Among these, noteworthy implementations include the recently introduced lightweight LB (LLB) and thread-safe LB (TSLB) methods \cite{tiribocchi2023lightweight,montessori2023thread,montessori_high-order}, which are based on the reconstruction of probabilistic distributions via Hermite projection. 

As in standard LB methods, the moment-based approach can be built starting from a set of distribution functions $f_i$ where each $f_i$  can be expanded around the equilibrium value 
\begin{equation}
\label{eq:expanded}
f_i = f_i^{(0)} + f_i^{(1)} + f_i^{(2)}+ f_i^{(3)}+\dots,
\end{equation}
being $f_i^{(0)} \stackrel{\triangle}= f_i^{eq}(\rho,{\bf u})$ \cite{kruger2009shear,latt2006,zhang2006efficient,ladd2001lattice}. All other components $f_i^{(k)}$ are
sought in the order $O(\epsilon^k)$, where $\epsilon$ is the Knudsen number with
$\epsilon\ll 1$.
Using a multiscale Chapman-Enskog expansion \cite{chapman1990mathematical}, it can be shown that only the first two terms, $f_i^{(0)}$ and $f_i^{(1)}$, are sufficient to recover (asymptotically) the Navier-Stokes equation \cite{halim,latt2008straight}.
Consequently, one can write $f_i = f_i^{(0)} + f_i^{(1)} + O(\epsilon^2)$
and identify 
$f_i^{(1)}= f_i - f_i^{(0)}+O(\epsilon^2) $ 
as the non-equilibrium hydrodynamic term, apart from $O(\epsilon^2)$ contributions. 
Therefore, if a relation between the term $f_i^{(1)}$ and the hydrodynamic fields is known, the distribution functions can be reconstructed without the direct storage of the probabilistic populations.

Such relation can be efficiently obtained by resorting to the regularization procedure \cite{latt2006}, whose  main 
goal is to convert $f^{neq}_i=f_i - f_i^{(0)}$ into a set of non-equilibrium distributions  
lying on a Hermite subspace spanned by the first three statistical moments  $\rho$, $\rho \mathbf{u}$ and $\Pi^{neq}$ \cite{zhang2006efficient}.
More specifically, by introducing Hermite polynomials and Gauss-Hermite quadratures, $f_i^{neq}$ is given by \cite{shan2006kinetic, montessori2015lattice}
\begin{equation} \label{hermite}
    f^{neq}_i=w_i \sum_n \frac{1}{n!}\mathbf{a}^n \mathcal{H}^n(\mathbf{c}_i),
\end{equation}
where $\mathcal{H}^n(\mathbf{c}_i)$
and $\mathbf{a}^n=\sum_i \hat{f}^{neq}_i \mathcal{H}^n(\mathbf{c}_i)$ 
are n-th order rank tensors. 
The former is the standard n-th order tensor Hermite polynomial
and the latter is the corresponding Hermite expansion coefficient. 
The non-equilibrium set of distributions, defined in Eq.(\ref{hermite}), contains information from hydrodynamic moments up to the second order of the expansion (for the Navier-Stokes level) and is free  
from higher-order fluxes \cite{zhang2006efficient,tucny2022kinetic}, thus can compactly be written as
\cite{latt2006,montessori2015lattice}
\begin{equation}\label{distnoneq}    f^{neq}_i(\Pi^{neq}_{\alpha\beta})=\frac{w_i}{2 c_s^4}(\mathbf{c}_i\mathbf{c}_i - c_s^2\delta_{\alpha\beta}):\Pi^{neq}_{\alpha\beta}+O(\epsilon^2),
\end{equation}
where $\Pi^{neq}_{\alpha\beta}=\Pi_{\alpha\beta}-\rho u_{\alpha} u_{\beta}$.

Also, note that 
the post-collision distribution $f^{pc}_{i}$ becomes \cite{succi2018lattice}
\begin{equation}
    f^{pc}_{i} = f_{i}^{pre} + \omega(f^{eq}_{i} - f_{i}^{pre})
    =f^{eq}_{i} + (1-\omega)f^{neq}_{i},
\label{LBpostmom}
\end{equation}
where $f_i^{pre}$ is the pre-collisional set of distributions
and the second equality stems from the fact that the full lattice distribution can be split into an equilibrium and non-equilibrium part as
$f_i = f_i^{eq}+f^{neq}_i$.
Thus, including the  streaming step,
Eq.(\ref{eq:lbm}) 
can be written in terms of three macroscopic quantities, i.e. density, momentum, and non-equilibrium stress tensor, as follows
\begin{eqnarray}
f_i({\bf x}+{\bf c}_i\Delta t,t+\Delta t)&&=f^{eq}_{i} \biggl( \rho({\bf x},t),{\bf u}({\bf x},t) \biggr) \nonumber\\&&+ \left( 1-\omega \right) f^{neq}_{i} \biggl( \Pi^{neq}_{\alpha\beta}({\bf x},t) \biggr),
\label{eq:llb-pull}
\end{eqnarray}
where $f^{eq}_{i}$ and $f^{neq}_{i}$ are computed via Eq.(\ref{disteq}) and Eq.(\ref{distnoneq}), which explicitly depend on the hydrodynamics fields. These fields are then updated using Eqs. (\ref{eq:rho}), (\ref{eq:u}) and (\ref{eq:Pi}).
 
\begin{figure*}[htbp]
\centering
\includegraphics[width=1.0\linewidth]{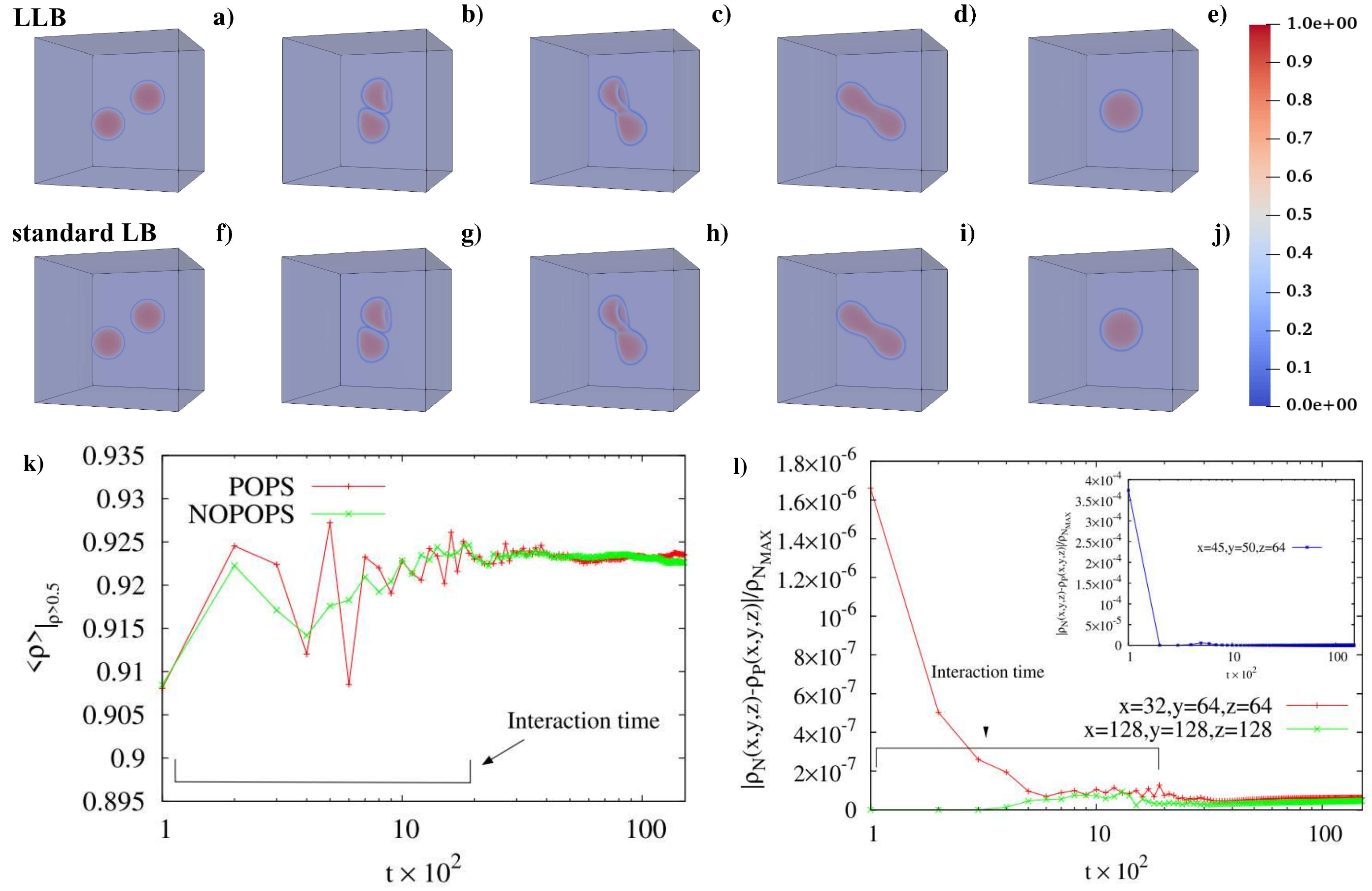}
\caption{A sequence of an off-axis collision between two fluid droplets simulated using the LLB scheme without populations (a)-(e) and a standard LB scheme with populations (f)-(j). The centers of mass are initially positioned at a distance where the droplets' mutual interaction is minimal. The droplets, starting as perfect spheres [(a)-(f)], acquire a bullet-like shape as they collide (b), (c) and (g), (h). They then separate (d)-(i) and regain a spherical shape once sufficiently far apart (e)-(j). There is no significant difference between the two methods. The color map shows the density field values, which range from 0 to 1. k) Time evolution of the fluid density $\rho$ computed using a standard LB (POPS) and the lightweight LB (NOPOPS), where $\langle...\rangle$ is a spatial average on lattice sites with $\rho > 0.5$ (i.e. within the droplets, red region). The interaction time
between the droplets is $10^2 \lesssim t \lesssim 15 \times 10^2$. l) Time evolution of the relative error $|\rho_N-\rho_P|/\rho_{N_{MAX}}$
at three different lattice sites, i.e. near the droplet interface $(x=32, y=64, z=64)$, within the droplet
$(x=45, y=50, z=64)$ and far from the droplets $(x=128, y=128, z=128)$. Here, $\rho_N$ is the density computed using the LLB and $\rho_P$ is the one from the standard LB. The figure is adapted with permission from Ref. \cite{tiribocchi2023lightweight}.}
\label{LLB-compare}
\end{figure*}

This procedure has been used to implement LLB 
and TSLB models described in Refs.\cite{tiribocchi2023lightweight,montessori2023thread} for single and two-component fluids.  
In Fig.\ref{LLB-compare} we show, for example,  an off-axis collision between equal-size fluid droplets in a cubic box, simulated using the LLB method presented in Ref.\cite{tiribocchi2023lightweight} (Fig.\ref{LLB-compare}a-e) and the standard LB method (Fig.\ref{LLB-compare}f-j). In both cases, the relative impact speed is fixed at 0.5 in lattice units, the Reynolds and Capillary numbers for the droplet (red fluid) are Re$_R$ = 450 and Ca$_R$ $\simeq 1.6$, and Re$_B$ = 50 for the surrounding medium (blue fluid). The time sequence of the collision plus the plots of density and relative error in Fig.\ref{LLB-compare} clearly show that both methods result in essentially similar dynamic behaviors.

The immediate advantage of the moment-based approach is that it allows the reconstruction of post-streamed and post-collided distributions using the local values of the hydrodynamic fields, without the need of streaming the distributions along lattice directions. 
This is expected to be particularly suited for the design of efficient LB models on shared-memory architecture, as recently demonstrated in Refs.\cite{montessori2023thread,montessori_high-order},  
since it eliminates memory dependencies emerging during non-local read and write operations and avoids race conditions potentially jeopardizing the memory access. In addition, the model may also hold interest in simulations on unstructured grids, precisely because the distributions can be reconstructed off lattice using the macroscopic fields. Finally, it decisively improves the computational performances of the LB method with respect to standard implementations (up to 40 percent of memory savings on large-scale simulations \cite{tiribocchi2023lightweight}), thus opening new avenues for the simulation of the 
multiscale physics of soft materials, 
where the necessity of minimizing data access and memory usage is often mandatory.

We conclude noting that, over the last two decades, several examples of moment-based models have been published, such as the CPU-based one presented in Ref.\cite{argentini2004efficiently} and more sophisticated releases implemented on GPU computers  
\cite{ferrari2023graphic,gounley2021propagation,vardhan2019moment}, 
where ad-hoc memory layouts are adopted for the hydrodynamics array to capitalize the full memory bandwidth of multi-GPU clusters \cite{ferrari2023graphic}. These techniques have been shown to enhance data continuity, reducing the impact of memory-bound dependence on the implementation \cite{bader2012space} with benefits in terms of computational performance. As a concrete example, Ferrari et al. \cite{ferrari2023graphic} measured a peak of 10.5 GLUPS on a single GPU NVIDIA A100 PCIe with 40GB RAM adopting a z-curve memory layout for a single component regularized-LB model with an increase of about 20\% compared to what observed by Lehmann et al. \cite{lehmann2022accuracy} for a single component BGK-LB model at the same floating-point precision. 

\subsection{Performance comparison between lattice Boltzmann and Navier-Stokes equation solvers}

We close this section by providing a quantitative comparison, in terms of GLUPS, between LB methods and Navier-Stokes solvers. 
We begin by considering the declared performance of the commercial computational fluid dynamic code ANSYS Fluent \cite{matsson2023introduction} based on the volume of fluid method, focusing on its GPU-accelerated version. For a single-component CFD simulation on a cubic grid of 126 lattice nodes per side (roughly 2 million nodes in total), the reported performance on a single GPU (NVidia A100) is 0.042 GLUPS, scaling up to 0.266 GLUPS when using 8 GPUs.
In comparison, a recent version of the LB method \cite{FluidX3D}, operating on a cubic grid of 256 lattice nodes per side (more than 16 million nodes in total), achieves approximately 8.5 GLUPS on a single GPU (see Table \ref{benchmark-lehmann}), roughly 200 times faster than the single-GPU performance of ANSYS code. It is worth noting that other non-commercial, GNU-licensed Navier-Stokes solvers, which are often optimized for the simulation of specific flow types (see \cite{COSTA2021502}, for example), show better performance than ANSYS, typically up to an order of magnitude faster. However, even in these cases, LB methods still demonstrate higher performance when it comes to updating a single computational node.

\begin{figure}[htbp]
\centering
\includegraphics[width=1.0\linewidth]{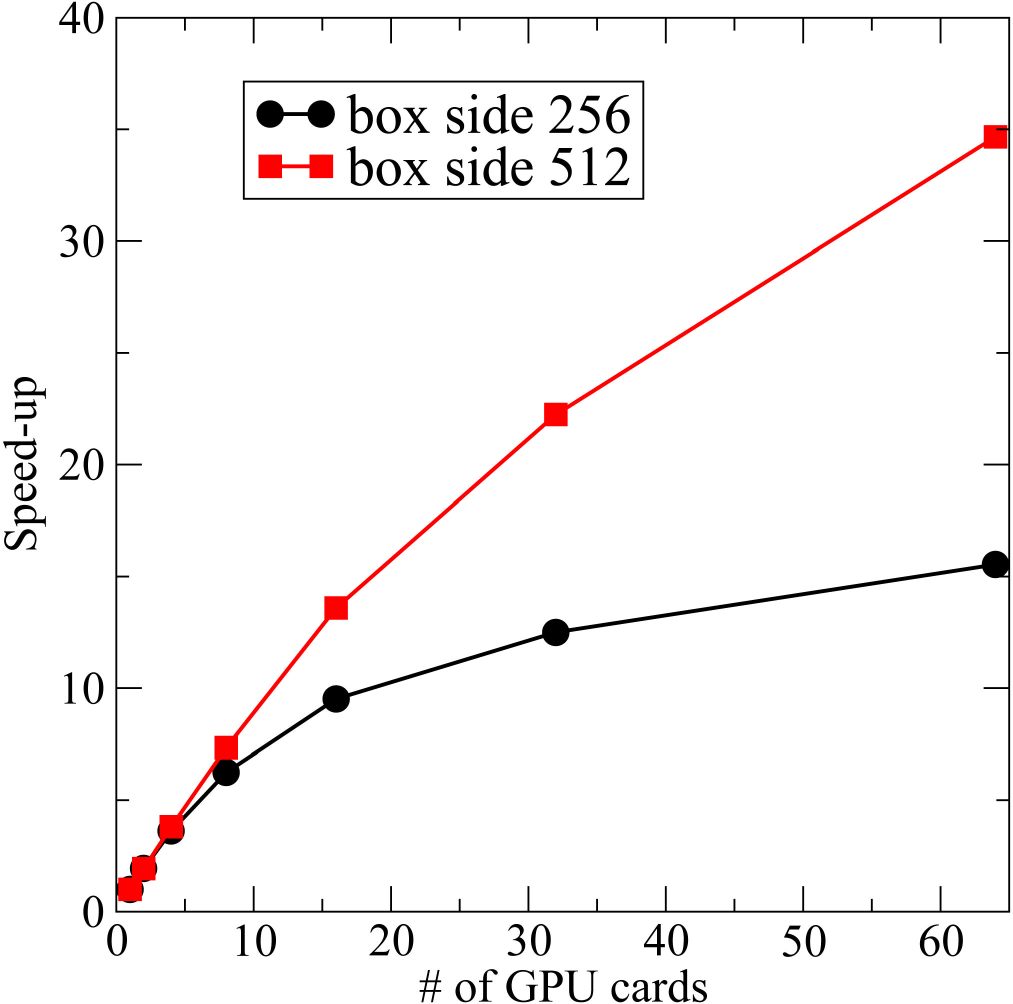}
\caption{Speed-up of the LBcuda code using different A100 GPUs (each with 80 GB of memory), tested on two cubic grids with side lengths of 256 and 512, respectively. The figure is adapted with permission from Ref.\cite{bonaccorso2022lbcuda}.}
\label{speed-up-lbcuda}
\end{figure}

Regarding massively parallel implementations, GNU-licensed Navier-Stokes solvers show
high scalability on multi-GPU clusters. The scalability can be evaluated in terms of the speedup, defined as the ratio between the CPU wall-clock time of a code executed on a single core and the one of the code executed in parallel mode. For instance,
the FluTAS solver, which  uses  finite difference methods (including a pseudo-spectral Poisson solver) to simulate multiphase flows with thermal effects, shows a  speed-up of about 83,
with performance ranging from 0.07 GLUPS on 8 GPUs NVidia A100 
to almost 0.73 GLUPS on 128 GPUs devices on a grid of $1024 \times 512 \times 256$ lattice points \cite{crialesi2023flutas}. 
On the contrary, a recent implementation  of the lattice Boltzmann, named LBcuda \cite{bonaccorso2022lbcuda}, 
proved a speed-up equal to 34 on 64 GPUs NVidia A100 devices
in a cubic box of linear size 512 (see Fig.\ref{speed-up-lbcuda}) and, more generally, a good scaling for increasing number of GPU cards (at least for a sufficiently large system size).
It is finally important to clarify that this comparison does not account for the ability of standard CFD methods to work with unstructured or non-uniform meshes, which remains an area where traditional Navier-Stokes solvers have certain advantages.

\section{Selected applications}\label{Applications}

In this section, we focus on specific applications which have mostly benefited from the LB approaches and where this method is currently playing a game-changing role, meaning by this that the computational modeling of these applications would have been more demanding, if possible at all, than other methods. We distinguish systems for which a validation against quantitative results has been successfully demonstrated, from simulations often inspired by experiments in contiguous fields of research and where the agreement is mostly built on qualitative grounds.
Rather than being a specific limitation, this conveys the idea that the method, besides being  capable of capturing some key features of the physics of highly complex systems, 
has the potential to predict behaviors not yet explored in the context of soft flowing materials.
We refer primarily to a variety of droplet motions in microfluidic devices, the rheology of confined dense emulsions and other complex phenomena which may lay the ground for a new class of droplet-based materials. More specifically, we identify the following five instances where, we believe, the LB method has shown a major impact: soft flowing crystals, soft granular media, dense emulsions under thermal flows, hierarchical multiple emulsions, active gel droplets in highly confined 
environments (such as pore-sized constrictions) and extreme flow simulations modeling macroscopic biological entities, such as deep-sea sponges. 

\subsection{Soft flowing crystals}\label{sfc}

We initially consider the case of fluid droplets flowing in a microfluidic channel, a system simulated using a color-gradient LB incorporating near-contact interactions, as described in section \ref{color-gradient} \cite{montessori_prf,montessori_phil}. Such droplets are usually produced through an emulsification process, in which their generation follows 
from the periodic pinch-off of the liquid jet of the dispersed phaseby  the stream of the continuous phase. Typical microfluidic platforms used for droplet generation are T-junctions, flow focusers and divergent channels.

\begin{figure*}[htbp]
\includegraphics[width=1.0\linewidth]{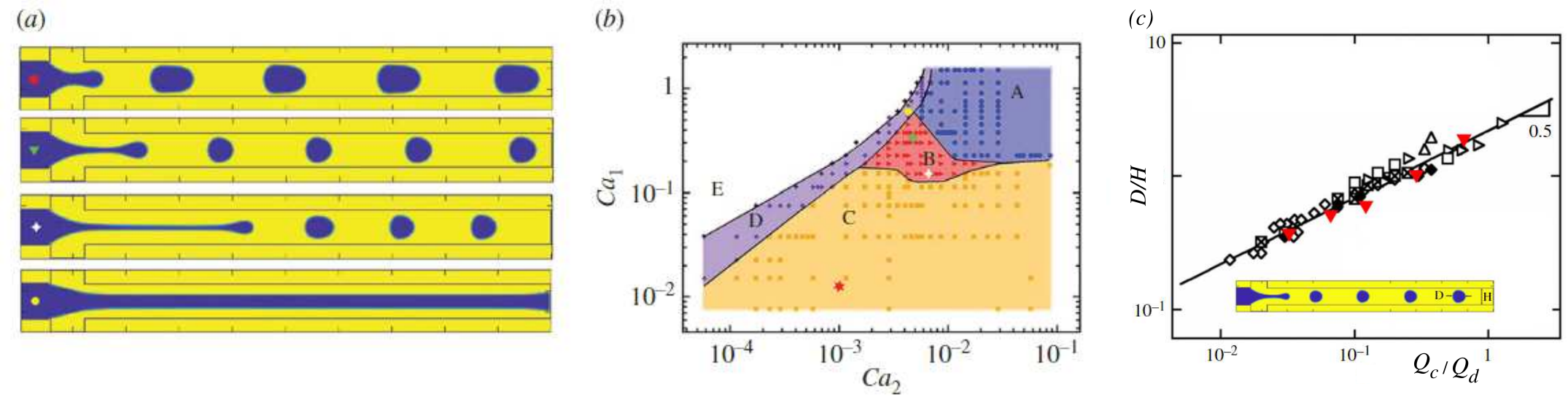}
\caption{(a) Different flow regimes captured by the color gradient LB. From top to bottom: dripping (red aster), jetting (green triangle and white diamond), and tubing (yellow circle). (b) Phase diagram of the different regimes on the $Ca_1-Ca_2$ plane, where $Ca_1$ is the capillary number of the dispersed phase and $Ca_2$ of the continuous one. (c) Plot of the normalized droplet diameter  $D/H$ versus the flow ratio $Q_c/Q_d$. Numerical results (red triangles) and experimental ones from Ref.\cite{cubaud} follow a power law of the flow rate with exponent $\simeq 0.5$. The figure is adapted, with permission, from Ref.\cite{montessori_phil_tr}.}
\label{fig_NEW_sfc}
\end{figure*}

\paragraph{T-junction.} As benchmark test, we start from a T-junction device where a droplet (blue region) is immersed in a continuous phase (yellow region) and flows within the microchannel (see Fig.\ref{fig_NEW_sfc}a). The two components have equal density while the viscosity of the continuous phase is approximately five times larger than that of the dispersed phase, in order to match experimental data of \cite{cubaud}. By changing the capillary number $Ca_1$ and $Ca_2$ of both phases, one can distinguish four flow regimes: a dripping regime, consisting of  elongated droplets, for $Ca_1<0.1$ and $Ca_2<0.1$,
where capillary forces dominate over inertial ones (top row in Fig.\ref{fig_NEW_sfc}a and $C$-zone in Fig.\ref{fig_NEW_sfc}b); two jetting regimes for $0.1<Ca_1<1$ and $10^{-3}<Ca_2<10^{-2}$, where low viscous forces of the continuous phase causes the formation of a thin jet of fluid delaying the droplet generation (second and third rows in Fig.\ref{fig_NEW_sfc}, and $B$-zone in Fig.\ref{fig_NEW_sfc}b); a tubing regime for $Ca_1\simeq 10^{-1}$ and $Ca_2<10^{-2}$, where a jet of the dispersed phase flows in parallel with the continuous one in the microchannel (fourth row in Fig.\ref{fig_NEW_sfc}a and $D$-zone in Fig.\ref{fig_NEW_sfc}b).

Besides reproducing with high accuracy the typical phase diagram of the experimental flow regimes (see Fig.\ref{fig_NEW_sfc}b), the color gradient LB quantitatively captures the behavior of the droplet diameter, which follows a power law of the flow rate over almost two decades.\cite{cubaud} (Fig.\ref{fig_NEW_sfc}c).

\paragraph{Flow focuser.} A more complex geometry is that of a flow-focuser, where an additional orifice separating the inlet and the outlet chambers causes the droplet generation. \cite{marmottant_soft,marmottant_prl,garstecki_apl}. A sketch of the device used in our simulations and the droplet arrangement observed in the channel are shown in Fig.\ref{fig_flow}, where the resulting emulsion is stabilized by dispersing surfactant molecules preventing coalescence.
These devices are particularly suited for the design of highly monodisperse droplets which are of relevance, for example, for the assembly of droplet-based soft templates with a well-defined structure \cite{weitz_loc,garstecki,costantini}.

A typical example are soft flowing crystals (also termed as microfluidic crystals \cite{marmottant_soft}), 
where monodisperse fluid droplets added in a self-repeating way are found to self-assemble in regular patterns, closely resembling the hexagonal order of solids. In LB simulations, droplet formation can be controlled by tuning the ratio of dispersed-to-continuous inlet flow, defined as $\chi=u_d/2u_c$ (where $u_d$ and $u_c$ are the speed of dispersed and continuous phases) and the dimensionless near-contact number ${\cal N}_c=A\Delta x/\sigma$ (see also Eq.\ref{near_cont_force}). Here, $A$ is a constant setting the magnitude of the repulsive force and $\Delta x$ is the minimum distance between two interfaces in close contact, generally ranging from $10$ to $100$ nanometers. If ${\cal N}_c\ll 1$ droplets coalesce, otherwise repulsive near-contact forces prevail and droplet merging is inhibited.

\begin{figure*}[htbp]
\includegraphics[width=1.0\linewidth]{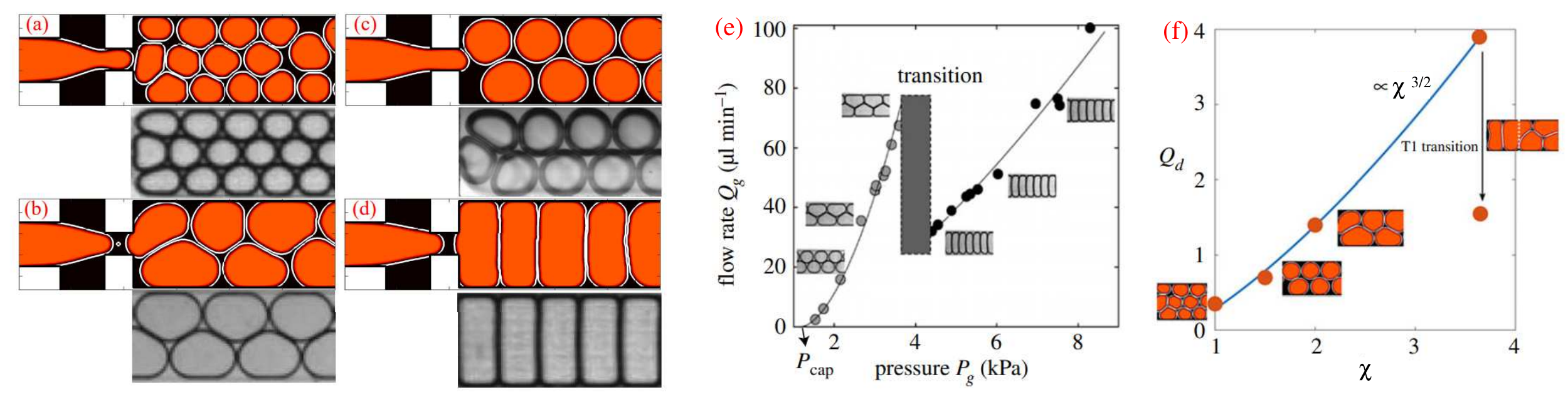}
\caption{(a)-(d) Simulated (top rows) and experimental (bottom rows) structures of flowing droplets in a microfluidic channel for $\chi=1$ (a) (hex-three), $\chi=1.5$ (b) (wet hex-two), $\chi=2$ (c) (dry hex-two) and $\chi=3.6$ (d) (hex-one). Here $\chi=u_d/2u_c$, being $u_d$ and $u_c$ the speed of dispersed and continuous phases. (e)-(f) Flow curves of experimental \cite{marmottant_soft} and simulated \cite{montessori_phil} soft flowing crystals.
In (e) $P_g$ and $Q_g$ are the gas pressure and gas flow rate. For values of $P_g$ lower than $4$kPa, the mixture is in the hex-two regime, where the flow rate scales as $Q_g \propto P^{\alpha}$ and $\alpha\simeq 3/2$. At $P_g\simeq 4$kPa, $Q_g$ displays a sharp transition towards the hex-one regime. In (f) $Q_d=\nu V_d$ is the flow rate of the dispersed phase, where $\nu$ is the frequency of droplet generation and $V_d$ is the volume of the generated droplet. 
The flow rate follows a power law with exponent $\simeq 3/2$ for low values of $\chi$, while for $\chi\simeq 3.5$ a decrease of $Q_d$ marks the transition from the hex-two to the hex-one phase. The transition front is indicated by a white dotted line in the top inset. The simulations are carried out on a 3D box of $420\times 80\times 20$ lattice nodes. The figures are reproduced with permission from Ref.\cite{montessori_prf,montessori_phil}.}
\label{fig_hex}
\end{figure*}

In Fig.\ref{fig_hex}a-d, we show a variety of crystal-like structures confined in microfluidic channels obtained from three-dimensional LB simulations \cite{montessori_prf,montessori_phil} and experiments \cite{marmottant_soft,marmottant_prl} of air bubbles in water. 
Despite the difference in density ratio, the two systems exhibit very similar features. In particular,
the emulsion displays a number of phases comprising a three-row configuration (hex-three, (a)), where a regular array of droplets accommodates along three parallel rows, two two-row ones (wet (b) and dry (c) hex-two) and a single-row structure (hex-one, (d)), in which highly-packed droplets arrange in a neat line. 
Besides being capable of describing the formation and ordering properties of fluid emulsions under confinement, LB simulations also capture the behavior of the experimental flow rate curve and, remarkably, the non-trivial transition from the hex-two to the hex-one regime (see Fig.\ref{fig_hex}e,f). Indeed, experiments show that, for a gas-liquid foam, $Q_g\propto P_g^{\alpha}$, where $Q_g$ and $P_g$ are flow rate and pressure of the gas, and $\alpha=3/2$ for small values of $P_g$ \cite{marmottant_soft}. On the contrary, for higher values of $P_g$, $Q_g$ displays a sharp decrease due to the augmented friction of the interfaces with the walls, an effect signaling the transition to the hex-one phase (Fig.\ref{fig_hex}e). These results are reproduced, with high accuracy, by LB simulations of fluid emulsions, where the flow rate $Q_d$ of the dispersed phase follows a power law of $\chi$ (which plays the same role as the gas pressure) with an almost equal value of the exponent. In agreement with experiments, simulations also show that the transition from the hex-two to the hex-one phase is marked by a T1 event \cite{graner_epje,graner_epje2,weaire} (corresponding to a topological rearrangement of neighboring droplets) and a transition front, separating the two phases and comoving with the flowing emulsion.

\begin{figure*}[htbp]
\includegraphics[width=1.0\linewidth]{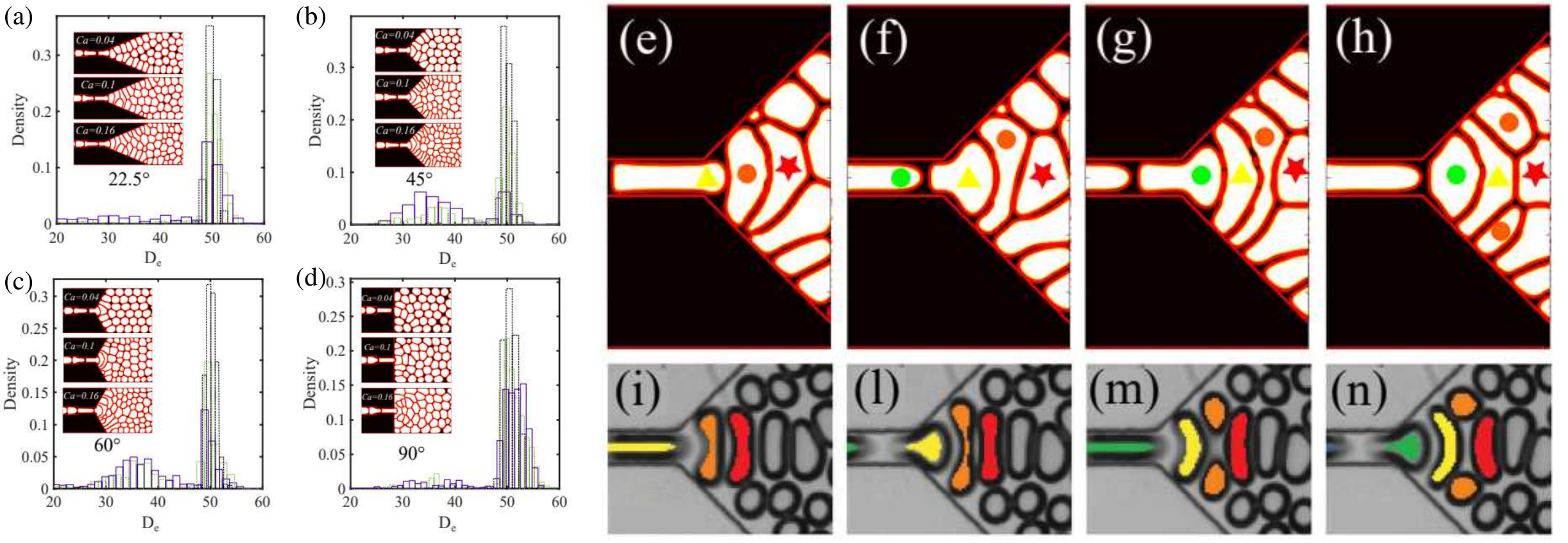}
\caption{(a)-(d). Droplet assembly in diverging channels of opening angle $\theta$ equal to $22.5^{\circ}$ (a), $45^{\circ}$ (b), $60^{\circ}$ (c), $90^{\circ}$ (d) and three different values of $Ca$. The histograms show the distribution of droplet diameters $D_e$ for $Ca=0.04$ (dashed line), $Ca=0.1$ (dotted line) $Ca=0.16$ (full line). Note that, for $45^{\circ}\lesssim \theta\lesssim 60^{\circ}$, the emulsion displays a bidisperse structure (disordered), while for values out of that range, the emulsion is essentially monodisperse (ordered). (e)-(n). Pinch-off mechanism from LB simulations (e-h) and experiments (i-n) at $Ca\simeq 0.08$ and $\theta=45^{\circ}$. The split of the orange droplet is caused by the confining effect of the yellow droplet (the "pincher") and the red one (the "wall"). The figure is reproduced with permission from Ref.\cite{montessori_wetdry}.}
\label{div_channel}
\end{figure*}

\paragraph{Divergent channel.} LB simulations of microflows have also been capable of describing morphology and assembly of droplets flowing in convergent and divergent channels \cite{montessori_prf,gai,tang,kluo,montessori_pof,montessori_wetdry}, as the ones shown in Fig.\ref{div_channel}, albeit in these cases mostly on a qualitative ground.
More specifically, numerical results suggest that the degree of monodispersity and the arrangement of a confined dense emulsion can be controlled by properly tuning the opening angle $\theta$ of a divergent channel and Capillary number $Ca$ (see Fig.\ref{div_channel}). Indeed, while for $\theta\lesssim 40^{\circ}$ the emulsion is generally monodisperse (Fig.\ref{div_channel}a) regardless of the values of $Ca$, for $45^{\circ}\lesssim\theta\lesssim 60^{\circ}$
the generated emulsion becomes bidisperse, an effect amplified 
at high values of $Ca$ (see Fig.\ref{div_channel}b,c) because breakup events are more likely to occur. 
The mechanics leading to droplet rupture is highlighted in Fig.\ref{div_channel}e-n. 
The orange droplet gradually stretches because of the combined effect of i) the normal force of the incoming droplet (yellow), ii) the extensional force of the channel expansion, and iii) the downstream droplet (red) acting as a "wall". Once a critical elongation is overcome, the orange droplet fragments produce two smaller droplets moving towards the upper and lower part of the channel. Then the process self-repeats; the yellow droplet fills the gap left by the orange one and acts as a "wall" droplet, while a newly generated one will stretch and finally break following a protocol similar to the orange one. 
Interestingly, further increasing $\alpha$ restores a monodisperse emulsion, once again regardless of the values of $Ca$ (see Fig.\ref{div_channel}d). This counter-intuitive behavior is due to the fast recovery of the circular shape in the main channel because of larger aperture angles. Such an expansion favors the transversal displacement 
of the incoming droplet, thus suppressing the squeezing effect. Note finally that, for all geometries considered, increasing $Ca$ generally leads to the formation of a foamy-like emulsion; this is because, in this regime, morphological deformations become easier and, once breakups occur, the resulting smaller droplets fill the voids among neighboring ones. 

A reliable description of foamy flows can also be obtained by using other numerical techniques, such as volume-of-fluids methods, which have recently demonstrated the capability of capturing the formation and ordering of air bubbles dispersed in water (thus high-density ratio mixtures) and flowing within microchannels \cite{koum_sci}. However, these methods have so far shown limited applicability to describe higher complex systems where close-packed deformable drops (or bubbles) are encapsulated within a soft shell, i.e. a soft granular medium with a double emulsion structure. 
In the next section, we discuss some recent advances in understanding the properties of these systems using LB simulations.

\subsection{Soft granular media}\label{sgm}

Soft granular materials abound in nature, with examples ranging from foams and dense emulsions \cite{datta,montessori_lan,guzowski_prl,bogdan_prl,adams,utada,abate} to biological tissues \cite{weaire,pitois,cuvelier}.  
The simplest realization in the lab consists of a high volume fraction of immiscible droplets encapsulated within a soft shell containing an oil phase, in turn, immersed in a bulk aqueous phase. 

\begin{figure*}[htbp]
\includegraphics[width=1.0\linewidth]{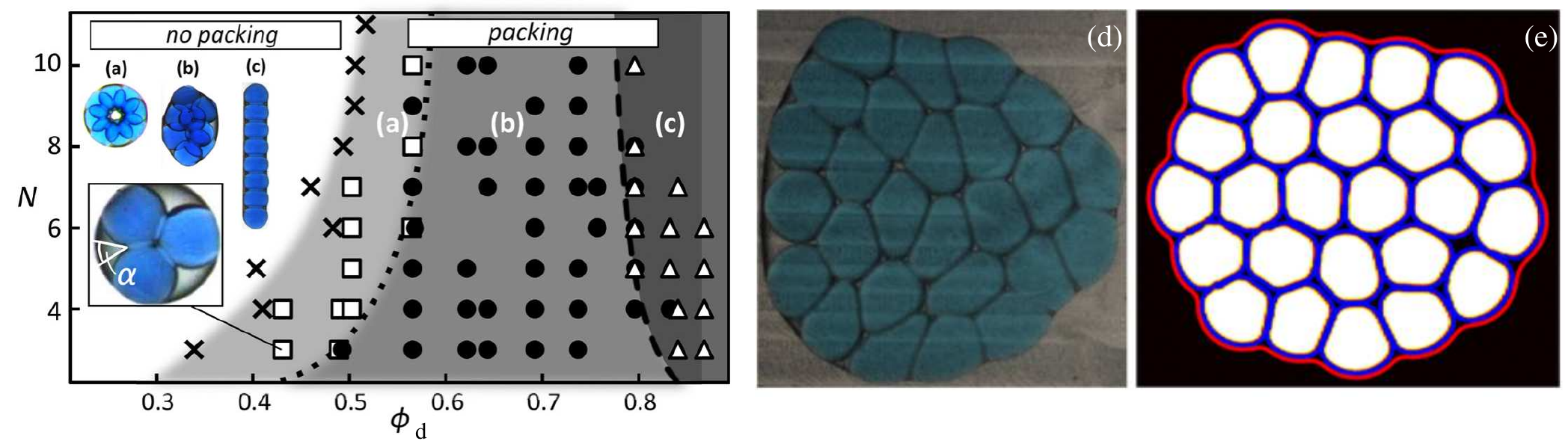}
\caption{Left panel, (a)-(c). Phase diagram of a soft granular medium, where $N$ is the number of cores and $\phi_d$ is the volume fraction they occupy. Insets (a), (b) and (c) represent three possible configurations, where (a) indicates weakly deformed shells, (b) highly deformed ones, and (c) a linear chain. Right panel, (d)-(e). Experimental realization of a soft granular medium (d) formulated using three Newtonian fluids. A result obtained from LB simulations (e) using three-component fluid, i.e. the dispersed phase (white droplets), the lubricating phase (blue), and an external fluid (black). Copyright granted under the \href{https://creativecommons.org/licenses/by/4.0/}{CC 4.0 International license} and \href{https://creativecommons.org/licenses/by/3.0/}{CC 3.0 license} respectively. No changes were made from the original figure from Ref.\cite{montessori_lan, guzowski_prl}.}
\label{hipde}
\end{figure*}

Recent experiments have shown that these materials exhibit a complex phase diagram, where multiple topologies can be obtained by varying the volume fraction $\phi_d$ and the number $N$ of the internal droplets (Fig.\ref{hipde}a-c). 
Indeed, one can identify a region where the packing mildly deforms the emulsion (for $\phi_d\simeq 0.5$) alongside a wider phase where the emulsion shows a significant departure from the spherical shape (for $0.6\lesssim\phi_d\lesssim 0.8$). 
In Fig.\ref{hipde}d we show a typical example of an experimentally realized soft granular medium displaying a high degree of monodispersity and in Fig.\ref{hipde}e a result from LB simulations, both at $\phi_d\simeq 0.8$. The emulsions consist of three immiscible fluids, i.e. droplet phase (also termed as cores), lubricating phase and outer phase, with  
surface tension of the order of $5$mN/m and viscosity ratio approximately equal to five between  outer and innermost fluid (the viscosity generally ranges between $1-5$ mPa$\cdot$s). For higher values of $\phi_d$, the internal droplets arrange into a stable linear chain encapsulated within an elongated shell. 

Such a picture decisively depends on the grain structure as well as on the deformability of the cores, features
controlling the formation of highly packed structures with a volume fraction typically well above the close packing limit of hard spheres. 

A far more intriguing scenario is observed in the presence of external flows, where the combination of granularity and deformability leads to complex multiscale dynamics, including plasticity \cite{graner_soft,dollet}, memory effects \cite{jiang,wang}, yield-stress transitions \cite{lulli,bonn_rmp} and glassy dynamics \cite{goyon,sbraga_soft}. 
If driven through narrow geometries, such as a constriction of a microchannel, these materials are usually subject to dramatic morphological transitions which may yield events like the rupture of the shell or the coalescence of the inner droplets, eventually jeopardizing the stability of the emulsion \cite{montessori_pof,montessori_lan,rosenfeld}. Hence, a deep understanding of the fluid-structure interactions governing the physics of these systems is fundamental for ameliorating their mechanical properties, also considering their relevance to biology 
as models to investigate the behavior of 
cell clusters crossing physiological constrictions \cite{tang,au_pnas}.

\begin{figure*}[htbp]
\includegraphics[width=1.0\linewidth]{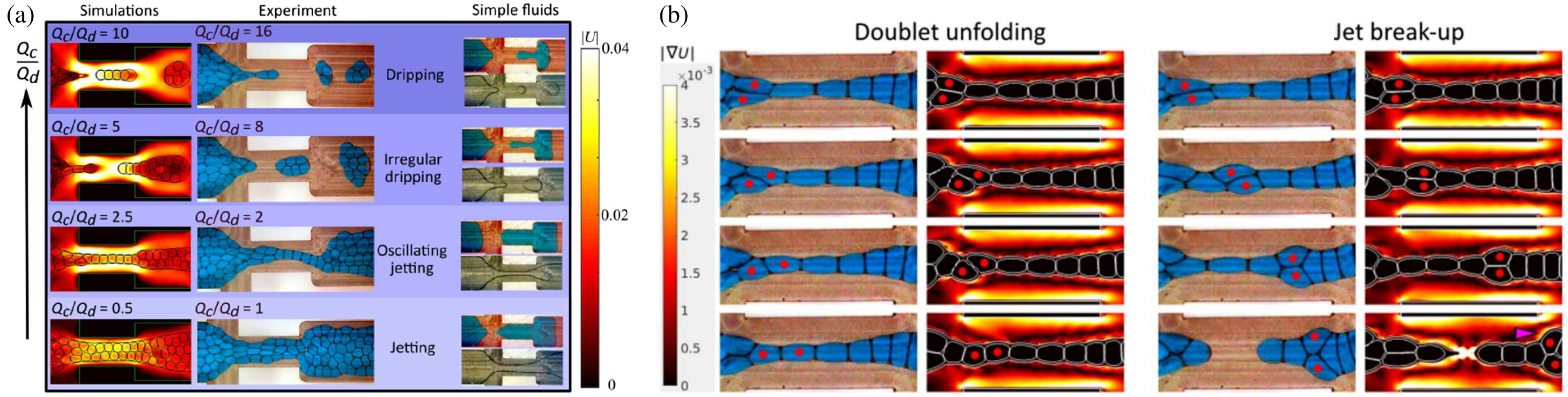}
\caption{(a) Dynamic patterns observed in a soft granular medium by varying $Q_c/Q_d$. The medium consists of a water-in-oil emulsion surrounded by a further immiscible phase (fluorinated fluid).
The left column shows the simulation results (on a 3d lattice of size $280\times 140\times 30$), the central one shows the corresponding experiments and the right one refers to a water (blue)-oil (transparent) mixture. The color bar refers to the fluid velocity in simulations. (b) Mechanism leading to the jet breakup. If a doublet enters the narrowing and unfolds, the droplet chain is stabilized. Otherwise, high-velocity gradients in the surrounding of the doubled (see the pink arrow) destabilize the chain. Copyright granted under the \href{https://creativecommons.org/licenses/by/4.0/}{CC 4.0 International license}. No changes were made from the original figure from Ref.\cite{bogdan_prl}.}
\label{double_em}
\end{figure*}

LB simulations prove, once again, instrumental to capture crucial aspects of the physics of such systems. In Fig.\ref{double_em}a, for example, 
a series of experiments about the formation of dynamic modes of a confined granular medium (a densely packed monodisperse emulsion focused by an external immiscible phase through an orifice) with $\phi_d\simeq 0.8$ are presented, while three dimensional numerical simulations are  performed using the color gradient LB described in section \ref{color-gradient}.
Upon varying the flow ratio $Q_c/Q_d$ between the continuous and the dispersed phases, both experiments and simulations show that one can distinguish 
four modes, consisting of (i) a thick, weakly oscillating jetting phase for $Q_c/Q_d\simeq 1$, (ii) a thin, strongly
oscillating jet with occasional breakups for $Q_c/Q_d\simeq 2$, (iii) an irregular dripping phase of
polydisperse double emulsions for $Q_c/Q_d\simeq 8$, and (iv) a regular dripping phase of monodisperse double emulsions for $Q_c/Q_d\simeq 15$.  It is worth highlighting that the experimental values of the flow ratio match almost one-to-one the numerical values (see Fig.\ref{double_em}a), with differences very likely caused by slightly different geometrical parameters (such as length and width of the constriction) and volume fractions of the cores.
In stark contrast, Newtonian liquids (such as water in oil mixture) only show two primary modes, i.e. dripping and jetting, although other regimes have been reported in experiments \cite{utada_prl}.

Besides describing at a high level of accuracy these complex structures, LB simulations also unveil the mechanism causing the jet breakup and leading to the dripping phase, as shown in Fig.\ref{double_em}b. This is essentially based on the droplet rearrangement upon entering the constriction and on the magnitude of the associated velocity gradients. Indeed, when a doublet (i.e. a pair of droplets) enters the narrowing, the surrounding continuous phase accelerates to conserve the flux. If the doublet unfolds into a chain, the velocity gradients remain weak and the chain is stabilized; on the contrary, if the doublet survives, increasing velocity gradients around it destabilize the jet leading to the breakup. 

These results suggest that the color gradient LB augmented with near-contact forces provides a robust numerical platform to study the hydrodynamics of different examples of confined soft granular media, ranging from soft flowing crystal to densely packed aggregates structured in a double emulsion configuration. 
Recently, it has been also shown that some aspects of the physics of dense emulsions can be efficiently described by using the Shan-Chen approach described in section \ref{lattice_pp}.  
In the next section, we discuss the dynamics of a dense emulsion subject to thermal convection simulated by this method. 

\subsection{Dense emulsion under thermal flows}

The Shan-Chen approach has been extensively used in applications, such as soft glassy systems \cite{benzi_jcp,benzi_soft,sbraga_soft}, or to simulate the coarsening dynamics \cite{pelusi_soft,tiribocchi_bijel}, where heat transfer is generally neglected.
However, heat transfer can be relevant in a variety of contexts, ranging from polymeric mixtures \cite{kramer_prl} and binary fluids \cite{gonnella_pre} to the convective motion of magma \cite{stein_geo}.  
In Refs.\cite{pelusi_cpc,pelusi_emu}, Pelusi and coworkers have extended the applicability of the Shan-Chen approach to the study of dense emulsions subject to thermal convection, where the temperature field is coupled with the momentum equation of the emulsion. More specifically, the dynamics of the scalar temperature $T$ is simulated by introducing an auxiliary set of distribution functions governed by a single relaxation time lattice Boltzmann equation
\begin{equation}\label{eq_giT}
g_i({\bf x}+ {\bf c}_i \Delta t,t+\Delta t)-g_i({\bf x},t)=-\frac{1}{\tau_g}(g_i({\bf x},t)-g^{eq}_i({\bf x},t)), 
\end{equation}
where $T({\bf x},t)=\sum_i g_i({\bf x},t)$ and $g_i^{eq}$ is the usual second-order expansion in the fluid velocity of the Maxwell-Boltzmann distribution. In the long-wavelength limit, Eq.(\ref{eq_giT}) approximates an advection-diffusion equation of the temperature
\begin{equation}
D_tT=\kappa_T\nabla^2 T,
\end{equation}
where $D_t=\partial/\partial_t+{\bf u}\cdot\nabla$ is the material derivative and $\kappa_T=c_s^2(\tau_g-1/2)$ is the thermal conductivity. 

\begin{figure*}[htbp]
\includegraphics[width=1.0\linewidth]{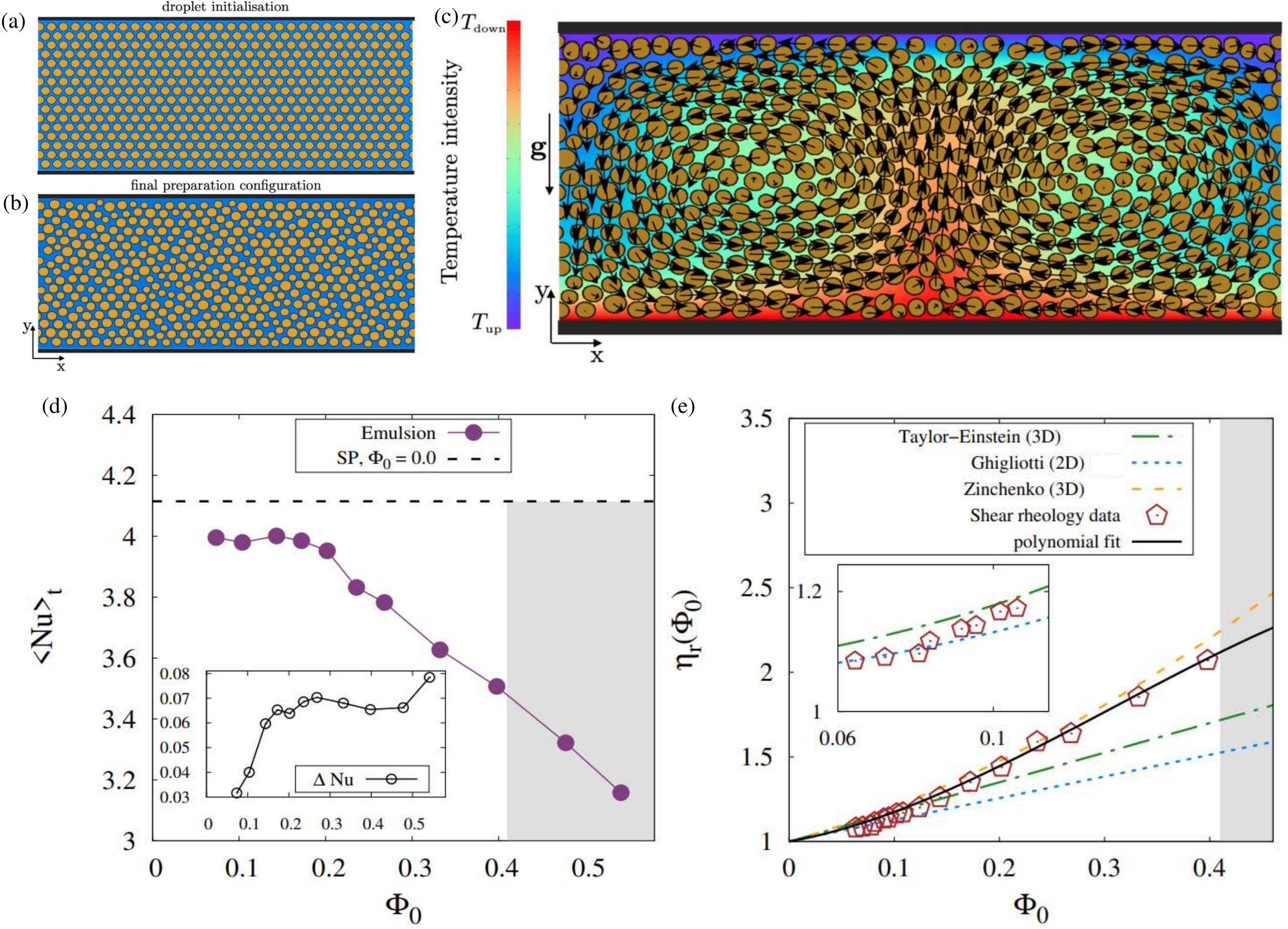}
\caption{(a) Initial setup of a dense emulsion arranged as a honeycomb structure. (b) Relaxed configuration, taken as a starting point for the convection simulation. In these configurations, the temperature does not affect the fluid flow. (c) Instantaneous configuration of a Rayleigh-B\'enard convection in a cell once the emulsion is heated from below. Black arrows indicate the direction of the droplet and the color map represents the values of the temperature. Finally, ${\bf g}$ is the gravitational acceleration. (d) Time-average of the Nusselt number for different values of $\Phi_0$. The dotted line represent the Nusselt number for solid spherical particles (SP) in the very dilute regime, where $\Phi_0=0$. The inset shows time-averaged fluctuations of
the Nusselt number $\Delta Nu = \langle (Nu(t)-\langle Nu\rangle_t)^2\rangle_t^{1/2}$.  The grey region indicates values of concentration where non-Newtonian effects emerge. (e) Comparison between the effective viscosity from simulations (red pentagons) and literature data from Taylor-Einstein \cite{taylor,einstein}, Ghigliotti \cite{ghiglio} and Zinchenko \cite{zinch}. The figure is reproduced with permission from Ref.\cite{pelusi_cpc}.}
\label{themal_conv}
\end{figure*}

A typical simulation setup is shown in  Fig.\ref{themal_conv}a, where a non-coalescing dense emulsion, initially structured as a honeycomb, is confined between two parallel plates on a two-dimensional lattice (the horizontal direction is periodic), while gravity acts along the wall-to-wall direction. At the boundaries, no-slip conditions are set for the velocity and Dirichlet conditions are imposed for the temperature. The emulsion is then relaxed and stabilized to a state taken as a starting configuration before the convection (Fig.\ref{themal_conv}b). A typical thermal effect is shown in  Fig.\ref{themal_conv}c, where convective rolls emerge as the emulsion is
heated from below and cooled from above (as in a Rayleigh-B\'enard convection \cite{rayleigh,benard,lohse_annu}). In these simulations, $Ca$ is kept lower than $0.01$ (thus shape deformations are negligible) and $Re$  lower than $100$.

A systematic investigation on the effect produced by varying droplet concentrations, from very dilute regimes (Newtonian emulsion) to dense ones, has been carried on in Ref.\cite{pelusi_emu}. 
It has been shown, for example, that the Nusselt number (measuring the heat transport efficiency), time-averaged over statistical steady states, remains below that of a single phase system (i.e. when  droplet concentration $\Phi_0$ tends to zero), is roughly constant for $\Phi_0\lesssim 0.2$ and progressively diminishes  for higher values of $\Phi_0$. 
At increasing concentration, it displays significant fluctuations around the mean value, in stark contrast with the dilute case where the fluctuations turn very weak and the transition to convection is accompanied by a steady flow (see Fig.\ref{themal_conv}d).
Further simulations have characterized the rheological properties via a Couette cell, by confining the emulsion (now at constant temperature) between flat walls placed at distance $h$ and moving along opposite directions with velocity $u$, thus under a shear rate $\dot{\gamma}=2u/h$. The numerical results show that
the measured relative viscosity $\eta_r=\eta_{eff}/\eta_{solv}$ (where $\eta_{eff}=d\Sigma/d\dot{\gamma}$ is the effective viscosity and $\eta_{solv}$ is the viscosity of the solvent) is in good agreement with values reported in literature for $\Phi_0\lesssim 0.12$, where $\eta_r$ follows a linear behavior proportional to $\Phi_0$ (for small droplet deformations) \cite{einstein,taylor}. For larger values of $\Phi_0$, numerical data deviate from the linear behavior and are found to agree with results shown in Ref.\cite{zinch}.
    
These results show that, alongside the near-contact LB method, the Shan-Chen approach can provide a further valuable platform to study the physics of confined dense emulsions where coalescence is inhibited. It would be of interest to extend these simulations to three dimensions where out-of-plane components of the fluid velocity are likely to play a non-trivial role. 

\subsection{Flowing hierarchical emulsions}

An alternative numerical strategy to study the hydrodynamics of dense emulsions with hierarchical structure (a broad class including soft granular media as well) is based on the hybrid free energy LB method using a multiphase field approach (see section \ref{free_energy}). In this model, a set of scalar field $\phi_n({\bf x},t)$, $n=1,...,N$ describes the density of each droplet (being $N$ the total number of droplets) and the vector field ${\bf u}$ accounts for the global fluid velocity. 
The dynamics of the scalar fields is governed by a set of advection-diffusion equations of the form of Eq.(\ref{multi_adv}), while the velocity field obeys the Navier-Stokes equation whose stress tensor includes forcing terms associated to each phase.   
Over the last few years the multi-phase field theory, often combined with LB methods,
has been used to simulate systems such as deformable droplets in microchannels \cite{foglino1,foglino2,negro_sc,tiribocchi_pof1} and cell monolayers displaying liquid crystal features \cite{loewe,yeomans_cell,giomi_natphys}.  
In this subsection, we discuss its use in the context of multiple emulsions flowing within a microfluidic channel, as presented in Ref.\cite{tiribocchi_nat}. 

\begin{figure*}[htbp]
\centering
\includegraphics[width=1.0\linewidth]{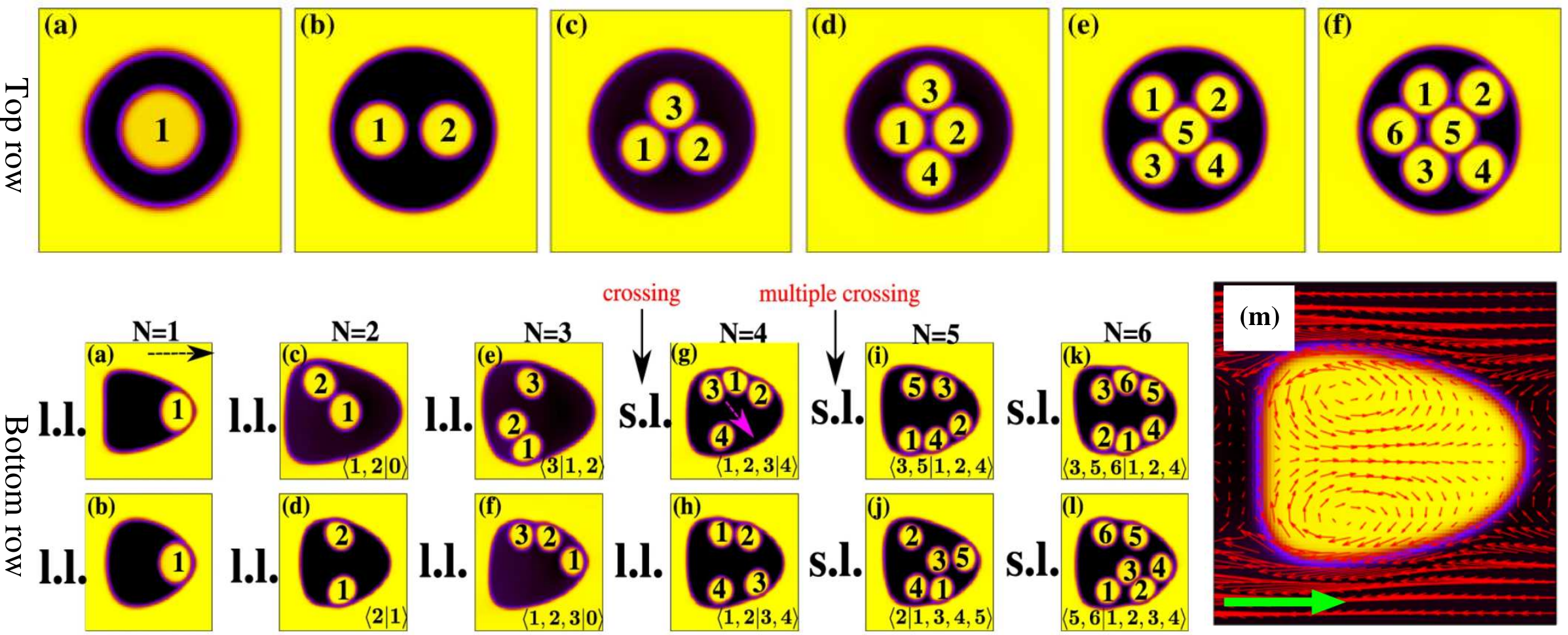}
\caption{Top row: Equilibrium configurations of a double emulsion (a), a two-core (b), a three-core (c), a four-core (d), a five-core (e) and a six-core (f) emulsion. Bottom row: 
(a)-(l) The figures show a list of non-equilibrium steady states of multiple emulsions under a pressure-driven flow. If the area fraction occupied by the cores is lower than $\sim 0.35$, the states are long-lived (l.l., corresponding to $N=1$, $N=2$, and $N=3$). The cores are either stuck at the front end or remain confined in the upper/lower part of the emulsion, often displaying a periodic motion along circular trajectories. If the area fraction is higher, the states turn to short-lived (s.l., corresponding to $N=4$, $N=5$, and $N=6$), and multiple crossings from one region towards the other one are observed. Here, Reynolds numbers range from $\sim 1$ to $\sim 3$ and Capillary numbers from $\sim 0.35$ to $\sim 0.85$. The dotted black arrow indicates the direction of motion of the emulsion, while the magenta one the direction of a crossing. In each snapshot, the steady states are classified in terms of an analogy with the occupation number formalism. (m) Typical double-vortex structure of the velocity field computed in the reference frame of a droplet subject to a pressure-driven flow.  The color map ranges from $0$ to $2$ and accounts for the values of the phase fields. Copyright granted under the \href{https://creativecommons.org/licenses/by/4.0/}{CC 4.0 International license}. No changes were made from the original figure from Ref.\cite{tiribocchi_nat}.}
\label{multiple}
\end{figure*}

Unlike the examples shown in sections \ref{sfc} and \ref{sgm} focused on droplet production, here the interest is in the long-term behavior of these systems. The emulsion is initialized by defining a set of phase fields $\phi_n$ which relaxes towards equilibrium for a relatively short simulation time (see Fig.\ref{multiple}, top row). At the boundaries, no-slip conditions are set for the fluid velocity and neutral wetting for the phase fields, although no wetting would essentially lead to similar results provided that the interaction between droplets and walls is negligible.
Once mechanical equilibrium is attained, a constant pressure gradient is applied across the longitudinal direction of the channel. 
As time progresses, a large variety of non-equilibrium states is found, ranging from long-lived ones at low values of area fraction of the cores to short-lived ones at higher values. In the former regime, the cores are found to display a correlated planetary-like motion, in which they cease each other while remaining confined either in the upper or lower part of the emulsion for long periods of time. In the latter one, the increase of area fraction triggers collisions 
and multiple crossings between the two regions, finally leading to a chaotic-like dynamics.

In the bottom row of Fig.\ref{multiple}, we show a selection of these steady states.
The first scenario is found at values of core area fraction lower than $0.35$ (corresponding to $N=1,2,3$), where the cores exhibit the periodic dynamics confined in a portion of the emulsion, while the second one, in which crossing events occur, appears at higher values (corresponding to $N=4,5,6$) up to approximately $0.5$. 
This complex behavior is decisively driven by the internal vorticity, consisting of two counter-rotating vortices within the external droplet. If the core area fraction is below $0.35$, the vortex structure remains essentially unaltered (such as the one shown in Fig.\ref{multiple}m), whereas it turns chaotic for higher values, due to the non-trivial coupling of the fluid velocity with the interfaces of the cores.Although a dedicated experiment on this complex dynamics is still missing, a behavior akin to the one observed for low core area fraction has been found in liquid crystal droplets moving inside microchannels \cite{weitz_prl}, where periodic director field changes are induced by the double-vortex flows inside the droplet. This effect, in turn, determines a cyclic motion of point defects which rotate in separate regions of the emulsion, closely resembling that of inner cores shown in Fig.\ref{multiple}.   

The existence of these steady states is generally guaranteed as long as the inner suspension is sufficiently monodisperse and the capillary number is lower than $1$. For highly polydisperse mixtures, for example,  the cores have been found to either get stuck at the front end of the emulsion or confine, separately, in the upper and lower region. Also, increasing the capillary number of the external interface induces large shape deformations which, in turn, squeeze the fluid vortices finally hindering the rotational motion \cite{tiribocchi_pre2}.
We finally note that a persistent rotating motion could also be produced under Couette flow, where two confining walls move along opposite directions \cite{tiribocchi_pof1,tiribocchi_pof2,tiribocchi_prf}. Unlike the previous case,  the motion would be triggered by a single vortex confined within the droplet and resulting from the sheared structure of the flow. 

\begin{figure*}[htbp]
\centering
\includegraphics[width=0.8\linewidth]{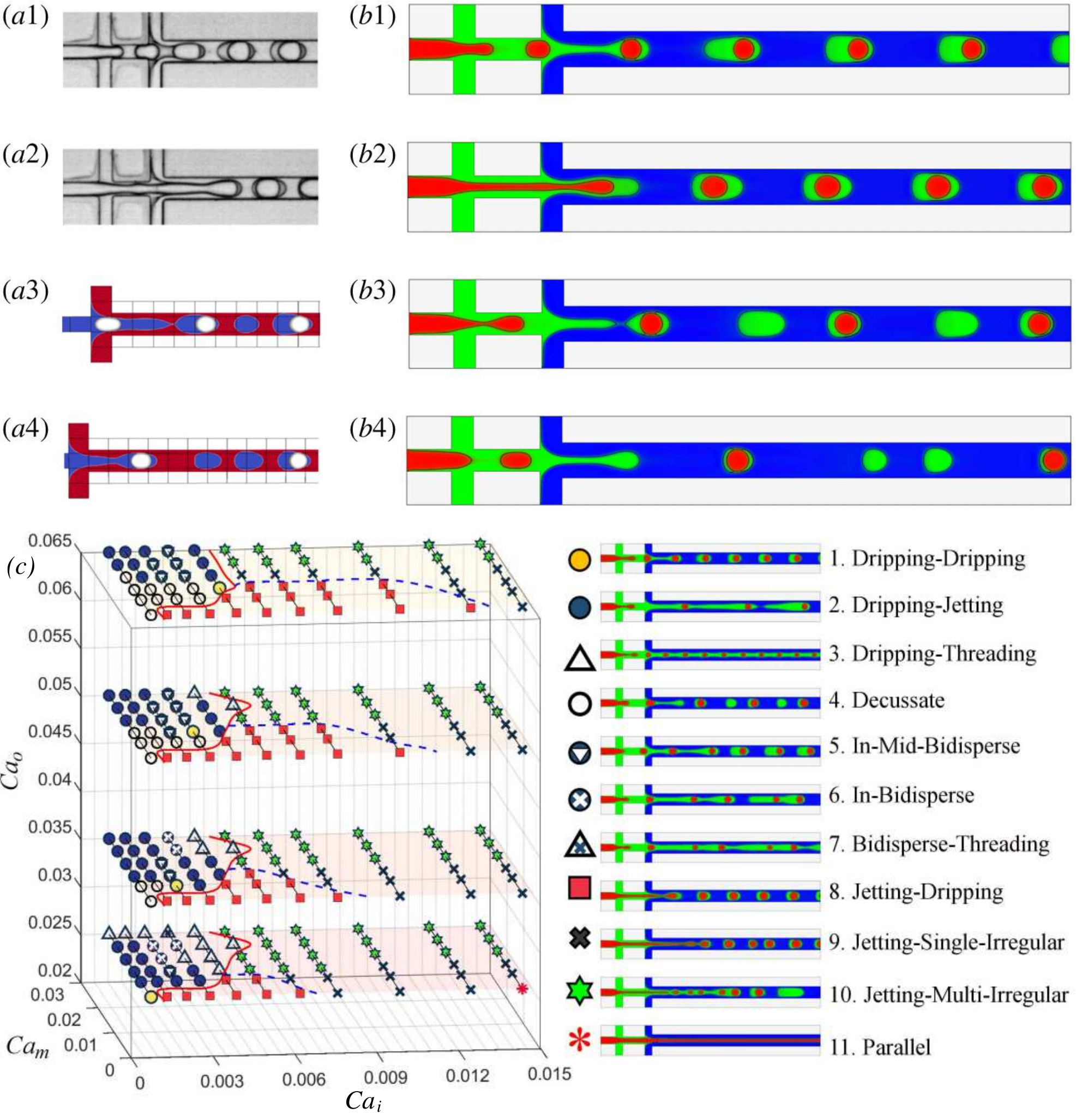}
\caption{(a-b) Flow regimes reported in experiments of Ref.\cite{abate2} (a1,a2) and simulation  of Ref.\cite{azarmanesh} (a3-a4) and  Ref.\cite{halim_jfm} (b1-b4). 
Capillary numbers reported in Ref.\cite{halim_jfm}
are: (b1) $Ca_i = 0.012$, $Ca_m = 0.011$
and $Ca_o = 0.035$; (b2) $Ca_i = 0.018$, $Ca_m = 0.011$ and $Ca_o = 0.035$; (b3) $Ca_i = 0.008$, $Ca_m =0.011$ and $Ca_o = 0.035$; (b4) $Ca_i = 0.008$, $Ca_m = 0.011$ and $Ca_o = 0.065$, where the subscripts $i$, $m$ and $o$ indicate inner, medium and outer fluid, respectively. (c) Three dimensional phase diagram of different flow regimes observed by varying $Ca_o$, $Ca_m$ and $Ca_i$. The red solid line marks the transition from the dripping-mode breakup (left side) to jetting-mode breakup (right side)  of the inner fluid. The blue dashed line separates a region (below the line) where the middle fluid breaks up in the dripping mode from a region (above the line) where the middle fluid breaks up in the jetting mode. Copyright granted under the \href{https://creativecommons.org/licenses/by/4.0/}{CC 4.0 International license}. The figures are adapted from Ref.\cite{halim_jfm}}
\label{halim_double}
\end{figure*}

Production and dynamics of a double emulsion in a planar flow focusing can be also described using a ternary free-energy LB\cite{halim_prl,halim_jfm} where, unlike the hybrid version previously mentioned, the Navier-Stokes equation and the two equations of the order parameters (i.e. the densities of the inner and outer droplets) are solved by means of three different sets of distribution functions, essentially one for each hydrodynamic field. 
A selection of flow regimes obtained using this method for different values of Capillary numbers are shown in Fig.\ref{halim_double}a-b, where a significant match with experiments is also reported. One can identify, for example,  the two-step formation regime, (a1-b1), the one-step one (a2-b2), the decussate regime with one empty droplet (a3-b3), and the decussate regime with two empty droplets (a4-b4). A systematic study of the breakup modes leading to different flow regimes is shown in Fig.\ref{halim_double}c, where a three-dimensional phase diagram illustrates a combination of breakup modes of inner and middle fluids obtained by varying $Ca_o$, $Ca_m$ and $Ca_i$. Dripping-dripping, dripping-jetting and jetting-dripping modes, for example, are found for the two-step and the one-step formation shown in Fig.\ref{halim_double}a1-a2, experimentally observed in two-cross junctions and glass capillaries \cite{abate2,kim2013}. Further modes, such as dripping-threading, decussate and bidisperse and irregular-jetting are reported more frequently in glass capillary devices \cite{nabavi,kim2013,shang}.

The method discussed in this subsection has been also adopted to describe active fluids where, unlike the previous cases, spontaneous flows emerge at a mesoscale level as a result of the action of smaller units that self-propel \cite{marchetti,ramaswamy}. Typical examples are dense suspensions of bacteria dispersed in a fluid, protein networks in living cells, and artificial microswimmers. In the next section, we present a recent application of the LB method to the dynamics of an active gel droplet migrating through a microfluidic constriction. 

\subsection{Migration of active droplets through constrictions}\label{cell_mig}

Active fluid droplets are a class of bio-mimetic self-propelled systems whose autonomous motion is powered by  an active gel often located within the droplet \cite{dogic,michielin,maass_epje,guillamat}. Experimentally realized active droplets typically consist of a water-in-oil emulsion containing, for instance, a dispersion of microtubules and kinesin \cite{dogic,guillamat,doost_natcomm,ruske_prx,dema_softmatter} or an actomyosin solution \cite{poincloux,tjhung_pnas,tjhung_pnas2,sakamoto}. Microtubules plus kinesin is an example of extensile material, in which the surrounding fluid is pushed away from the center of mass of the particle, whereas the opposite holds for actomyosin, belonging to contractile systems. 
These active droplets hold interest as model tools for studying some aspects of the dynamics of living cells \cite{aranson}, such as swimming \cite{tjhung_pnas}, crawling \cite{tiribocchi_nat3} and spontaneous division \cite{giomi}, as well as for the design of biomimetic soft materials of relevance in pharmaceutics, for drug delivery \cite{pagonabarraga}, and material science, for tissue engineering \cite{yeomans_nat}.

\begin{figure*}[htbp]
\includegraphics[width=1.0\linewidth]{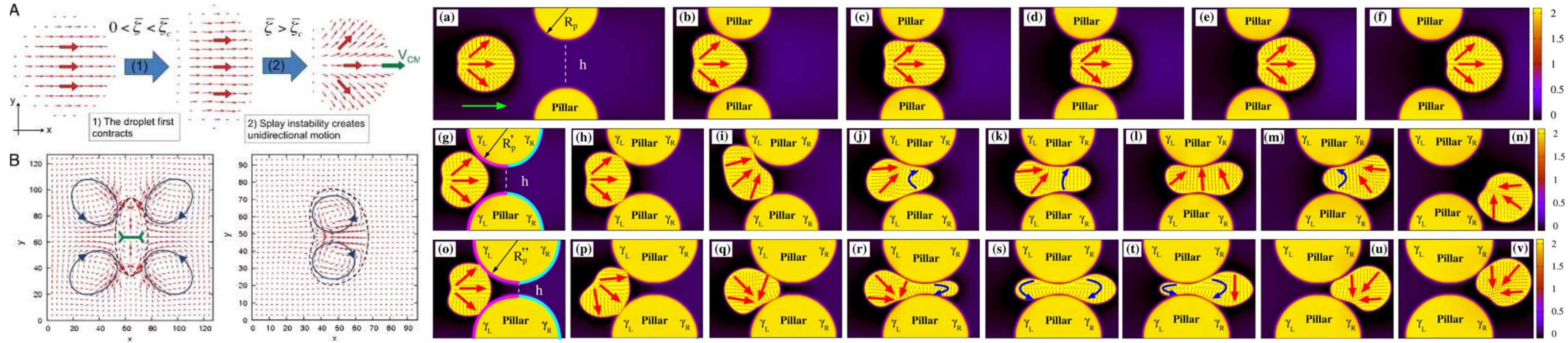}
\caption{Left panel, (A)-(B). In (A), the figure shows the initial configuration of the active droplet (left), the intermediate non-motile state (middle), and the motile state (right). Red arrows indicate the direction of the polar field, which turns from a uniform orientation at equilibrium to a splay deformation at the steady state. In (B), the figure shows the steady-state velocity fields for the intermediate configuration (left) and the motile one (right). The former consists of a four-roll mill flow, while the latter exhibits two counter-rotating vortices propelling the droplet. Here, red arrows indicate the direction of the velocity field.
Right panel, (a)-(v). This panel shows the crossing of an active fluid droplet of diameter $D$ through a constriction of height $h$. In (a)-(f) $h/D\simeq 0.8$, in (g)-(n) $h/D\simeq 0.5$ and in (o)-(v) $h/D\simeq 0.2$.
Decreasing the ratio $h/D$ yields larger shape deformations, some closely resembling experimentally observed structures, such as ampule-like (j,k,r,t) and hourglass (s) configurations. For $h/D\lesssim 0.5$, larger adhesive forces (pink layer in (g) and (o)) at the entrance of the pore and weaker ones (cyan layer) at the exit are crucial to 
favor the crossing. Here, $\gamma_L$ and $\gamma_R$ indicate the adhesion coefficient at the left and right sides of the pillars, large red arrows highlight splay deformations, and blue ones indicate bend distortions. The color map ranges from $0$ to $2$ and accounts for the values of the phase fields. The left panel is reproduced with permission from Ref.\cite{tjhung_pnas}. Copyright granted under the \href{https://creativecommons.org/licenses/by/4.0/}{CC 4.0 International license}. No changes were made from the original figure from Ref \cite{tiribocchi_nat2}.}
\label{active_drop}
\end{figure*}

The theoretical framework capturing the behavior of these systems at a continuum level (described in Section \ref{free_energy}) combined with the free-energy LB method has been particularly fruitful in simulating the dynamics of active fluid droplets in a highly diluted regime and under confinement. 
In Fig.\ref{active_drop}A-B (left panel) we show, for example, the structures of the polar and velocity field of a contractile (i.e. $\zeta<0$) fluid droplet
in an unbounded medium. The droplet is initialized as a circular region where the polarization vector ${\bf p}$ is uniform and unidirectional. At increasing values of $\zeta$, two regimes can be distinguished: a non-motile one where a four-roll mill flow stretches the droplet longitudinally and a motile one where, once $\zeta$ overcomes a critical value, the internal flow acquires a double-vortex pattern propelling the droplet forward. This structure breaks the inversion symmetry of the polar field, leading to an emergent elastic splay and motion along the direction of ${\bf p}$. 

This 2D contractile droplet can be viewed as a simplified {\it in silico} version of an {\it in-vitro} cell extract, typically used in cell motility experiments. In this respect, the contractile-induced propulsion and the related flow structure could mimic the myosin-induced intracellular flow activating ketarocyte cells crawling on glass \cite{yam_biol}. LB simulations also show that contractility can trigger autonomous motion of three-dimensional droplets \cite{tjhung_pnas}, a result that could be relevant for experiments of  quasi-spherical tumor cells  moving inside an elastic gel where the locomotion is shown to be caused mainly by myosin contractility \cite{poincloux,hawkins_bio}.

A similar theoretical description can be used to study the dynamics of a contractile droplet through constrictions, a phenomenon of relevance in a number of biological processes such as cancer spreading and wound healing \cite{davidson1,au_pnas,cao_biophys,bruckner}. In Fig.\ref{active_drop}a-v, we show the motion of a droplet of diameter $D$ through a pore, whose design draws inspiration from experiments
of  living cells (such as a mouse fibroblast) migrating within a microfluidic channel hosting constrictions \cite{davidson1,davidson2}.
The pore is modeled by means of two static and fluid-free phase fields (the pillars) placed at distance $h$ and glued to opposite flat walls. 
Although a crude approximation of a realistic constriction, this design allows for a simple implementation of mesoscale effects, such as repulsion and adhesion between the droplet and pillars. Adhesion, in particular, can be included by adding a contribution of the form $\sum_{n,m,n<m}\gamma_{n,m}\nabla\phi_n\nabla\phi_m$ to the free energy, where $\gamma_{n,m}$ is an adhesion constant between different phase fields.

Importantly, adhesive forces are found to be crucial in enabling the crossing, especially through narrow interstices \cite{tiribocchi_nat2}. Indeed, 
while for large pores (Fig.\ref{active_drop}a-f, $h/D\simeq 0.8$) the crossing generally occurs if the speed (which is controlled by the activity $\zeta$) is high enough, for smaller pores (Fig.\ref{active_drop}g-v, $h/D\simeq 0.5$ and $h/D\simeq 0.2$) the sole active stress is not sufficient to enable the process. In this case, the droplet would either get stuck at the pore's entry or bounce back into the microchannel. Simulations show that, under these conditions, non-uniform adhesive forces (i.e. higher at the entry of the pore and lower at the exit) are necessary for a successful crossing, provided that inertia, surface tension, and adhesion are carefully balanced. Suitable dimensionless numbers assessing the importance of these effects are the adhesion number $A=\gamma_L/\gamma_R$, which is the ratio between adhesion forces at the entry and the exit of the pore, and the inertia-over-adhesion number $I_{A_{L,R}}=\rho v^2R^2/\gamma_{L,R}$, where $v$ is the droplet velocity and $R$ is the droplet radius. For $\lambda\simeq 0.5$, for example, one has $2\lesssim A\lesssim 10$, $I_{A_R}\simeq 0.25$ and $0.025\lesssim I_{A_L}\lesssim 0.06$. The values of the last two numbers, in particular, mean that inertial forces are weaker than adhesive ones at the entry (L) of the pore to enhance the connection between the droplet and pillars. A lower adhesiveness at the exit (R) allows for an easier detachment from the pore, finally favoring the crossing. 

Although the match to experiments is mostly qualitative since this model remains distant from a living cell in many aspects (such as the lack of the nucleus, which represents a limiting factor in the crossing, and the absence of a membrane, replaced by a thin interface), these results may suggest that some features of cell motility can be reasonably captured by minimal mesoscopic models built upon a restricted number of dynamic parameters and hydrodynamic fields.  

\subsection{Peta-scale simulations of deep-sea sponges}\label{sponge_sim}
We finally discuss a recent application where the LB method is adopted to study the fluid dynamics of the deep-sea glass sponge {\it Euplectella aspergillum} \cite{falcucci_nature}, a soft living organism that,  besides representing a fascinating example of life under extreme conditions  (at depth of $100$-$1000$ meters in the Pacific Ocean and Antarctic area with no ambient sunlight), 
displays intriguing structural properties due to the trabecular arrangement of its skeletal system \cite{monn_pnas,fernandes_natmat}.

The interest of this study lies in the fact that, despite the use of a single-component interface-free LB (thus its easiest realization), the detailed three-dimensional reconstruction of the skeleton of the sponge and the computation of the fluid flow in which this organism is immersed need to leverage two of the main advantages of the method, namely the excellent scalability on parallel architectures and the capability of handling complex geometries. 
Indeed, to reproduce the sponge's living conditions, the simulations feature over fifty billion grid points spanning approximately four spatial decades, while a complex mesh structure is used to simulate the skeleton.

\begin{figure*}[htbp]
\centering
\includegraphics[width=0.7\linewidth]{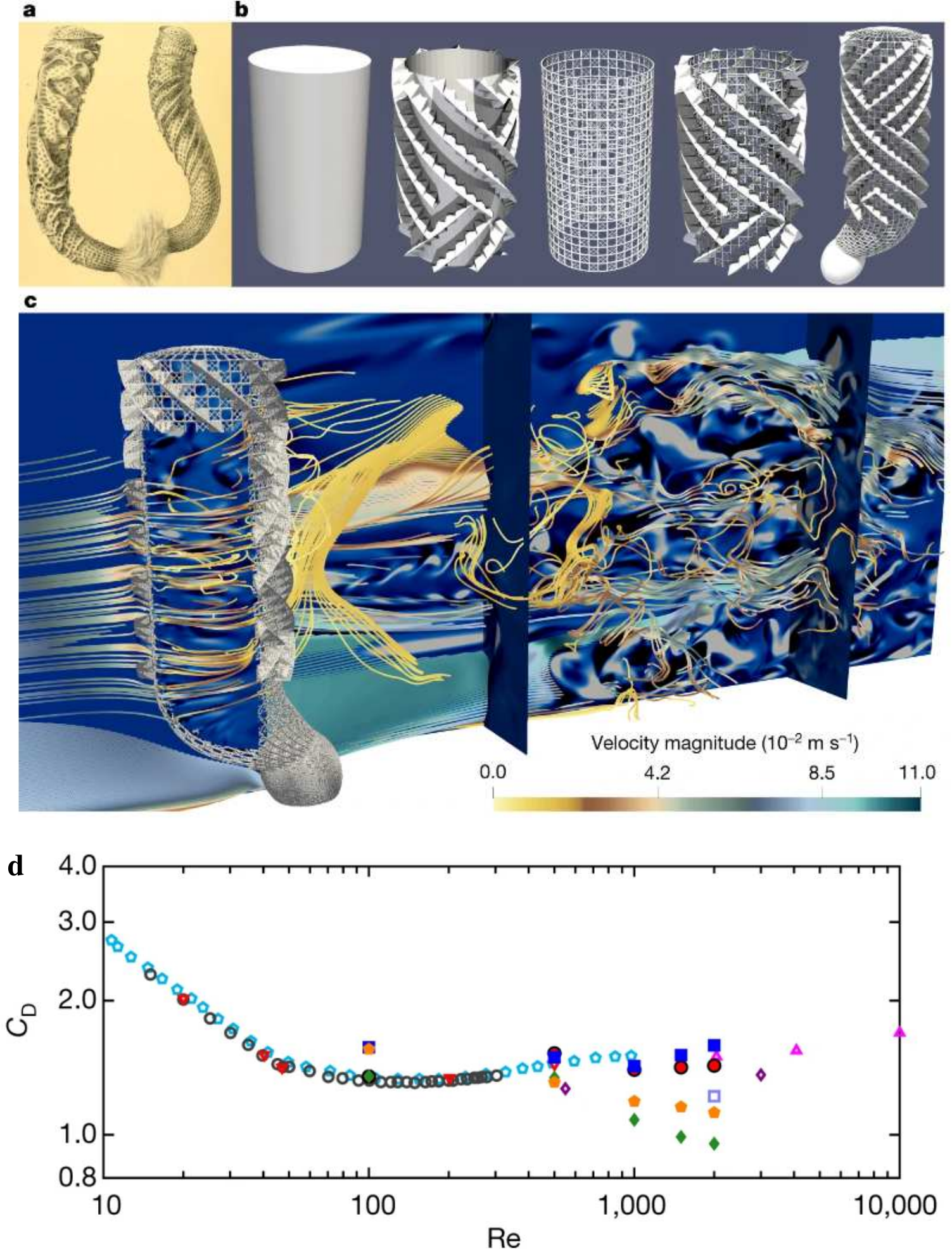}
\caption{(a) Drawing of a deep-sea sponge E. aspergillum. (b) A three-dimensional computational model of the organism showing increasing complexity.
From left to right: solid cylinder, solid cylinder with helical ridges, porous cylindrical lattice, porous cylindrical lattice with helical ridges, and a complete model. (c) Simulation of a deep-sea sponge immersed in a hydrodynamic flow at $Re\simeq 2000$, where contours of helicity ${\cal H}$ (defined as ${\cal H}={\bf u}\cdot {\bm \omega}$, where ${\bm \omega}$ is the vorticity)
and streaklines of the flow are shown. (d) Drag coefficients at various $Re$ for the simplified models shown in (b) (red circles for solid cylinder, blue squares for solid cylinder with ridges, green diamonds for porous cylinder and orange pentagons for a porous cylinder with ridges) and compared to the following values from literature for cylinders: Ref.\cite{kawamura} pink triangles, Ref.\cite{hanci} violet diamonds, Ref.\cite{sahin} red triangles, Ref.\cite{fujisawa} empty purple squares, Ref.\cite{henderson} empty light blue pentagons, Ref.\cite{posdz} empty gray circles.  
The figure is reproduced with permission from Ref.\cite{falcucci_nature}.}
\label{euplectella}
\end{figure*}

In Fig.\ref{euplectella}, we show the computational model and the hydrodynamic flow of  the sponge
at $Re\simeq 2000$, where $Re=uD/\nu$, being $u$ the water velocity, $\nu$ is its viscosity and $D$ is the sponge diameter at the top section. 
The sponge structure comprises an anchoring bulb, a section connecting the bulb to the body, the main body and the final sieve plate located at the top (see Fig.\ref{euplectella}a,b). The body, in particular, is simulated using two intersecting patterns of lattices, the first composed of filaments orthogonally crossing each other and the second made by smaller ligaments placed at $45^{\circ}$ with respect to the first one. The resulting cylindrical structure finally bends into a cone connecting with the anchoring bulb. Also, 
no-slip conditions are imposed at the internal boundaries of the sponge and at the seafloor.

The simulations show, for example, that a considerable decrease in the flow speed occurs inside the sponge's body (where vortices are formed) and downstream of the organism, while intermittent fluid patterns emerge several diameters further away (see Fig.\ref{euplectella}c). These features are supposed to play a role in feeding through a selection mechanism where nutrients are absorbed in the sponge chambers and inorganic particles are discarded. The  quiescent region downstream, in particular, is also expected to reduce the hydrodynamic load,  an effect quantified in terms of the drag coefficient $C_D=2\bar{F}_{drag}/A\rho_{inlet}u^2_{inlet}$, where $F_{drag}$ is total drag force, averaged at the steady state and along the flow direction, $\rho_{inlet}$ and $u_{inlet}$ are density and speed at the domain inlet and $A$ is the area of the transverse section of the model, perpendicular to the fluid flow.
Indeed, in agreement with previous studies, a considerable drag reduction occurs at $Re> 500$ for porous cylindrical models,  while the presence of ridges is found to mildly mitigate this effect at large $Re$. More recent simulations have also demonstrated the capability of describing
a further aspect of the fluid  dynamics inside the body cavity, i.e. the existence of an organized vertical flow toward the oscular aperture, 
with values of flow speed $\sim 0.83$ cm/s \cite{falcucci_prl}, in agreement with experimental results on living {\it Hexatinellida}, reporting values around $1$ cm/s \cite{leys_plos}.
Finally, note that the presence of the sponge's tissue has been neglected in these simulations although it could impact the structure of the fluid flow, because the tissue may form a barrier with low permeability over the sponge wall \cite{leys_sponge}. 

\subsection{Open challenges.}
Despite the fundamental role of the simulations (and specifically LB methods) in the understanding of the properties of these systems, a number of theoretical problems remain open.

In soft flowing crystals and soft granular media (including hierarchical emulsions), for example, a "comprehensive" multiscale simulation fully covering the spectrum of six orders of magnitude in space and twice as many in time is presently still unfeasible.  The reason of such a challenge  is that 
it would require a combination of innovative GPU-based HPC techniques and grid refinement strategies to deal with
the necessary computational resources and the increasing algorithmic complexity plus, very likely, new coarse-grained models to improve existing multi-scale theories. 
Such simulations would be highly desirable for establishing, for example, the extent to which microscopic scale effects (such as interface fluctuations, electrostatic forces, and steric interactions) condition morphology and stability of an emulsion under flow. 
This is particularly important in systems with high droplet packing fractions (typically higher than the close-packing limit of hard spheres), where interfacial effects often dominate 
the physics and where refined numerical techniques would be crucial to cope with inevitable numerical instabilities. 
In these systems, a 3D rheological characterization pinpointing, for example, the role of topological transitions, is currently missing, especially when a highly-packed emulsion crosses narrow gaps, inducing dramatic morphological deformations. These studies, besides being relevant for the design of new soft materials, may also help to describe complex biological processes, such as the collective migration of circulating tumor cells in microchannels  \cite{au_pnas} or the coordinated movement of hundreds of epithelial cells observed, for example, in wound healing \cite{poujade} and gastrulation \cite{mahadevan}. 

Intriguing perspectives may also be envisaged for confined dense emulsions subject to thermal flows. 
For example, it would be interesting to understand the role of heat fluctuations when moving from two to three-dimensional systems, as well as when the number of convective rolls in the Rayleigh-B\'enard cell increases.
A further open problem deals with the effect of thermal convection in yield stress materials, where the volume fraction of the droplets is very large and the emulsion exhibits non-Newtonian features \cite{pelusi_int}. In this respect, a detailed study of the interplay between the microscopic constituents of these systems and the non-linear rheology is still missing.

In the context of active matter, the study of the physics of self-propelled fluid droplets surely represents a promising field of research. As previously discussed, these systems have been shown to capture a number of features typical of cell motility, such as swimming and crawling. 
In this respect, a more realistic model than that presented in section \ref{cell_mig} would include a nucleus, whose deformability is known to be a limiting factor for a successful crossing through a micro pore. A minimal mechanistic model could be built using a double emulsion, where the intermediate layer would contain an active liquid crystal while the encapsulated droplet would represent the nucleus.  
Replacing the solid pillars of the pore with soft ones could be a viable route to adapt this system to the study of more complex biological processes, such as the extravasation in which cancer cells breach the soft barrier of endothelial cells to infiltrate tissues \cite{dirusso}.

LB simulations have been also successfully used in the study of biological systems, such as deep-sea glass sponges, for which {\it in vivo} experimentation is basically unfeasible. Starting from the results discussed in section \ref{sponge_sim}, future works may consider the effect of the soft tissue covering the outside surface of the sponge. 
This condition could be partially reproduced by modeling the sponge as a porous medium with an effective resistance to the flow or through the use of phenomenological boundary conditions capturing the complexity of the combined structure (skeleton plus tissue). However, additional experiments and empirical
data are needed to test more complex models than that presented in Refs.\cite{falcucci_nature,falcucci_prl}.

\subsection{From applications to engineering design}

The set of selected applications presented in the previous sections 
is meant to convey the idea 
of the flexibility and computational efficiency of the LB methods 
for soft-flowing matter across different regimes of motion.
Such flexibility and efficiency are expected to find profitable 
use in various sectors of industrial and materials design, such as food science, chemical-pharmaceutical (drug design and delivery), tissue engineering
and many others. 
While the details of each and every specific application
must necessarily be worked out on a case-by-case basis, in the following 
we wish to convey the flavor of the potential of computational 
design in the above sectors based on the LB methods discussed in this review.

For the sake of concreteness, let us refer to a soft material, say 
a dense emulsion flowing in a box device of height 
$H=0.1$, width $W=1$ and length $L=10$, all in millimeters.
With a lattice spacing $\Delta x =10^{-3}$ (1 micron), the LB
simulation consists of $G=10^{2+3+4}= 10^9$ lattice cells.
Such volume contains about a thousand cylindrical droplets of 
diameter $D=0.1$ mm (100 microns) and height $H$.
At a processing speed of $100$ GLUPS, i.e. a hundred billion lattice sites per second, the LB code updates 
the full box domain a hundred times
in a single second of computer time. 
With a timestep of $\Delta t = 10^{-9}$ (1 ns), one million timesteps
cover $1$ millisecond of physical time. This duration corresponds to  
$10^4$ seconds, namely about three hours of elapsed time over an HPC cluster containing several GPU-based devices \cite{bonaccorso2022lbcuda}.

At this point it should be observed that, in one million time steps, the LB populations complete a hundred longitudinal tours along the domain, which sounds very satisfactory. 
However, the fluid moves much slower than 
the populations; at a speed of $u = 10$ mm/s, in one millisecond, the fluid covers a distance of just ten microns.
Hence, the fluid speed must be artificially accelerated by some three orders of magnitude for the fluid to make at least one tour around the device (1 cm long).
This is common practice in many simulations, not just LB.
The idea is to artificially increase the fluid speed while keeping the relevant dimensionless numbers, in our case the Capillary number $Ca$ and the near-contact number $\mathcal{N}$, unchanged. 
This can be done by boosting the fluid speed $u$, the surface tension $\sigma$
and the strength of the near-contact potential $E_{nc}$ by the same factor, so as to
keep the ratios $u/\sigma$ and $\alpha/\sigma$ unchanged.

With all this said and done, LB updates a millimeter 
cube of material consisting
of about one thousand droplets at a rate of about ten millisecond/day.
The above figures appear consistent with the requirements of engineering design.   
     
\section{Whither LB? Outlook and future perspectives}

The directions for future LB developments aimed at applications in soft matter research are far beyond the scope of any single review. Therefore, in the following, we restrict to three directions which might bear a special interest in the coming years: quantum nanofluidics, machine learning and prospects for the quantum simulation of soft flowing systems.

\subsection{Towards quantum nanofluidics}

The general trend of modern science and engineering towards miniaturization has placed a strong premium towards the study of fluid phenomena at nanometric scales \cite{bocquet1,bocquet2,bocquet_annurev,bocquet_nature}.
This tendency draws from many technological sources, including biology, biomedical, chemical-pharmaceutical, and energy/environment.  
Given the vast amount of energy lost on frictional contacts, low contact friction is paramount to the optimal design of most micro and nano-mechanical devices involved in these applications.

According to continuum mechanics, the 
pressure gradient to push a given mass flow across a channel of diameter $D$ scales like $D^{-4}$. 
Indeed, the centerline velocity of a Poiseuille flow across a channel of diameter $D$
reads as follows
\begin{equation*}
u_c= const. \; \frac{\nabla_x p D^2}{\mu}, 
\end{equation*}
where $const.$ is a geometry-dependent constant.
The mass flow rate, $\dot M$,  is then obtained by multiplying the
centerline speed by a factor proportional to the area of 
the cross-section, which provides an additional $D^2$ 
contribution, whence the $\nabla_x p \sim {\dot M} D^{-4}$ dependence.
Such relation speaks clearly for
the difficulty of pushing flows across miniaturized 
devices: with all other parameters fixed, a ten-fold 
decrease in radius demands a ten-thousand 
fold increase in pressure.

The above scaling derives from the assumption that the 
fluid molecules in contact with
solid walls do not exhibit any net motion (the 
so called no-slip condition), because they remain trapped in local corrugations of the solid wall.
This assumption is no longer valid whenever the size of the channel becomes
comparable with the molecular mean free path and, more generally, whenever the
fluid-solid molecular interactions cannot be described in terms of simple mechanical collisions. 

Whatever the driving mechanism, the onset of a non-zero fluid velocity
at the wall (slip flow), $u_s \ne 0$,  is a much-sought effect, as it turns
the hydrodynamic $D^{4}$ barrier into a much more  
manageable $D^{2}$ dependence.
Slip flow is typically quantified in terms of the so called slip length 
\begin{equation*}
L_s=u_s/(du/dy)_w,
\end{equation*}
where
$(du/dy)_w$ is the velocity gradient at the wall.
Under no-slip conditions, $L_s$ is of the order of 
the molecular mean free
path, i.e. about 1 nm for water, while suitably treated 
(geometrically or chemically) walls
can reach up to $L_s \sim 10 \div 100$ nm,  which is 
comparable to the size of the
nanodevice, leading to a very substantial decrease of the effective
viscosity (roughly speaking a factor $D/L_s$).

Achieving large slip lengths involves nano-engineering 
of fluid-wall interactions such
as to prevent fluid molecules from being trapped by nano-corrugations.
This is usually pursued by clever geometrical and/or chemical coatings which
promote near-wall repulsion between fluid and solid molecules (hydrophobic coating) \cite{bocquet_annurev}.

\begin{figure*}[htbp]
\centering
\includegraphics[width=0.9\linewidth]{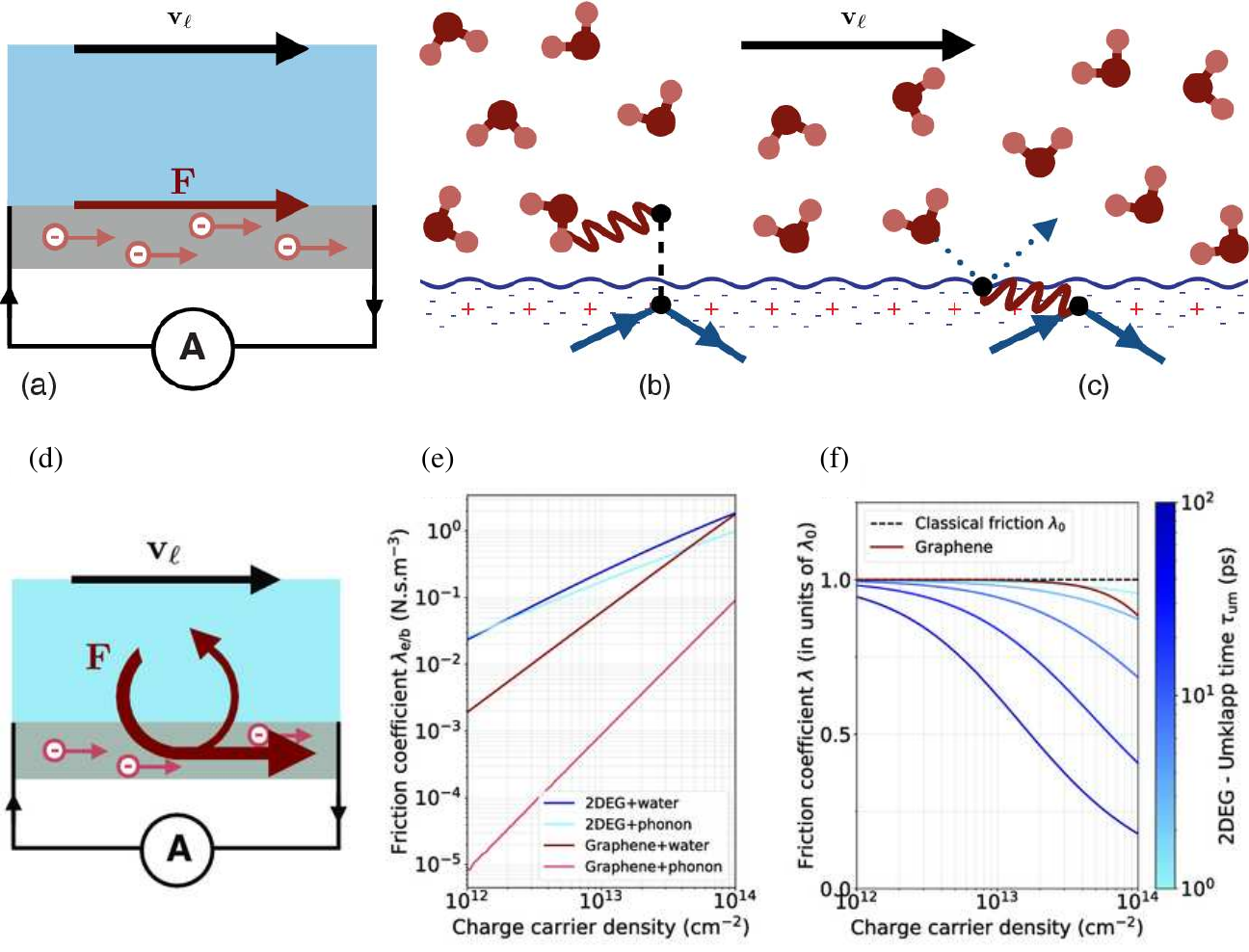}
\caption{Quantum friction scenarios. (a)-(c) A nanofluidic device exhibiting reduced friction through two different electronic current-drive  mechanisms: charged fluctuations in liquid water
imparting momentum
to the electrons in the solid through Coulomb interactions across the liquid-solid interface (b);
electrons in the solid driven by phonons, which are 
excited by water molecule collisions with the solid molecules (c).
(d) A sketch of the electron feedback
to the liquid. (d) The electron-phonon friction coefficient as a function of the electron density in the solid for four different scenarios: 2D electron gas (2DEG) coupled to water, 2DEG and phonons coupled to water, Graphene coupled to water, graphene with phonons coupled to water. (f) The overall friction coefficient in graphene as compared to classical friction as a function of the electron density. Copyright granted under the \href{https://creativecommons.org/licenses/by/4.0/}{CC 4.0 International license}. No changes were made from the original figure from Ref.\cite{bocquet2}.}\label{quantum_frict}
\end{figure*}

However, in recent years, it has been argued and experimentally shown that
unanticipated quantum-electro-mechanical phenomena can also lead to spectacular
reduction of frictional losses \cite{bocquet2,bocquet3}.
For instance, it has been shown that slip flow 
in carbon nanotubes   is
largely underestimated by MD
simulations, indicating that standard
force-field procedures fall short of describing the actual physics
of the water-graphene interactions, pointing to a crucial role of electronic
degrees of freedom in the solid \cite{kavokine_nat}. Interestingly, ab-initio MD, besides being even more
unpractical computing-wise, would also fall short of capturing the basics of quantum friction because
the electrons in the solid are non-adiabatically 
coupled to charge fluctuations in the liquid.  
Therefore, it would be extremely interesting to enrich 
the LB formalism
in such a way as to include quantum interfacial interactions.
This is indeed possible by exploiting recent results obtained
from quantum non-equilibrium Keldysh analysis of interfacial transport in nanofluids \cite{bocquet2}. 
The Keldysh formalism accounts for non-equilibrium quantum transport phenomena
and, in this respect, it necessarily involves very elaborate analytical
calculations based on advanced and retarded Green functions.
The final outcome, though, comes in the form of Langevin-like friction terms
which could be easily incorporated within an LB-extended framework.
For the sake of concreteness let us refer to the 
case of a charged ionic liquid, in
which charge fluctuations couple to phononic and 
electronic degrees of freedom in the solid walls.
The Keldysh analysis shows that, under suitable geometrical and physical
conditions, such interfacial quantum coupling can lead to a reduction
of the wall friction, because part of the momentum imparted to solid electrons
is recycled back to the liquid, an effect known as "negative quantum friction" (see Fig.\ref{quantum_frict}).  
Clearly, the inclusion of "quantum friction" alone is not 
enough since, based on the fluctuation-dissipation 
theorem, quantum fluctuations must be accounted for as well.
This is a genuinely new theoretical and computational 
challenge to LB since, depending
on their energy, the quantum fluctuations may last longer 
then the LB timestep so that, at variance with 
standard versions of the fluctuating LB formalism \cite{FLB1,FLB2},
the correlator of the fluctuating force extends beyond 
a single LB time-step.

In summarizing, one would write three LB equations for the three species in play, hydrons, electrons, and phonons, coupled via screened Coulomb interactions,
frictional terms and fluctuating forces.
Such a model is best implemented on higher-order lattices so as to capture 
stronger non-equilibrium effects, stronger fluctuations, and also mitigate
spatial non-local effects.
It is argued that such Keldysh LB formalism 
might open an entirely new avenue for the 
computational design of quantum-controlled low-friction 
nanodevices.

\subsection{Machine learning}

Machine learning (ML) techniques based on artificial neural networks are increasingly 
employed in microfluidics to perform various tasks, both in experimental and modeling contexts. 
While in the former case ML is used mainly as an analyzer tool to extract information from 
images or movies (for instance, to track droplets, bubbles \cite{durve2023benchmarking,durve2022droptrack} and cells \cite{gardner2022deep}) or to infer physical quantities, such as shear and deformation rate \cite{ning2023application,sathyan2020modelling,nazar2023evolutionary}, in the latter ML has recently shown its capability to enhance numerical approaches by improving their accuracy. 

Although ML is essentially a data-driven approach, it can be enriched and reinforced with
physical foreknowledge of the problem under investigation. 
This concept is crucial, for example, in Physics-Informed Machine Learning techniques \cite{karniadakis2021physics}, in which data and mathematical models (such as partial differential or integral equations) are integrated and then implemented through neural networks or
kernel-based regression networks \cite{wang2022and,cai2021physics}.
Thus, ML can be used not only as a heuristic tool driven by the statistical inference of large datasets but also as a system that embeds, in the learning process, the physical laws governing a specific data set.  In this framework, the artificial neural networks in the ML approach could act as a class of physical-informed models satisfying universal approximation theorems while preserving the physical ingredients and, concurrently, enhancing the overall computational performance. 

In the following, however, we shall focus on a specific item: the use of machine learning to enhance the physical accuracy and computational efficiency of LB simulations of complex soft matter flows, including near-contact interactions acting at the scale of a few nanometers. 

The task of reaching out to scales of experimental and engineering devices 
(mm-cm) accounting for near-contact interactions (1-10 nm) entails at least six 
decades in space and about twice as many in time, resulting in a computational demand which far 
exceeds the capabilities of current leading-edge near-exascale computers. 
The standard practice is then to represent and parameterize the effects of
the unresolved fine scales on the resolved ones, by suitable coarse-grained models.
The success of such strategy hinges heavily on scale separation arguments, meaning by this
a sufficiently weak coupling between the fine and coarse scales.
Such an assumption may or may not apply depending on the specific problem at hand,
hence it is highly desirable to devise non-perturbative methods capable of handling strong-coupling regimes as well.
Machine learning is a good candidate to reach this goal, as it
can be employed to learn optimal models from experimental or
highly-resolved simulation data, without resorting to any weak-coupling 
assumption, by now a common practice in many areas of computer simulation.
In our specific case, we shall focus on two related and yet distinct strategies:
i) Learning collision operators for ideal and non-ideal fluids and ii) Learning coarse-grained pseudo-potentials. 

\subsubsection{Learning collision operators for ideal fluids}

Collision operators for LB simulations of simple fluids can be
derived analytically, based on the constraints imposed by the 
mass-momentum-energy conservation laws.
Formally, the mapping from the actual distribution $f$
to the corresponding local equilibrium $f^{eq}$ for
the standard single-time BGK collision operator can be
written in mode-coupling form as follows
\begin{equation}
\label{LEQF}
f_i^{eq} = A_{ij} f_j + B_{ijk}f_j f_k/\sum_l f_l,    
\end{equation}
where $A_{ij} = c_{i\alpha} c_{j\alpha}/c_s^2$ and $B_{ijk}=Q_{i\alpha\beta} c_{j\alpha}c_{k\beta}/c_s^4$
are constant matrices to be learned by the machine.
The denominator at the right-hand side is the fluid density,
but since the neural network, in principle, is not aware of it
we have left this association unspoken.
Formally, the above expression is a highly nonlinear function of the
$b$ variables $f_i$ ($i=1,...,b$), not only because of the quadratic term $f_j f_k$
but especially because of the denominator $\sum_l f_l$, which gives rise
to a non-polynomial nonlinearity.
The task of learning the multivariate nonlinear expression of Eq.(\ref{LEQF}) is greatly facilitated by supplementing the neural network
with the conservation constraints
\begin{eqnarray}
\sum_i f_{i}^{eq} = \sum_i f_i = \rho,\\
\sum_i f_{i}^{eq} c_{i\alpha} = \sum_i f_i c_{i\alpha} = J_{\alpha}.
\end{eqnarray}
With such assistance, the neural network should be able to learn the
local equilibrium in its "mean-field" form, namely
\begin{equation}
\label{LEQM}
f_i^{eq} = w_i (\rho + {J_{\alpha}}{c_{i\alpha}} + \frac{J_{\alpha}J_{\beta}}{\rho} Q_{i\alpha\beta}),      
\end{equation}
where the discrete speeds are normalized by the sound speed $c_s$.
This mean-field form is obviously easier to learn, as it depends only
on four independent variables, $\rho$ and $J_{\alpha}$.
To the best of our knowledge, this is how machine learning of LB
equilibria has been developed so far \cite{lou2021physics,corbetta_epje}.

However, many variants of Eq.(\ref{LEQM}) have been developed in the last decades
in order to achieve better stability or incorporate additional macroscale 
and microscale physics, whose derivation is usually based on informed ad-hoc 
phenomenological assumptions \cite{succi2018lattice}. 
The path to these phenomenological models is generally not unique because
no systematic procedure is available to derive them from 
first principles, due to the fact that one is dealing with flowing
systems far from equilibrium.
This opens up a major scope for machine learning collision operators based on 
training from experimental data or high-resolution simulations.

Essentially, one is presented with three possible scenarios.
The first and most pessimistic one is that the machine 
fails to learn the existing models for non-ideal fluids, such as 
Shan-Chen, free-energy or chromodynamic models.
The second, less pessimistic, possibility is that the
machine learns exactly the same models derived on phenomenological grounds.
This would be a success for machine learning but pointless
to the purpose of enhancing the LB simulators.
The third, and definitely most exciting possibility, is that 
the machine learns {\it new} models which have not been discovered yet.
In this case, machine learning would literally teach us new physics, thereby
providing a new generation of LB schemes capable of accessing
strongly-coupled regimes out of reach for the 
current LB models for non-ideal fluids.

\subsubsection{Learning generalized collision operators for non-ideal fluids}

So much for the general scenario. To date,
machine learning for LB is still literally in its infancy, and the work so far has been directed to ideal fluids with no potential energy. 

In the field of turbulence, for example, Bedrunka et al. \cite{bedrunka2021lettuce} have presented 
a PyTorch-based LB code, where a neural collision model (a more accurate version of classical collision models) is trained on a shear layer flow and then applied to a decaying turbulence flow with satisfactory numerical results. 
Also, convolutional neural network (CNN) and gated recurrent unit neural network (GRU) have been combined with the LB method to 
reduce computational time and improve the efficiency of turbulent simulations up to Reynolds number of 4000 \cite{zhao2023improvement}.

An intriguing perspective is that offered by the physics-informed neural networks (PINNs). In Ref. \cite{lou2021physics}, for example, they are used to learn the BGK collision operator of various benchmark flows, such as continuous Taylor-Green flow and rarefied micro-Couette flow. 
Following a similar footprint, Corbetta et al. \cite{corbetta_epje} have shown that a neural network architecture combined with suitable physical properties (such as conservation laws and symmetry) allow for a precise reconstruction of collisional operators of the LB and accurate dynamics of standard fluids (such as Taylor-Green vortex and lid-driven cavity flows). 

\begin{figure}[h]
\centering
\includegraphics[width=1.0\linewidth]{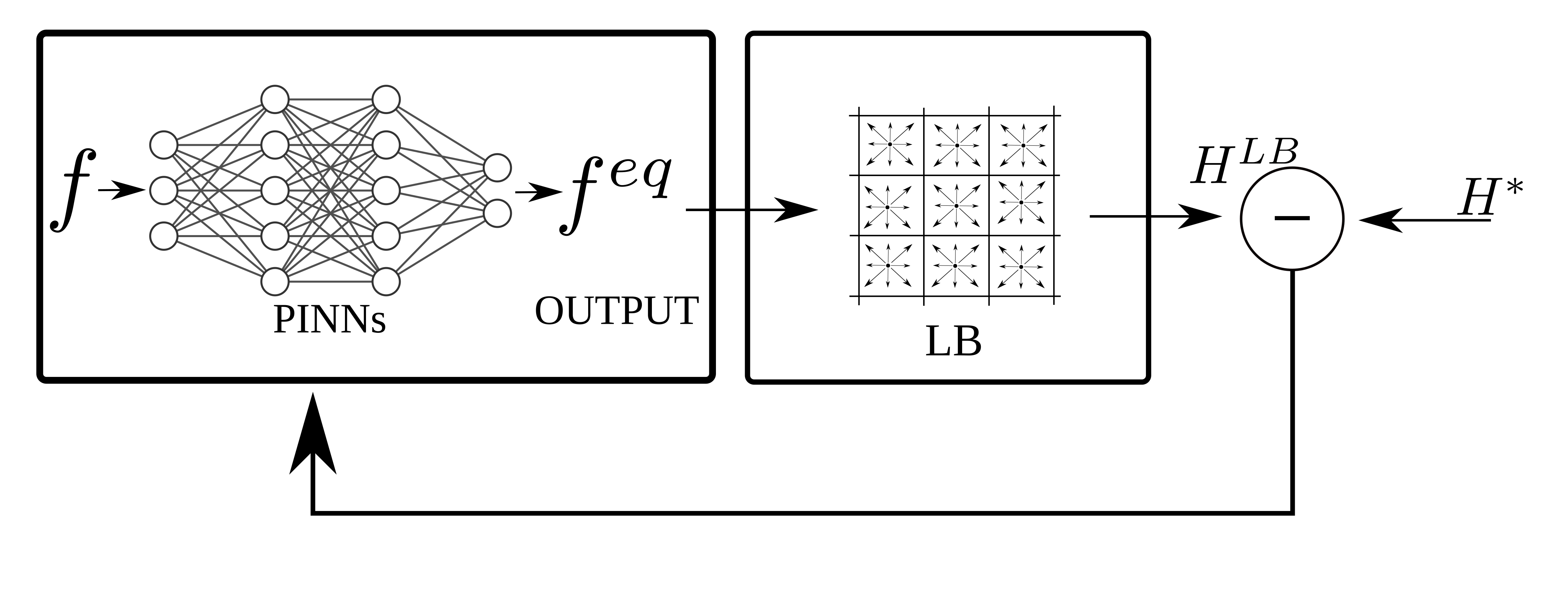}
\caption{A sketch of the procedure aiming at learning the collision operators of non-ideal fluids. 
A set of initial distribution functions $f$ trains a neural network which delivers a provisional set of equilibrium distribution functions $f^{eq}$, given by Eq.(\ref{feq_pinn}), as output. These ones feed a LB simulation which  provides the hydrodynamic
 observable (such as mass and momentum) to be compared with experimental data. If the match is accepted within a given tolerance, the search ends otherwise the weights $W$ are updated by a gradient-descent method and the algorithm restarts.}
\label{ML_collision_operator}
\end{figure}

The challenge we wish to briefly discuss in this review is 
the extension of some of these approaches to the case of collision operators
for multiphase and multicomponent flows, including near-contact interactions. 
A possibility is to develop PINNs incorporating the additional constraints associated
with non-ideal momentum-flux tensor inclusive of non-ideal interaction, i.e. 
the Korteweg tensor.  
Here we outline the basic steps of the procedure, shortly sketched in Fig.\ref{ML_collision_operator}.

{\it Step 1}: The NN is presented with the actual value $f_i(x)$ of the discrete 
distributions as an input, and delivers a provisional local equilibrium $f_i^{eq}(x)$
as an output, where $x$ denotes the full set of spatial locations.
Symbolically,
\begin{equation}\label{feq_pinn}
f^{eq} = \sigma^L (Wf-b),
\end{equation}
where $\sigma^L$ denotes the activation function as recursively applied to $L$
hidden layers composing the (deep) neural network, $W$ is the set of weights and 
$b$ are the associated biases.

{\it Step 2}: The LB simulation is run with the provisional 
set of local equilibria derived in Step 1.

{\it Step 3}: The results, namely a set of hydrodynamic 
observables $H^{LB}$, are compared with the "truth" $H^*$, either 
experimental data or highly-resolved simulations,
to deliver the current value of the loss function 
\begin{equation}
\mathcal{L}=dis(H^{LB},H^*),
\end{equation}
where "dis" indicates some suitable metric in the space of the functions $H(x)$,
for instance the Euclidean distance $\sum_x (H^{LB}(x)-H^*(x))^2$ or similar ones.

{\it Step 4}: If the distance $dis(H^{LB},H^*)$ falls below a given 
tolerance, the search is stopped. Else, the weights are updated by some form
of gradient descent, 
\begin{equation}
\delta W = -\alpha \partial \mathcal{L}/\partial W,
\end{equation}
and then back to Step 1.

It is known that this search is greatly accelerated by augmenting the loss function
with physically inspired constraints. In the case of a standard LB, these are the mass-momentum and momentum-flux constraints.
In compact four-dimensional notation, they read as follows:
\begin{equation}
\mathcal{L}_{\mu,\nu}^{PI} = dis (P_{\mu,\nu}^{kin}, P_{\mu,\nu}^{eq}),
\; \hspace{0.5cm}\mu,\nu=0,d
\end{equation}
where we have defined  
$P_{\mu,\nu}^{kin} = \sum_i f_i^{eq} c_{i,\mu} c_{i,\nu}$.
In the above, we have set $c_{i0}=1$, so that $P_{0,0}^{eq} =\rho$, 
$P_{0,a}^{eq}=P_{a,0}^{eq}=J_a$
and $P_{ab}^{eq} = \rho u_a u_b + p\delta_{ab}$, where 
Latin indices running along the spatial dimensions.
The loss function associated with these physical constraints  
is then the sum over the single components $\mu,\nu=0,d$.
Such physics-informed loss function is then added to the standard loss function discussed above, as essentially done by
previous authors.

For the case of non-ideal fluids, the procedure stays basically the same, with the key proviso that the momentum flux must include the non-ideal component, namely 
the Korteweg tensor $K_{ab}$, whose kinetic expression is
\begin{equation}
K_{\mu,\nu}^{kin} = \sum_i f_i^{eq} F_{i,\mu} d_{i,\nu},
\end{equation}
where $d_{i0}=0$ and $d_{ia} = c_{ia} \Delta t$ and $F_{ia}$
is the central force acting between two lattice sites $x_a$ and $x_a+d_{ia}$.

Further on, in the presence of near-contact interactions, the Korteweg tensor must be augmented with these
contributions, as explicitly derived in \cite{montessori_jfm}.
This procedure requires knowledge of the augmented Korteweg tensor, hence it is expected to recover existing local equilibria or equivalent forcing
terms currently used in LB simulations.
However,  it cannot be ruled out that 
new generalized equilibria might emerge from the ML procedure 
discussed above. 
This would mark a decisive contribution of machine learning to
LB simulations of complex soft matter flows.
Although we are not aware of any detailed work in this 
direction, this is a decidedly interesting topic for future
research in the field.

\subsubsection{Learning near-contact pseudo-potentials}

In Ref.\cite{montessori_jfm} an effective way to incorporate near-contact interactions within a chromodynamic LB model for multicomponent flows 
is presented, which stands as an extreme instance of coarse-graining of the 
complex physics arising when 
two fluid interfaces covered by a surfactant come into close contact. 
Such a picture is typical of soft many-body flowing systems, such as dry foams and dense emulsions. As mentioned in Section \ref{basic_physics}, such near-contact interactions aim at condensing 
a plethora of effects due to the onset of forces arising at nanometric scales, such 
as electrostatic double-layer, dispersion forces and 
Casimir-like interactions \cite{derjaguin,verwey}, within a simplified, one-parameter density functional. 

The problems with such a treatment may be summarized through two main points: 
a) the physical fidelity of the upscaled mathematical formulation is hard to 
assess a priori and can only be judged ex-facto;
b) there is an intrinsic uncertainty related to the choice of the parameters  
controlling the magnitude of such interaction forces.

The hard way to solve such issues is to systematically "down-scale" the 
problem, by studying the details of the interactions through micro-scale 
techniques, such as molecular dynamics and perhaps even electronic structure methods in case electronic degrees of freedom take an active role \cite{bocquet1,bocquet2}. 
The main problem of such a microscopic approach is the lack of computer power to reach out to scales of experimental and engineering design interest, say centimeters.
The alternative, top-down approach, is to augment mesoscale forces and pseudo-potentials with additional terms which would retain the essential 
microphysics within a viable computational framework.
Translated into formulas, the near-contact force per unit volume
along direction $x_a$ is given by
\begin{equation}
F_{a,nci}= \nabla \cdot \Sigma_{nci}= 
\nabla \cdot \left(A_{nci}[z;\kappa_{int},u_{rel}] \frac{\nabla\phi}{|\nabla \phi|}\right)\vec{n}\delta_{ab},
\end{equation}
where $z$ is the spatial coordinate along the normal $\vec{n}$ of a
given interface, $u_{rel}$ is the relative velocity 
and $\kappa_{int}$ is the local curvature of the interface.

By dimensional arguments, one could generically write
\begin{equation}
A(z;\lambda) = \frac{\epsilon(\lambda)}{z^4},
\end{equation}
where $\epsilon$ is the pseudo-potential to be machine-learned as a
function of the "hydrodynamic" parameters, symbolically denoted by $\lambda$. 
\begin{figure}[h]
\centering
\includegraphics[width=1.0\linewidth]{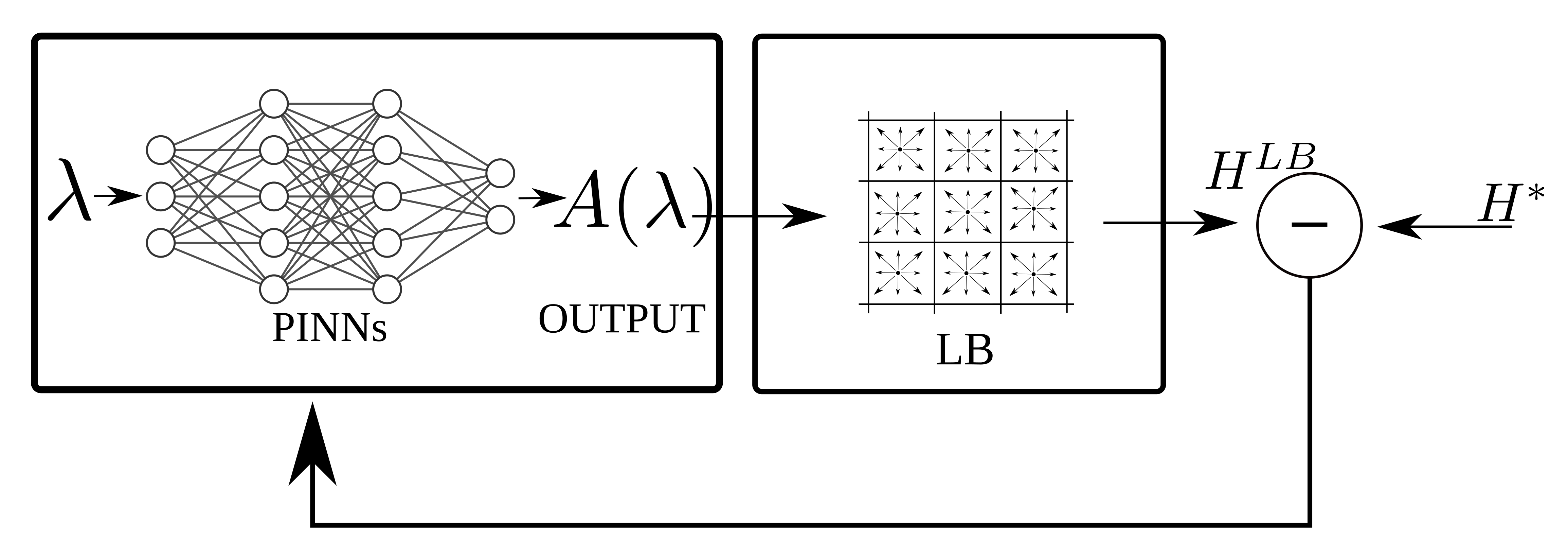}
\caption{A schematic procedure aiming at learning the functional form of near-contact forces to be included in LB simulations. The training of the PINNs follows the same procedure as the one described in Fig.\ref{ML_collision_operator}, except that the input is a list of parameters $\lambda(x)$ and the output is the function $A(\lambda)$.
}\label{ML_A}
\end{figure}

The actual procedure would look pretty much like the one discussed earlier, except that the input
to the PINN is no longer the actual distribution $f_i(x)$ but the set of parameters
$\lambda(x)$ which govern the near-contact physics at position $x$ in space (as sketched in Fig.\ref{ML_A}).
Once the PINN finds the functional form of $A$, the near-contact interactions can be introduced within the LB framework through the usual shift of the equilibria or via direct forcing. Then, a LB simulation is run with this temporary form of $A$ to finally produce a list of hydrodynamic variables. Once this step is completed, the subsequent ones proceed following steps 3 and 4 discussed in the previous paragraph.

\subsection{Machine learning for the deformable many-body problem}

Yet another application of machine learning we discuss in this review is the derivation of effective equations of motion for soft suspended bodies based on the geometrical parameters of the experimental device.

The first step along this direction is to automatize the identification and tracking of the droplet trajectories. Object tracking constitutes the automated process of identifying and monitoring multiple objects within a series of images. Typically, these tasks are managed by distinct algorithms. The first category encompasses object detection algorithms, which aim to identify multiple objects within a single image. Conversely, the second category, referred to as object trackers, is responsible for assigning unique identity numbers to each detected object, maintaining consistency across successive frames. These persistent unique identifiers serve as the basis for tracking objects across sequential images, thereby generating trajectories.

\begin{figure*}[htbp]
\includegraphics[width=1.0\linewidth]{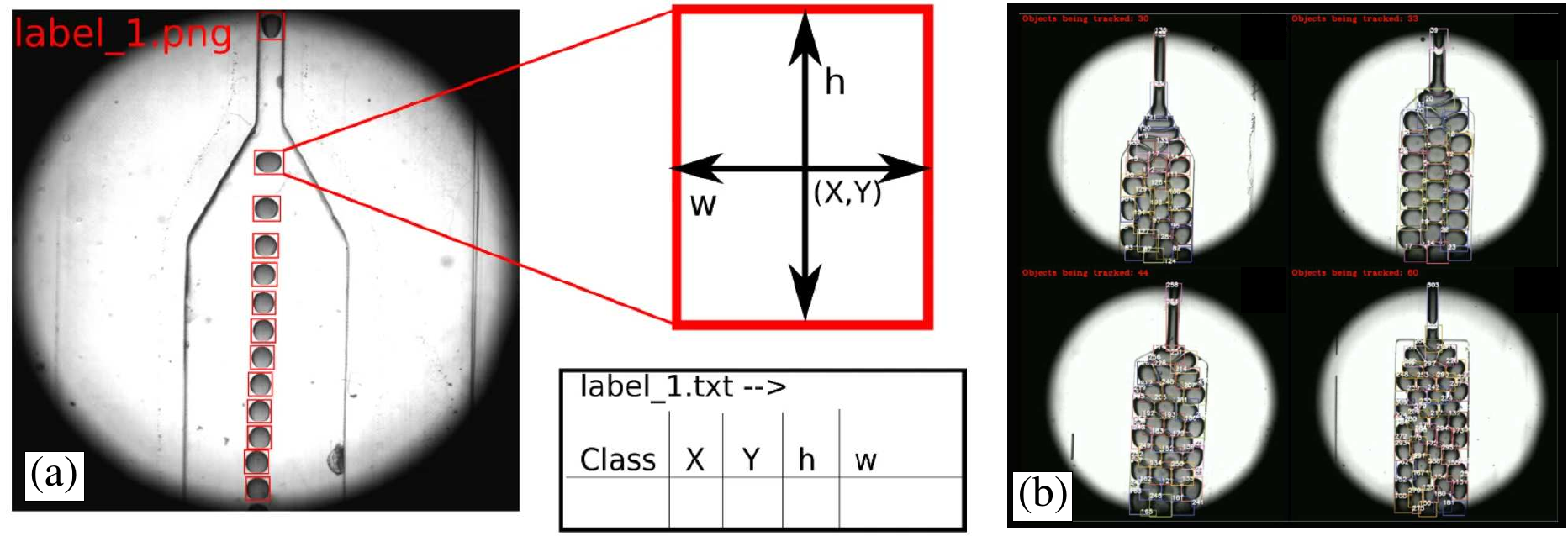}
\caption{(a) A picture of a sequence of monodisperse droplets in a divergent microfluidic channel. This represents an example of a training dataset for YOLO. Each droplet is identified by a bounding box, whose dimensions are noted in an associated file.  (b) Examples of DropTrack's outputs, which identify and track droplets of a dense emulsion flowing in microchannels of different geometry. Copyright granted under the   \href{https://creativecommons.org/licenses/by/4.0/}{CC 4.0 license}. No changes were made to the original figures in Ref. \cite{durve2023benchmarking}.
}\label{yolo_fig}
\end{figure*}

Object detection algorithms primarily undertake two key tasks: (a) object localization, involving the determination of object positions within an image and the delineation of their boundaries, and (b) classification of the detected objects into predefined categories. Recent years have witnessed significant advancements in the development of object-localization algorithms, such as Region-Based Convolutional Neural Networks (RCNN) \cite{Girshick_2014_CVPR} and their iterations (Fast R-CNN \cite{Girshick_2015_ICCV}, Faster R-CNN \cite{ren2016}, Cascade R-CNN \cite{Cai_2018_CVPR}), You Only Look Once (YOLO) \cite{yolo1} and its subsequent versions \cite{yolo3,yolo4,yolo5,yolo7,yolo8}, Single Shot MultiBox Detector (SSD) \cite{ssd}, Single-Shot Refinement Neural Network for Object Detection (RefineDet) \cite{ssrnn} and Retina-Net \cite{ratina_net}, to name a few ones.
Conversely, object tracking algorithms are focused on correlating detected objects across consecutive frames and tracking them by learning their distinctive features. Among many examples, DeepSORT \cite{deepsort} stands out as one of the most prominent object-tracking algorithms which incorporates deep learning
into the tracking process. 
Interestingly, it has been recently combined with YOLO in order to perform the two aforementioned tasks, i.e. 
detect and track droplets within experimental setups. 
This software, named "DropTrack" \cite{durve2022droptrack,durve2023benchmarking,durve2}, is capable of inferring trajectories of individual droplets (by analyzing over 30 frames per second using GPU-accelerated hardware) and of extracting pertinent physical observables, such as droplet counts, size distributions, degree of droplet arrangements, and packing fraction, from its output (see Fig.\ref{yolo_fig}).

\subsubsection{Learning shape dynamics} 

In many soft matter applications there is substantial scope for controlling not only 
the position and velocity of the single droplets, but also their shape changes as such 
droplets move along the flow. 
Since this adds a significant number of internal degrees of freedom, machine learning 
can find profitable use to accomplish this task as well. 
For instance, Li et al. \cite{math12010109} used machine learning techniques to predict the shape of bubble droplets 
to be used in a multiphase LB simulations using the pseudopotential method. 
A regularized feedforward deep neural network, made of a first hidden layer of fifty neurons 
and a second one of ten neurons, was then trained on this data 
with a gradient descent method, 
providing a R-coefficient ranging between 0.9965 and 1. 
However, the study is limited by the narrow range of surface tension (0.01 and 0.02) and viscosity ratios (0.05 and 0.10), as well as by the bi-dimensionality of the simulation.
Recently, a class of machine learning methodologies, called autoencoder, has been employed to acquire minimal droplet shape descriptors, subsequently applied in forecasting droplet breakup occurrences \cite{Khor2019, durve2024shape}, a process crucial in many industrial applications \cite{rosenfeld2014,gai2016}.  

As a general goal, one may think of machine-learning as a useful tool to devise the effective
equations of motion of suspended bodies, including internal degrees of freedom 
associated with the deformability of the soft bodies. 

The corresponding set of effective dynamic equations 
for a set of $N_d$ droplets would take the following form. 

{\it 1}. For the external degrees of freedom, i.e. position $R_i$ and velocity $V_i$, the equations are:
\begin{eqnarray}
\dot{R}_i&=&V_i\\  
M_i\dot{V}_i&=&\sum_{j=1}^{N_d}f(R_i,R_j,u_i,u_j,a_i,a_j) \equiv \sum_{j=1}^{N_d} f_{ij},
\end{eqnarray}
where $u_i$ is the flow velocity at location $R_i$, and $a_i$ is the sequence of Fourier coefficients (a vector) describing 
the shape of the $i$-th droplet.

{\it 2}. For the internal degrees of freedom, i.e. shape coefficients $a_i$, one has:
\begin{equation}
\dot{a}_i=\sum_{j=1}^{N_d} g(R_i,R_j,u_i,u_j,a_i,a_j) \equiv \sum_{j=1}^{N_d} g_{ij}
\end{equation}

{\it 3}. Finally, for the flow fields, the equations are:
\begin{equation}
\dot{u}_i=\sum_{j=1}^{N_d} h(R_i,R_j,u_i,u_j,a_i,a_j) 
\equiv \sum_{j=1}^{N_d} h_{ij}.
\end{equation}

The above is a formidable problem since it involves functions of 
several thousands of variables: thousand droplets with, say, ten internal 
shape parameters each, generate $(6+10) \times 1000 = 16000$ dimensions.
However, this is precisely the ground where machine learning maybe expected
to offer insights outside the reach of analytics and also large-scale numerics.

Incidentally, we do not expect reciprocity, that is 
$f_{ij} \ne f_{ji}$, $g_{ij} \ne g_{ji}$ and $h_{ij} \ne h_{ji}$
since the flow field, as well as the deformations, break 
both translational and rotational invariance.
This is reminiscent of active-matter behavior and can indeed be related to
self-propelling effects, along the lines pioneered by 
Shapere and Wilczek, where the problem of self-propulsion at low Reynolds 
number is formulated in terms of a gauge field over the space of shapes \cite{wilczek1,wilczek2}.
In this respect, machine learning could be useful to inform a statistical theory of
shape dynamics, with potential applications outside the realm of soft matter.

\subsection{Prospects for quantum computing of soft fluids}
We finally discuss a further perspective about potential links between LB methods and quantum computing. 
Indeed, quantum computing is one of the most vibrant topics of modern science, holding promises
of spectacular applications far beyond the reach of classical
electronic computers, if only for a limited set of applications \cite{q_sup,nisq}.
The main point is that qubits, the quantum analog of classical bits, 
represent an arbitrary superposition of the two fundamental states $|0>$ (ground state)
and $|1>$ (excited state), so that a collection of $Q$ qubits spans an exponentially
large Hilbert space, consisting of $2^Q$ classical states.
The blue-sky scenario for fluids is mind-boggling.
Given that a fluid flow at Reynolds number $Re$ consists of about $Re^3$ active
degrees of freedom, the number of qubits required to represent 
such state-space is given by
\begin{equation}
Q(Re) = 3 Log_2 Re \sim 10 Log_{10} Re.    
\end{equation}
This expression shows that a flow at $Re=10^8$, basically the state-of-the-art
of current electronic supercomputers, can be represented by mere $80$ qubits, which
is well within the {\it nominal} capabilities of current quantum hardware \cite{QIBM},
now offering several hundred physical qubits.

However, realizing this exponential advantage faces a number of steep technological
and conceptual challenges. First, qubits decohere quite fast, in a matter of microseconds
even with the best current-day quantum technology. Second, even when qubits are in a coherent superposition,
quantum updates can still fail due to quantum noise, with a typical error rate
around $10^{-3}$, to be contrasted with a classical error rate around $10^{-18}$!
The end result, to date, is that the number of effective qubits that can be actually
used to perform reliable simulations is at least a factor ten below the nominal value.
Hence, a more realistic estimate is 
\begin{equation}
Q(Re) \sim 100 Log_{10} Re, 
\end{equation}
which means that reaching the exascale performance will require of the order
of thousands effective qubits.

Realizing such potential for fluids meets with two additional challenges: nonlinearity
and dissipation. Indeed, while quantum mechanics is linear and conservative, the physics
of fluids is typically neither, hence additional procedures have to be devised to
formulate quantum computing algorithms for fluids.
Dissipation can be dealt with in several ways, for instance by adding a reservoir to the
original quantum states, such as the union of the two conserves energy \cite{mezzacapo_scirep}.
However, the non-unitary update of the system implies a non-zero probability of failure
which accumulates in time, leading to a deterioration of efficiency. 

Nonlinearity is a more fundamental problem.
Several strategies are available to deal with it 
(for a recent review see  \cite{ITANI2}), but in the following we 
shall briefly focus on one which has received particular attention in the last few years, namely
Carleman linearization \cite{CARLE}.

\begin{figure*}[htbp]
\centering
\includegraphics[width=1.0\linewidth]{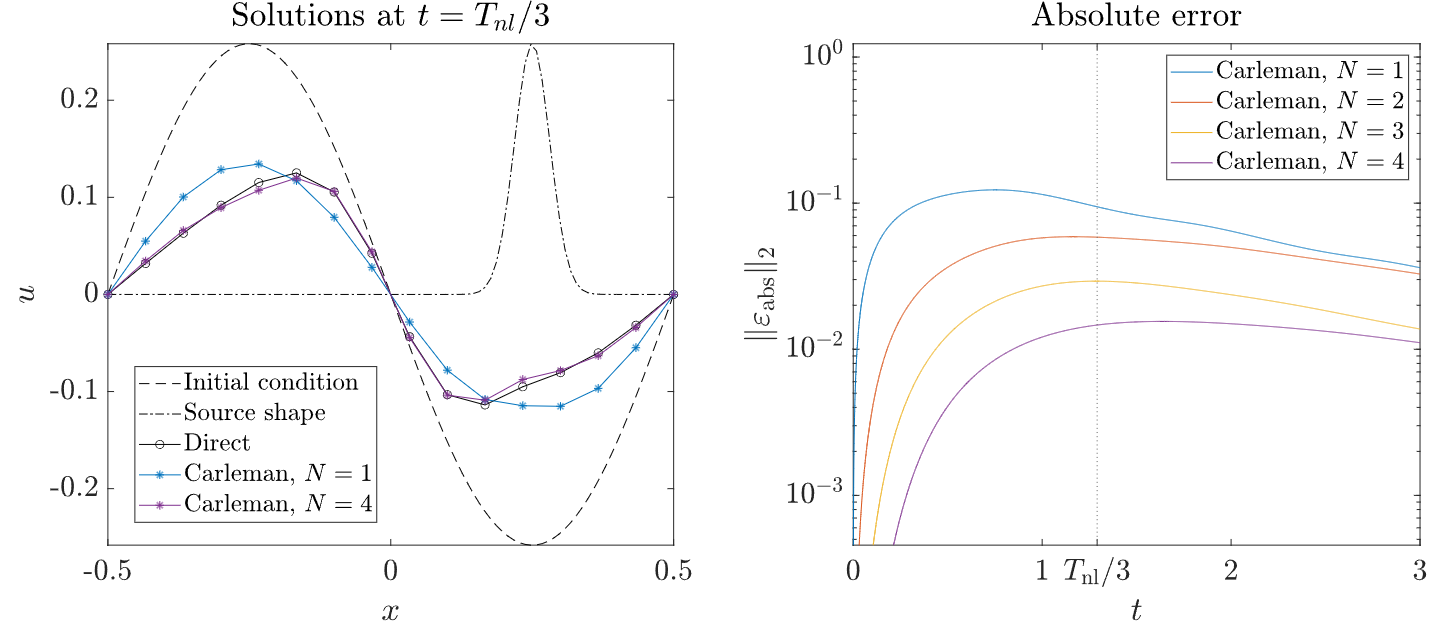}
\caption{Left panel: Initial conditions, source and solutions of the velocity field of the Burger equation computed at Re=20 with a spatial discretization of 16 grid points and run on a classical computer. Right panel: $l_2$ 
norm of the absolute error between different Carleman truncation levels. 
Note that the error of the Carleman method converges exponentially with $N$. 
Copyright (2021) National Academy of Sciences Ref.\cite{childs}.}
\label{carleman}
\end{figure*}

The idea is very simple and best illustrated by means 
of a zero-dimensional nonlinear system, namely the logistic equation
\begin{equation}
\frac{dx}{dt} = x(1-Rx), \;\;\;   
x(t=0)=x_0,
\end{equation}
where $R$ measures the strength of the nonlinearity.
The Carleman procedure consists in renaming $x \equiv x_1$ and $x^2 \equiv x_2$, so that
the logistic equation takes the following form:
\begin{equation}
\frac{dx_1}{dt} = x_1 -R x_2,
\end{equation}
This is formally linear, but open, hence it requires an additional equation for $x_2$.
This is easily derived by differentiating $x^2$, which yields
\begin{equation}
\frac{dx_2}{dt} = 2(x_2 -R x_3),
\end{equation}
which is again linear but in need of a closure for $x_3 \equiv x^3$.
Reiterating the procedure up to a generic level $k$, we obtain
\begin{equation}
\frac{dx_k}{dt} = k(x_k -R x_{k+1}),
\end{equation}
which can be closed by setting $x_{k+1}=f(x_k)$, the typical case
being truncation $x_{k+1}=0$.
The idea is clear, one trades nonlinearity in finite dimensions 
(just one variable in the logistic case) for a linear problem in infinite dimensions,
in the hope that a low-order truncation can nevertheless provide 
an accurate approximation to the original problem.
For the case of quantum computing for fluids, the point is to turn the 
nonlinear equations of fluids into a linear set of the form
\begin{equation}
\frac{dV}{dt} = CV,
\end{equation}
where $V$ is the set of fluid variables and $C$ is the associated Carleman matrix.
Such a linear system could then {\it in principle} be solved by QLAS, Quantum Linear Algorithm
Solver \cite{HHL}.
The reality tells however a much more complicated story.
First, the number of Carleman variables grows exponentially with the level of the Carleman truncation. For instance at level $k=1$ the relevant 
Carleman set of variables is $\lbrace \rho,J_a,J_a J_b/\rho \rbrace$ , 
namely $1+3+6=10$ in dimension $d=3$.
On a $1000^3$ grid, this makes $10^{10}$ variables, requiring $Q = 100 Log_2 10 \sim 330$
effective qubits. At the next Carleman level, we have all possible cross-couplings of the
former terms which arise as one constructs the dynamic equation 
for $K_{ab} \equiv J_a J_b/\rho$, including coupling to 
the non-local dissipative term $\partial_a J_b$.
This leads to $O(100)$ Carleman variables, hence $Q \sim 370$ qubits, which looks fine.
However, this hinges on two additional assumptions: First, 
all couplings remain local; second, that the Carleman procedure shows satisfactory
convergence already at Carleman level 2.
By a suitable choice of the discrete scheme, the first requirement can be met. 
However, based on actual evidence, no satisfactory convergence is observed at moderate
Reynolds numbers, $Re \sim O(10)$, over a significant period of time \cite{SANAVIO}.
To date, the only partially successful application is the simulation of a one-dimensional
Burgers flow, with $16$ grid points over a few thousands time-steps \cite{childs}, truncated
at the fourth Carleman level (see Fig.\ref{carleman}).
Note however that the Carleman procedure applied to the Burgers 
equations is far simpler than Carleman applied to the Navier-Stokes equations.

It should be mentioned that much better convergence can be obtained by applying the
Carleman procedure to the LB equation \cite{ITANI,ITANI2,SANAVIO}.
Unfortunately, the corresponding Carleman-LB algorithm is non-local, which reflects
into an unviable depth of the corresponding quantum circuit.
As a result, with the current state of affairs, the implementation of 
a quantum algorithm for fluid flows is still open.
Coming back to soft flowing matter, the good news is that the Reynolds 
number is generally moderate, often below $100$, hence, once the aforementioned
problems are solved (assuming they will), about a hundred effective qubits 
will suffice to quantum simulate complex states of soft flowing matter. Work along this line is currently in progress.

\section{Conclusions}

Summarizing we have discussed a series of LB methods to simulate microscale multi-component
soft flows under strong geometric confinement.
Such methods leverage the flexibility of the LB method to efficiently 
incorporate mesoscale physics beyond hydrodynamics and yet in no 
need of detailed molecular specificity.
The inclusion of near-contact interactions via 
a simple repulsive force stands as a paradigmatic 
example of this approach and has proven successful for a broad variety of complex 
rheological applications which would be very hard to treat with 
different methods, particularly in the case of dense confined emulsions.

This success hinges on what we have called {\it Extended Universality} (EU),
meaning by this the dependence on a series of dimensionless parameters 
rather than on the specifics of the near-contact interactions.  
In our case, near-contact interactions have been represented by a one-parameter
repulsive force, but it is not hard to imagine situations where multi-parameter
representations would be needed, if not a fully atomistic description.
The quantum-nanofluidic applications discussed in the previous section is 
a likely candidate in this respect.
Likewise, biological applications present many instances of broken EU, for 
instance it is known that cells develop major protrusions, called philopodes, whose 
task is to explore the surroundings and provide crucial feedback for 
the cell motion \cite{bray_book}. Philopodes are nanometric structures 
whose shape and dynamics provides {\it specific} feedback to the cell, hence 
it is hard to imagine how they could be realistically modeled by a 
coarse-grained approach such as the one discussed in this review.

This indicates that there is much room for further refinement 
and enhancement of  the techniques discussed in this work aimed towards
a new generation of LB schemes with increasing molecular/chemical specificity.

For instance, most of this work relies on the direct 
simulation of about three decades in space, say millimeters
to microns, leaving another two to three decades
to a coarse-grained formulation of sub-micron interactions. 
In the future, the availability of exascale computers will allow for the direct simulation of four spatial decades, thereby reducing the range that needs to be covered by coarse-graining.
Furthermore, the resort to machine learning can lead to
more elaborated and specificity-aware coarse-grained models,
such as those discussed in Section \ref{modernLB}, possibly combined
with high-order-lattice implementations including
statistical fluctuations and higher-order pre-hydrodynamic effects.

The concurrent progress of the three lines of 
development mentioned above is expected to take the next 
generation of specificity-aware LB models for 
soft flowing matter down to the ten nanometers 
scale, thus greatly facilitating the coupling to microscopic models.

Substantial work shall be needed to develop a robust
and efficient specificity-aware LB models, but the 
prospects look decidedly bright and the range of applications 
extremely rich and challenging.

\section*{Acknowledgements}
A.T., M.D., M.L., A.M., and S.S. acknowledge funding from the European Research Council under the European Union's Horizon 2020 Framework Program (No. FP/2014-2020) ERC Grant Agreement No. 739964 (COPMAT) and ERC-PoC2 grant No. 101081171 (DropTrack). A.T. and M.L. acknowledge the support of the Italian National Group for Mathematical Physics of INdAM (GNFM-INdAM). J.M.T thanks the FRQNT ``Fonds de recherche du Qu\'ebec - Nature et technologies (FRQNT)'' for financial support (Research Scholarship No. 314328).

\bibliography{bibliography}
\end{document}